\documentclass[fleqn,usenatbib]{mnras}
\usepackage{newtxtext,newtxmath}
\usepackage[T1]{fontenc}
\usepackage{ae,aecompl}
\usepackage{multirow}
\usepackage{lineno}

\usepackage{amsmath}	
\usepackage{arydshln} 

\usepackage{graphicx}
\usepackage{subfigure}
\newcommand\msun{{\,M_\odot}}

\newcommand\zsun{{\rm \,Z_\odot}}
\newcommand\lsun{{\,L_\odot}}
\newcommand{\unit}[1]{\ensuremath{\, \mathrm{#1}}}
\usepackage{aecompl}
\usepackage[T1]{fontenc}
\usepackage{threeparttable}
\usepackage{caption}
\usepackage{times}
\usepackage{ifpdf}

\usepackage{float}

\newcommand{\cMpc}{~\mbox{comoving}~\mbox{Mpc}}

\newcommand{\cmci}{~\mbox{cm}^{-3}}

\newcommand{\Msun}{~\mbox{M}_{\odot}}

\usepackage{tikz}
\usetikzlibrary{shapes.geometric, arrows.meta, positioning}

\tikzset{
  startstop/.style = {rectangle, rounded corners, minimum width=3.5cm, minimum height=1cm, text centered, draw=black, fill=gray!10},
  process/.style   = {rectangle, minimum width=3.5cm, minimum height=1cm, text centered, draw=black, fill=blue!10},
  decision/.style  = {diamond, minimum width=3.5cm, minimum height=1cm, text centered, draw=black, fill=orange!20},
  arrow/.style     = {thick, ->, >=Stealth}
}

\title[IMF Sampling and SN feedback Effects on UFD Galaxies]{Exploring effects of IMF Sampling and SN feedback injection on Star Formation and Metallicity in Ultra-Faint Dwarf Galaxies}

\author[M. Jeon and M. Go]{Myoungwon Jeon$^{1,2}$\thanks{E-mail:
myjeon@khu.ac.kr} and  Minsung Go$^{1}$\\
$^{1}$School of Space Research, Kyung Hee University, 1732 Deogyeong-daero, Yongin-si, Gyeonggi-do 17104, Korea\\
$^{2}$Department of Astronomy \& Space Science, Kyung Hee University, 1732 Deogyeong-daero, Yongin-si, Gyeonggi-do 17104, Korea
}

\date{Accepted XXX. Received YYY; in original form ZZZ}

\pubyear{2024}

\begin{document}
\label{firstpage}
\pagerange{\pageref{firstpage}--\pageref{lastpage}}
\maketitle

\begin{abstract}
We examine the impact of various Initial Mass Function (IMF) sampling and supernova (SN) feedback injection methods on the star formation and metal enrichment histories of Ultra-Faint Dwarf (UFD) galaxy analogs. These analogs, characterized by $M_{\rm vir}\approx10^8\msun$ and $M_{\ast}\lesssim10^{4.5}\msun$ at $z=0$, are simulated using high-resolution cosmological hydrodynamic zoom-in simulations with a gas particle mass resolution of $\sim63\msun$. We compare three approaches: the burst model, stochastic IMF sampling, and individual IMF sampling. These methods differ not only in how star particles are sampled following the IMF, but also in how SN feedback energy is injected---specifically in the degree of temporal and spatial discreteness, with the individual IMF sampling method being the most discrete SN feedback and thus the most physically realistic. Our findings indicate that, despite variations in sampling and SN feedback injection, the final stellar masses across methods are generally similar. However, star formation is notably more continuous in the individual sampling runs due to the weaker suppression from discrete SN events, which enables star formation in denser environments. This sustained star formation leads to more frequent self-enrichment of star-forming gas, resulting in stellar metallicities that are 0.2 to 0.5 dex higher in individual sampling runs compared to burst and stochastic models. These findings highlight the importance of considering both IMF sampling and SN feedback implementation when modeling the star formation and chemical evolution of UFD galaxies.

\end{abstract}




\section{Introduction}
Dwarf galaxies ($M_{\rm vir}\lesssim10^9\msun$) are considered excellent laboratories for studying galaxy formation, as they are essential components in the hierarchically evolving Lambda cold dark matter ($\Lambda$CDM) Universe (reviewed in \citealp{Simon2019}; see also \citealp{Tolstoy2009, McConnachie2012}). Unlike massive systems such as the Milky Way (MW) galaxy, low-mass dwarfs are vulnerable to baryonic feedback because of their shallow gravitational potential wells, which allow gas to escape easily into the intergalactic medium (IGM). The simplicity of ultra-faint dwarfs (UFDs) - the smallest ($M_{\ast}\lesssim10^5\Msun$) and most metal-poor ($\rm [Fe/H]<-2$) galaxies in the Universe -provides valuable insights into how stellar feedback regulates the star formation histories (SFHs) of low-mass galaxies (e.g., \citealp{Sales2022, Collins2022}).

UFDs tend to exhibit bursty star formation activity, followed by early quenching during or even before cosmic reionization, leading to relatively brief SFHs (e.g., \citealp{Bullock2000, Bovill2009, Brown2014, Weisz2014, Gallart2021, Savino2023, Azartash-Namin2024}). Specifically, photoionization heating by stars and outflows caused by supernova (SN) explosions are primarily responsible for this bursty star formation (e.g., \citealp{Simpson2013, Jeon2017, Zhang2024}). Consequently, UFDs are likely to cease star formation as the gas within their haloes evaporates, with little chance of re-infall due to global heating during reionization. Moreover, the host environment—such as tidal interactions or ram pressure stripping as UFDs fall into their host halo—also plays a crucial role in suppressing star formation (e.g., \citealp{Simpson2018, Carlin2019, Akins2021}). Thus, UFDs are more sensitive to baryonic feedback, necessitating sophisticated subgrid models to describe star formation and the accompanying stellar feedback, which are essential for accurately simulating UFD analogs.

Numerous theoretical studies have explored the formation and evolution of UFD analogs, both as satellite galaxies of the MW or M31 (e.g., \citealp{Sawala2011, Wetzel2015, Applebaum2021}) and as isolated systems (e.g., \citealp{Simpson2013, Onorbe2015, Wheeler2015, Wheeler2019, Jeon2017, Rey2019, Agertz2020, Sanati2023, Goater2024, Rey2025, Brauer2025}), using hydrodynamic simulations. In the context of assembling a MW-mass galaxy ($M_{\rm vir}\approx10^{12}\msun$), the mass of individual dark matter (DM) or baryon particles has been too large to resolve low-mass systems, including UFDs. Due to this limitation, high-resolution simulations have often been conducted for isolated UFDs, with recent advances in numerical techniques allowing for gas mass resolutions as low as a few tens of $\msun$ (e.g., \citealp{Wheeler2019}; \citealp{Agertz2020, Rey2020, Jeon2021a}). As numerical resolution improves, it becomes crucial to assess whether the subgrid recipes previously adopted remain valid (e.g., \citealp{Revaz2016, Hu2017, Su2018, Emerick2019, Applebaum2020, Andersson2023, Brauer2025}).

In most galaxy simulations to date, star particles are typically represented as single stellar populations (SSPs), with stars distributed according to a specified initial mass function (IMF) (e.g., \citealp{Katz1992, Navarro1993}). This method averages the effects of stellar feedback across the IMF, influencing the interstellar medium (ISM) and the subsequent evolution of the galaxy. Although this approach is effective for low-resolution simulations of larger galaxies, it may not be suitable for high-resolution simulations aimed at resolving UFDs, where gas particle masses range from a few tens to a thousand solar masses ($10-1000\msun$). As noted by \citet{Revaz2016}, stars cannot fully populate an IMF when the gas particle mass is below $1000\msun$ (also see e.g., \citealp{Hu2017}). This creates a significant challenge when stars form from low-mass gas particles: the number of SNe occurring within a single numerical timestep might be less than one. As a result, the energy from a single SN event is distributed over multiple timesteps, significantly reducing the effectiveness of SN feedback compared to scenarios where SN energy is released explosively in a single timestep (e.g., \citealp{Stinson2007, Durier2012}).

Beyond the conventional IMF-averaged approach, several alternative methods have been developed to manage SN energy release during the formation of low-mass galaxies. One such method is the burst model, which addresses the previously mentioned issue by releasing all the SN energy from an SSP simultaneously into the surrounding environment while still continuously distributing stars according to a given IMF. This ensures a substantial impact of SN explosions on the ISM, although it might be excessively intense, potentially impeding further star formation. An alternative approach involves maintaining a continuous IMF distribution while utilizing a stochastic method to sample SN based on a time-dependent SN rate (e.g., \citealp{Stinson2010, Hopkins2014, Revaz2016, Smith2018}). This method calculates the expected SN rate as a function of time, applying time delays as stars evolve along the main sequence, and stochastically triggers individual SNe. As a result, the number of high-mass stars that end as SNe varies for each star particle, affecting the intensity of SN feedback and shaping the SFHs of UFD analogs. This approach is particularly effective for simulating UFD analogs because it reflects the discrete nature of SN events, allowing individual explosions with time delays during a single starburst, and thus releases SN energy in a discrete manner.

Recently, with enhanced resolution and star particle masses under ($m_{\ast}<100\msun$), a novel IMF sampling method has been proposed, involving the individual extraction of stars from specified IMFs. In this approach, each massive star capable of becoming an SN is represented by a single star particle (e.g., \citealp{Emerick2019, Gutcke2021, Andersson2023, Deng2024, Brauer2024}). Unlike the SSP method, this allows individually formed star particles to better reflect the specific conditions of the gas, such as the immediate aftermath of previous SNe and the metallicity of the gas.

In our previous papers (\citealp{Jeon2017, Jeon2021a, Jeon2021b}), we studied the formation of dwarf galaxies ($M_{\rm vir}\approx10^8-10^9\msun$) by conducting cosmological hydrodynamic zoom-in simulations, tracking their evolution from the time of the first stars, known as Population III (Pop III) stars, to the present day ($z=0$). These simulations incorporated crucial baryonic processes, including SN feedback and the effects of reionization. For IMF sampling, we adopted the standard approach, in which stellar masses are continuously assigned according to the IMF within a single SSP. SN feedback was implemented using the burst model, where all SN energy from the SSP is released into the ISM simultaneously, without any delay time. This significantly affected the surrounding environment and potentially suppressed subsequent star formation excessively. Despite this, we found that the physical characteristics of the simulated dwarfs (i.e. $M_{\rm vir}-M_{\ast}$ and $M_{\ast}-\rm [Fe/H]$ relationships) closely matched those observed in dwarf spheroidals (dSphs) within the MW galaxy. However, as mentioned previously, the physical properties of the simulated dwarfs can vary depending on IMF sampling and SN feedback injection methods.

While the choice of these methods plays a critical role in high-resolution simulations, there has been limited research comparing various IMF sampling methods in the context of dwarf galaxy formation (see, e.g., \citealp{Revaz2016, Hu2017, Su2018, Applebaum2020}). For example, \citet{Applebaum2020} explored two IMF sampling strategies, stochastic IMF sampling and the IMF-averaged method, by implementing them in cosmological hydrodynamic simulations of MW-like galaxies and isolated dwarf galaxies at $z=6$. They suggested that the discrete SN feedback from stochastic IMF sampling tends to induce more bursty star formation compared to simulations that adopt continuous SN feedback from a fully populated IMF. This can cause earlier suppression of star formation, especially in smaller dwarf galaxies ($M_{\rm vir}\lesssim10^{8.5}\msun$), leading to lower stellar masses in dwarf galaxies. In addition, \citet{Smith2021a} investigated the impact of SN feedback, photoionization heating by stars, and photoelectric heating of dust grains by assuming two different models: stochastic IMF and IMF-averaged rate. They simulated idealized isolated dwarf galaxies with $M_{\rm vir}\sim10^{10}\msun$ and emphasized that the choice of IMF sampling method has a significant effect on the results, particularly when considering photoionization heating by stars.

Stellar metallicity is another crucial property of UFD analogs, which is also influenced by the method of star sampling and the way SN energy is distributed into the ISM. High-resolution simulations of UFD analogs show that it is challenging to evolve a galaxy to match the observed stellar mass-metallicity relation (MZR) (e.g., \citealp{Kirby2013}). For example, \citet{Agertz2020} suggested that the lower metallicity compared to observational estimates could be due to the powerful SN feedback that disperses a metal-rich gas or the lack of metal enrichment from Pop~III stars. On the other hand, the inability of any subgrid model to reproduce the observed MZR in the UFD regime could be due to environmental effects (e.g., tidal stripping), which are generally not accounted for in high-resolution simulations. However, \citet{Applebaum2021} argued that there is no environmental difference in the MZR by comparing satellite UFDs and near-field UFDs in MW-like simulations. They predict a lower metallicity than the observed MZR at low luminosity ($\lesssim10^4\lsun$) and explain this by the long timescale of Type Ia SNe, which delays the release of iron, while star formation quenching occurs too early, leading to stars with low $\rm [Fe/H]$ values.

In this study, we explore the combined impact of varied IMF sampling and corresponding SN feedback implementation on the evolution of UFD analogs using cosmological zoom-in simulations of UFD analogs ($M_{\rm vir}\approx10^{8}\msun$) with enhanced resolution, characterized by a gas particle mass of $m_{\rm gas}\approx63\msun$. We specifically compare the unique properties of the simulated UFD analogs employing three different IMF sampling and SN injection methods: the burst model, stochastic IMF sampling, and individual IMF sampling. Additionally, we examine how the cosmic metal enrichment histories of the UFD analogs vary depending on the chosen methods. The paper is organized as follows. In Section~\ref{sec:2}, we describe the numerical methodology used. In Section~\ref{sec:3}, we present and discuss the detailed simulation results, focusing on star formation and the history of metal enrichment, and compare our findings with other relevant studies in Section~\ref{sec:4}. The main conclusions of this work are summarized in Section~\ref{sec:5}. Unless otherwise stated, all distances are provided in physical (proper) units for consistency.

\begin{table}
\centering
\begin{tabular}{c c c c c} 
\hline
Halo & $M_{\rm vir}$ & $M_{\ast}$ & $\rm \langle[Fe/H]\rangle$ & IMF \\
\hline
{\sc halo1-Burst}  & 8.0 &  14.6 & -3.05 & {\sc Burst} \\
\hline
{\sc halo1-Simf}  & 8.0 &  12.8 & -3.19 & {\sc Simf} \\
\hline
{\sc halo1-Indiv}  & 8.0 &  12.2 & --2.79 & {\sc Indiv} \\
\hline
{\sc halo2-Burst}  & 5.0 &   10.7 & -3.16 & {\sc Burst} \\
\hline
{\sc halo2-Simf}  & 5.0 &   8.2 & -3.35  & {\sc Simf} \\
\hline
{\sc halo2-Indiv}  & 5.0 &   6.4 & -3.04  & {\sc Indiv} \\
\hline
{\sc halo3-Burst}  & 2.0 &   0.7 & -2.75 & {\sc Burst} \\
\hline
{\sc halo3-Simf}  & 2.0 &   0.7 & --3.01 & {\sc Simf} \\
\hline
{\sc halo3-Indiv}  & 2.0 &  0.1 & -2.84 & {\sc Indiv} \\
\hline
\end{tabular}
\caption{Physical characteristics of the simulated UFD analogs at $z=0$. Column (1): Run name. Column (2): Viral mass (in units of $10^8\msun$ at $z=0$). Column (3): Stellar mass (in $10^3\msun$). Column (4): Average stellar iron-to-hydrogen ratios. Column (5): IMF sampling method.}
\label{table:simul}
\end{table}

\section{Numerical methodology}
\label{sec:2}

\subsection{Simulation Setup}
\label{sec:2.1}
We carry out a suite of cosmological hydrodynamic zoom-in simulations using a significantly extended version of the N-body smoothed particle hydrodynamics (SPH) code GADGET (\citealp{Springel2001, Springel2005, Schaye2010}). The $\Lambda$CDM cosmological parameters used in this study are: a matter density parameter $\Omega_{\rm m}=1-\Omega_{\Lambda}=0.265$, baryon density $\Omega_{\rm b}=0.0448$, present-day Hubble expansion rate $\rm H_0 = 71\unit{km\, s^{-1} Mpc^{-1}}$, spectral index $n_{\rm s}=0.963$, and normalization $\sigma_8=0.8$ (\citealp{Komatsu2011,planck2016}). We create the initial conditions by running the cosmological initial conditions code MUSIC (\citealp{Hahn2011}). Each target halo with a virial mass $M_{\rm vir}\approx10^8\msun$ at $z=0$ is identified from a preliminary DM-only simulation carried out with $128^3$ particles in a periodic box of linear size $L=3.125 \ h^{-1} \cMpc$ box. We then apply four consecutive refinements to the target halo region, confined by $3\ R_{\rm vir}$, where $R_{\rm vir}$ is the virial radius at $z=0$, giving rise to DM and gas masses of $m_{\rm DM}\approx500\msun$ and $m_{\rm gas}\approx63\msun$, respectively, in the most refined area. The softening lengths for the DM and the star particles are fixed $\epsilon_{\rm DM}\approx20$ pc at all redshifts. For gas particles, we use an adaptive softening length proportional to the SPH kernel length, with a minimum value of $\epsilon_{\rm gas, min}=2.8$ pc.

The baryonic physics used in this work is similar to that used in \citet{Jeon2017}. For detailed descriptions, we refer the reader to that paper, but we provide a brief overview of the critical implementations here. Notably, in this study, we further revise the star formation and SN feedback modules to utilize the stochastic and individual IMF sampling methods, as described in detail in Section~\ref{sec:2.3.2} and Section~\ref{sec:2.4.2}. To investigate the effects of IMF sampling on the evolution of UFD analogs, we performed three sets of comparison simulations. Each pair of runs starts from the same initial condition, targeting one of the UFD analogs ({\sc Halo~1}, {\sc Halo~2}, and {\sc Halo~3}), but adopting three different IMF sampling methods, resulting in a total of nine runs. To assess the impact of stochasticity, we performed two additional runs with different random seeds for each halo and each method, increasing the total number of simulations to 27. However, our discussion primarily focuses on the main set of 9 simulations. We designate the runs with burst modeling, stochastic IMF sampling, and individual IMF sampling as {\sc Burst}, {\sc Simf}, and {\sc Indiv}, respectively. Key physical characteristics of the simulated UFD analogs at $z=6$ are listed in Table~\ref{table:simul}.

\begin{table}
    \centering
    \setlength{\tabcolsep}{2pt} 
    \small 
    \begin{tabular}{c c c c} 
        \hline
        \textbf{Name} & \textbf{{\sc Burst}} & \textbf{{\sc Simf}} & \textbf{{\sc Indiv}} \\
        \hline
        \textbf{IMF Sampling} & Averaged & Stochastic & Individual \\
        \hline
        \multirow{4}{*}{\textbf{Form as}}
         & & & SSP-NoSN ($63\msun$) \\
         & SSP ($500\msun$) & SSP ($500\msun$) & [$0.1-8]\msun$ \\
        \cdashline{4-4}[0.5pt/1pt] 
         & $[0.1-100]\msun$ & [$0.1-100]\msun$ & Individual star\\
         & & & [$8-100]\msun$ \\
        \hline
        \multirow{2}{*}{\textbf{$Z_{*}$}} & \multirow{2}{*}{Uniform} & \multirow{2}{*}{Uniform} & \multirow{1}{*}{Uniform} \\ \cdashline{4-4}[0.5pt/1pt]
        & & & \multirow{1}{*}{Individual} \\
        \hline
        \multirow{2}{*}{\textbf{SN feedback}} & All at once & Individually & Individually \\
        & ($3.4\times10^{51}$ erg) & ($10^{51}$ erg) & ($10^{51}$ erg) \\
        \hline
        \multirow{2}{*}{\textbf{Delay Time}} & \multirow{2}{*}{Immediate} & Delayed & \multirow{2}{*}{Immediate} \\
         & & by lifetime & \\
        \hline
        \textbf{Number of SNe} & \multirow{2}{*}{5} & \multirow{2}{*}{0 -- 10} & \multirow{2}{*}{1} \\
        \textbf{per star paticle} &  &  &  \\
        \hline
    \end{tabular}
    \caption{Comparison of the properties of the {\sc Burst}, {\sc Simf}, and {\sc Indiv} methods.}
    \label{tab:imf_methods}
\end{table}

\begin{figure*}
  \centering
  \includegraphics[width=180mm]{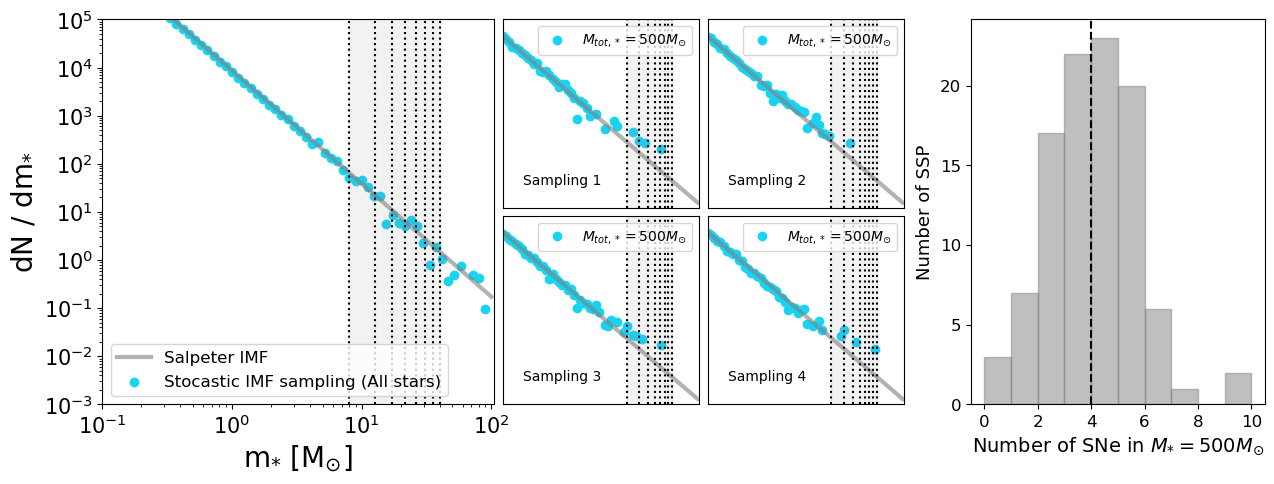}%
   \caption{Reconstructed IMF of stars using stochastic IMF sampling. {\it Left-most panel:} The IMF of all Pop~II stars from the {\sc Halo1-Simf} run shows good agreement with the assumed Salpeter IMF. For efficient SN modeling (indicated by the gray-shaded region), we store information only for stars within the mass range of $8-40\msun$, divided into eight logarithmic bins indicated by the vertical dotted lines. {\it Middle four panels:} Example of star sampling from a single SSP ($m_{\rm Pop II}=500\msun$), illustrating the stochastic nature in the number of Type~II SN progenitors. {\it Right-most panel:} The distribution of the number of SNe within a single Pop~II cluster, showing that the most probable number of Type~II SN progenitors is 4, with an average of 3.7, and a maximum reaching 10.}
   \label{fig1}
\end{figure*}

\subsection{Chemistry, Cooling and UV background}
\label{sec:2.2}
We solve the coupled, non-equilibrium rate equations for primordial species ($\rm H, H^{+}, H^{-}, H_{2}, H^{+}_2, He, He^{+}, He^{++}, e^{-}, D, D^{+}$, and HD) at each timestep, accounting for all relevant primordial cooling processes (e.g., H and He collisional ionization, excitation and recombination cooling, bremsstrahlung, inverse Compton cooling, and collisional excitation cooling of $\rm H_{2}$ and HD). In addition, we incorporate gas cooling by metal species such as carbon, oxygen, silicon, magnesium, neon, nitrogen, and iron, with cooling rates determined using tables from the photoionization package CLOUDY \citep{Ferland1998}. These cooling tables are constructed element by element, rather than using only the total metallicity. In our implementation, following \cite{Glover2007}, only carbon, oxygen, and silicon species are included in the metal cooling network, even though we track seven elements (C, N, O, Mg, Si, Ne, and Fe). We also consider the photodissociation of molecular hydrogen ($\rm H_{2}$) and deuterated hydrogen (HD) by Lyman-Werner (LW) radiation. To give an effect of reionization, we introduce a cosmic UV background (\citealp{Haardt2012}) starting at $z=7$, and linearly increase its strength until $z=6$, which corresponds to the expected completion of cosmic reionization (e.g., \citealp{Gunn1965, Fan2006}).

\subsection{Star formation}
\label{sec:2.3}
When the gas density exceeds the critical threshold of $n_{\rm th}=100 \cmci$, the gas particles become eligible for star formation. According to the Schmidt law \citep{Schmidt1959}, star formation occurs stochastically at a rate described by $\dot{\rho}_{\ast}=\rho/\tau_{\ast}$, where $\tau_{\ast}=\tau_{\rm ff}/\epsilon_{\rm ff}$ is the star formation timescale. Here, $\tau_{\rm ff}=[3\pi/(32G\rho)]^{1/2}$ denotes the free-fall time at a density $\rho$, and $\epsilon_{\rm ff}$ represents the star formation efficiency per free-fall time. In this study, we assume a star formation efficiency of $\epsilon_{\rm ff}\sim0.01$ for both Pop III and Pop II stars, which is typical for star formation efficiency in the local Universe (e.g., \citealp{Leroy2008}). We should emphasize that the efficiency must be adjusted according to how stars are sampled, even though most UFD galaxy simulations do not consider these factors. In this study, we use the same star formation efficiency across three different methods, but, as discussed in detail in Section~\ref{sec:5}, caution is needed when considering efficiency based on the star sampling method.

During each numerical timestep $\Delta t$, gas is converted into a collisionless sink particle only if a randomly generated number between 0 and 1 is less than the smaller of $\Delta t/\tau_{\ast}$ and 1. In our simulations, we find that the typical timestep value is about 0.01 Myr, with a maximum timestep value of about 0.5 Myr, ensuring that $\Delta t \ll \tau_{\ast}$ even in the densest star-forming regions ($n_{\rm H}\sim10^{3}\ \mathrm{cm^{-3}}$). Consequently, the star formation timescale is expressed as:
\begin{equation}
\tau_{\ast}=\frac{\tau_{\rm ff} (n_{\rm H})}{\epsilon_{\rm ff}}\sim400 {\rm Myr} \left(\frac{n_{\rm H}}{100 \cmci}\right)^{-1/2}.
\end{equation}

\subsubsection{Pop III star}
\label{sec:2.3.1}
It is well known that, due to the lack of metals, Pop III stars typically have masses a few 10 - 100 times greater than those of stars in the present-day Universe (e.g., \citealp{Abel2000, Bromm2001a}). However, the exact final masses of Pop III stars are still a matter of ongoing debate (e.g., \citealp{Bromm2013, Hirano2014, Hirano2015, Klessen2023}). For instance, \citet{Hirano2014} conducted 2-D hydrodynamic simulations of several hundred cases of Pop III star formation and suggested that their masses could vary significantly, from a few tens of solar masses to $\sim$1000 $\msun$, depending on their formation environment. In our study, we take advantage of our high mass resolution to represent Pop III stars as individual star particles, unlike the approach in \citet{Jeon2017}. However, this work primarily aims to examine how different IMF sampling methods for Pop II stars affect the evolution of UFD analogs. Therefore, to minimize the random effects associated with the mass variability of Pop III stars, we fixed the Pop III star mass at $m_{\rm PopIII}=20\msun$. We note that we do not enforce strict mass conservation when converting a $63 M_{\odot}$ gas particle into a $20 M_{\odot}$ Pop III star particle, resulting in a residual mass difference of $43 M_{\odot}$ per event. However, the overall impact on the mass budget is negligible because Pop III stars contribute to less than 2\% of the total stellar mass. This rarity arises from the rapid transition in star formation from Pop III to Population II (Pop II) stars. Each SN triggered by a Pop III star rapidly enriches the surrounding medium with metals, quickly initiating Pop II star formation and thereby halting further Pop III star formation.

\subsubsection{Pop II star - Burst and SIMF}
\label{sec:2.3.2}
Pop~II stars emerge from gas clouds enriched with metals expelled by earlier Pop~III SNe. Although the density threshold for Pop~II star formation is identical to that of Pop~III stars, an additional criterion requires that the gas possesses a metallicity greater than the critical level $Z_{\rm crit}=10^{-5.5}\zsun$, as suggested by dust-continuum cooling (e.g., \citealp{Omukai2000, Schneider2010, Safranek2016}). In both the {\sc Burst} and the {\sc Simf} runs, we assume that Pop~II stars form as stellar clusters with a mass of 500$\msun$. Therefore, once a gas particle satisfies the two conditions, $n_{\rm th}$ and $Z_{\rm crit}$, it is replaced by a sink particle that accretes surrounding gas until reaching $m_{\rm PopII}=500\msun$, meaning that the seven nearest gas particles lose all of their mass instantaneously. Note that when a star particle forms, it inherits the metallicity of the gas from which it was created.

For Pop~II stars, we use the Salpeter IMF (\citealp{Salpeter1955}), which is defined by the form $dN/d\log m \approx m^{-\alpha}$, with a slope $\alpha=1.35$ over the mass range of $[0.1-100]\msun$. In the {\sc burst} approach, once a Pop~II stellar cluster is identified, we treat a star particle as a SSP where stars are continuously distributed according to the given IMF. In {\sc Simf} runs, on the other hand, stochastic IMF sampling is performed immediately upon the formation of a Pop~II cluster by extracting individual stars from the given IMF. Specifically, we utilize the inverse transform sampling method using the IMF function as the probability density function. The sampling continues until the cumulative mass of the extracted star masses reaches the mass of the stellar cluster, $m_{\rm PopII}=500\msun$ (e.g., \citealp{Haas2010}; \citealp{Applebaum2020}). 

However, due to the stochastic nature of the sampling, it is not always possible to match the target mass of 500$\msun$ exactly. To address this, we adopt the stop-nearest method introduced by \cite{Haas2010}. In this scheme, if the addition of the last sampled star causes the cumulative mass to exceed 500$\msun$, we compare the deviations from the target with and without the star, and retain the option that is closer to 500$\msun$. This procedure introduces only minor fluctuations of the final stellar particle mass around the target value, thereby mitigating systematic mass non-conservation over multiple sampling events. Fig.~\ref{fig1} shows the IMFs of all stars from the {\sc Halo1-Simf} run on the left, demonstrating good agreement with the assumed Salpeter IMF. Among the stars sampled, we only retain information for stars within the mass range of $8-40\msun$ (indicated as the shaded region in Fig.~\ref{fig1}), since these correspond to the masses of Type~II SN progenitors (\citealp{Heger2003}). This mass range is further divided into eight bins, denoted by dotted vertical lines in Fig.~\ref{fig1} for the sake of efficient SN explosion modeling.

We note that, when applying stochastic IMF sampling, the number of SN progenitors from a SSP with $m_{\rm PopII}=500\msun$ varies due to randomness. This variability is illustrated in the four middle small panels of Fig.~\ref{fig1}. For instance, in the first sampling shown in Fig.~\ref{fig1}, there are 4 SN progenitors, while in the second sampling, the number drops to 2. The rightmost panel of Fig.~\ref{fig1} presents a histogram of the number of SNe in the SSP. It reveals that the most probable number of Type~II progenitors is 4, with an average of 3.7 per Pop~II cluster of $500\ M_\odot$ and the maximum number of stored stars can reach 10. In an extreme case, there could be no progenitors, meaning that the star cluster would not experience SN events. For comparison, the {\sc Burst} model assumes a fixed number of 3.4 SN events for the same cluster mass ($500 M_{\odot}$). Thus, the {\sc SIMF} runs inject around 9\% more SN energy per unit stellar mass than the {\sc Burst} model. However, this small difference in the injected energy budget does not affect our main conclusions, given the relative trends among the IMF sampling methods.

\subsubsection{Pop II star - Indiv}
\label{sec:2.3.3}

\begin{figure}
\centering
\includegraphics[width=80mm]{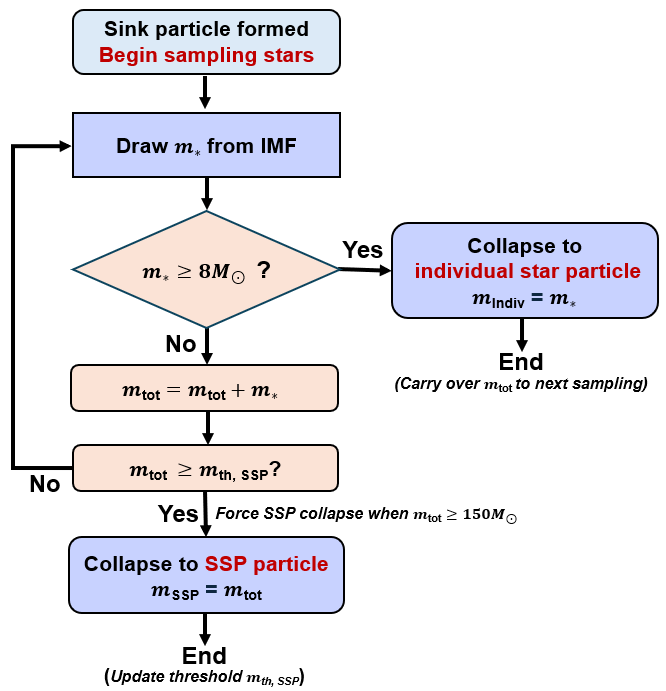}%
\caption{Flow chart of IMF sampling method for {\sc Indiv} run.}
\label{fig:imf_flowchart}
\end{figure}

In the individual sampling method, the goal is to model massive stars ($\gtrsim 8\msun$) as individual particles, thereby allowing discrete SN events. To achieve this, we employ two types of star particles: (1) a SSP particle representing a population of low-mass stars, and (2) an individual star particle representing a single massive star above 8$\msun$. Fig.~\ref{fig:imf_flowchart} illustrates the flowchart of this sampling procedure. The process begins when a gas particle is converted into a sink particle. During sampling, if a massive star ($\gtrsim 8\msun$) is drawn, the sink particle immediately collapses into a star particle with $m_{\rm star}=m_{\rm *, \ Indiv}$. 
Simultaneously, the cumulative mass of sampled low-mass stars up to that point is carried over to the next sampling sequence for the subsequently formed star particle.

To conserve mass between gas and star particles, we introduce the threshold mass, $m_{\rm th, \ SSP}$. The sink particle is converted into a SSP particle only when the total mass of sampled low-mass stars exceeds this threshold, rather than a fixed value of 63$\msun$. The motivation for this adaptive threshold is that the collapse of an individual massive star from gas particle into star particle---typically with $m_{\rm *, \ Indiv} < m_{\rm gas}$---breaks mass conservation between star and gas particles. Thus, at each star formation of an individual star particle, the threshold mass is updated according to 
\begin{equation}
m_{\rm th,\ SSP} = m_{\rm th,\ SSP} + (63 - m_{\rm Indiv}),
\end{equation}
where $m_{\rm th, \ SSP}$ is initially set to the gas particle mass ($m_{\rm gas}=63\msun$)\footnote{When $m_{\rm Indiv} \gtrsim 63\msun$, the threshold mass for the subsequent SSP can fall below the nominal gas particle mass. Note also that $m_{\rm th,\ SSP}$ is not a constant but varies each time an individual star forms.} However, repeated sampling of massive stars can cause $m_{\rm th, \ SSP}$ to diverge, making it impossible for the accumulated low-mass stars to reach the required threshold. To prevent this, we set an upper limit, $m_{\rm limit, \ SSP}=150\msun$, above which the accumulated mass is forced to collapse into a SSP particle. Consequently, the {\sc Indiv} method produces SSP particles with non-uniform masses.

\begin{figure}
\centering
\includegraphics[width=80mm]{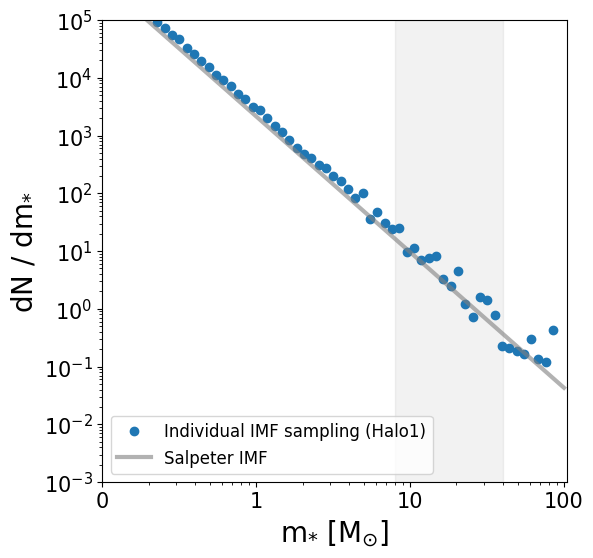}%
\caption{The derived IMF of stars adopting individual IMF sampling. The data from the {\sc Halo1-indiv} run, depicted as blue circles, shows a strong agreement with the assumed IMF (grey solid line).}
\label{fig:imf_indiv}
\end{figure}

Despite our efforts to conserve mass, the method does not perfectly achieve mass conservation because star formation can cease before the remaining low-mass stars---those sampled but not yet collapsed---are fully converted into SSP particles. For example, if 10 gas particles are converted into stars, the total stellar mass should be $63\times10=630 \msun$. However, in practice, the total mass of the formed stars could be around 550$\msun$, with the remaining 80$\msun$ left as remnants, waiting to collapse during the sampling process. Here, the mass-loss fraction represents the discrepancy between the expected total stellar mass and the actual mass of the formed stars. In this example, it would be calculated as $(630-550)/630 \times 100=12.7\%$. Nonetheless, the resulting mass-loss fraction is negligible in most cases, and the overall evolutionary trends are preserved. Across the three halos, the degree of mass conservation decreases as the total stellar mass decreases: the mass-loss fraction increases from $\sim4$\% in {\sc Halo1} to $\sim$44\% in {\sc Halo3}. Fig.~\ref{fig:imf_indiv} shows the resulting IMF from the {\sc Indiv} run of {\sc Halo1}, where the sampled individual stars, depicted as blue circles, are included. The distribution confirms that the method successfully reproduces the expected IMF shape, shown as the gray solid line in Fig.~\ref{fig:imf_indiv}, while explicitly resolving individual massive stars.


Furthermore, to ensure consistency within each halo, stars formed in the same region are sampled from a common IMF realization. To achieve this, we pre-generate arrays of stellar masses following our IMF sampling scheme and assign them sequentially to star particles during simulations. During runtime, the spatial information of newly formed stars is used to maintain locality in sampling: if a star forms within $d_c = 1.5$ kpc of any previously formed Pop~II star particle, it is assigned to the same sampling group and shares the same pre-generated mass array; otherwise, it initiates a new progenitor halo group. The choice of $d_c = 1.5$ kpc reflects the typical separation between halos in our simulations at high redshift ($z > 6$), and thus effectively distinguishes stars belonging to different progenitor halos. This approach guarantees that stars within a progenitor halo are sampled coherently, without contamination from neighboring halos, while ensuring that the resulting stellar populations follow the intended IMF. Stars with masses between $8-40~M_\odot$ are designated as Type II SN progenitors, whereas those between $40-100~M_\odot$ are assumed to collapse directly into black holes without feedback.

\subsection{Stellar feedback}
\label{sec:2.4}
During a SN explosion, energy is transferred as thermal energy to the nearest gas particles, significantly increasing their temperature to levels that exceed $T=10^6-10^8$ K. However, this energy can quickly dissipate if the numerical resolution is inadequate to accurately model the evolution of the SN ejecta (e.g. \citealp{Stinson2007}). To tackle this well-known overcooling issue, we adopt the numerical method proposed by \citet{Vecchia2012}, which reduces the number of neighboring gas particles to achieve a temperature increase of $\Delta T>10^{7.5}$ K, a necessary condition for effective SN feedback. In our implementation, the strategy for coupling energy varies according to the feedback method employed. In the {\sc Indiv} and {\sc Simf} runs, each SN transfers its thermal energy to the single nearest gas particle. In contrast, in the {\sc burst} run, the total energy of an SSP, which amounts to $3.4 \times 10^{51}$ erg, is evenly distributed among three neighboring particles, with each receiving approximately $1.13 \times 10^{51}$ erg. This method ensures that each gas particle receives a consistent amount of energy, allowing for a direct comparison of IMF sampling and the associated feedback effects across all methods. Furthermore, we utilize a timestep limiter and adjust the timestep to ensure that the ratio of the surrounding SPH particles does not exceed 4 (e.g. \citealp{Saitoh2009}), allowing the gas particles to respond appropriately to the sudden input of SN energy (e.g., \citealp{Vecchia2012}).

\subsubsection{Pop~III stars}
\label{sec:2.4.1}
The ultimate fates of Pop~III stars are primarily determined by their initial mass (e.g., \citealp{Heger2010, Yoon2012}). For example, metal-free stars with masses ranging from $10\msun \lesssim m_{\ast} \lesssim 40\msun$ are anticipated to undergo conventional core-collapse supernovae (CCSNe), whereas those with masses between $140\msun$ and $260\msun$ are likely to result in pair-instability supernovae (PISNe). In this study, with the fixed mass of Pop~III stars set at $m_{\rm Pop~III}=20\msun$, the SN explosion energy for an individual Pop~III star is established at $E_{\rm CCSN}=10^{51}$ erg for a CCSN. Also, we utilize the metal yields of Pop~III SNe provided by \citet{Heger2002, Heger2010}. Given that the mass resolution of this work is high enough to represent an individual massive Pop~III star, the SN explosion and the release of metals are treated as a single event whenever a Pop~III SN is triggered. 

\begin{figure}
  \centering
  \includegraphics[width=80mm]{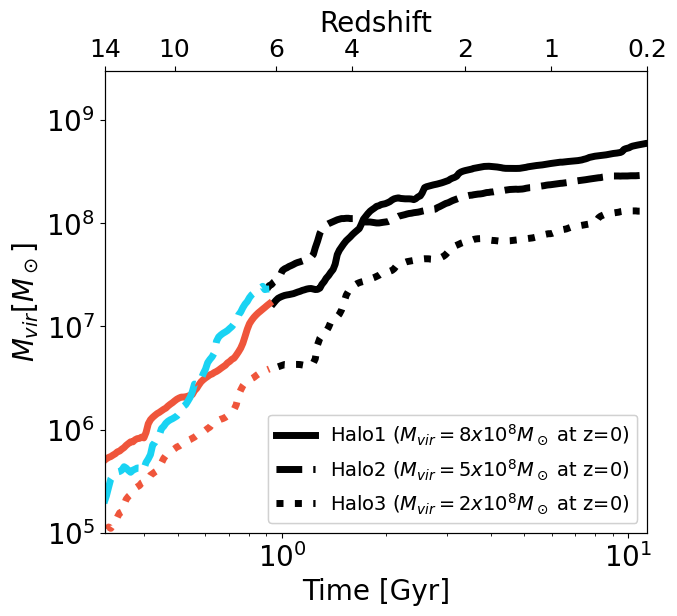}%
   \caption{Dark matter assembly histories of the simulated UFD analogs. Initially, all haloes begin as minihaloes with masses around $M_{\rm halo} \approx 10^{5} \msun$ at redshifts $z \gtrsim 14$. Then, they gradually grow to reach masses of about $M_{\rm halo} \approx 10^{8} \msun$ by redshift $z = 0$.}
   \label{fig4}
\end{figure}

\subsubsection{Pop~II stars}
\label{sec:2.4.2}
For the stellar feedback associated with Pop~II stars, we consider conventional Type~II SN, Type Ia SN, and Asymptotic Giant Branch (AGB) feedback. However, the most critical factor influencing the evolution of UFDs might be the Type~II SNe from massive Pop~II stars. As previously mentioned, this study seeks to examine whether the evolution of UFD analogs can vary depending on the IMF sampling method employed and the associated SN feedback. To address this question, we use three distinct approaches for Type~II SN explosions.

First, under the {\sc Burst} approach, the total energy of SNe is released simultaneously into the surrounding medium during simulations. This method has been used in our previous studies (e.g. \citealp{Jeon2017, Jeon2021a, Jeon2021b}). We define the energy of a single SN as $E_{\rm SN, single}=10^{51}$ erg. Therefore, the total energy is calculated as $E_{\rm SN, total}=N_{\rm SN}\times E_{\rm SN, single}$, where $N_{\rm SN}$ represents the IMF-averaged number of Type II SN progenitors, about \textcolor{blue}{3.4} when using the Salpeter IMF. As a result, \textcolor{blue}{$E_{\rm SN, \ total}=3.4\times10^{51}$ erg} is deposited each time a Pop~II SSP forms in the {\sc Burst} run. The metal yields of these SN events are also averaged according to the IMF and released simultaneously, based on metallicity-dependent yield tables (\citealp{Portinari1998}). It is important to note that, due to the absence of photoionization heating by stars, the thermal energy and metals from the SN events are designed to be deposited immediately after the formation of Pop~II stars. This approach prevents the overproduction of stars during their main-sequence phase.

Secondly, in simulations with stochastic IMF sampling, the SN energy is released through discrete events rather than in a single burst. As detailed in Section~\ref{sec:2.3.2}, after randomly sampling Pop~II stars from a specified IMF, we retain only the information of massive stars, categorized into eight mass bins. If the age of a selected star surpasses the main-sequence lifetime of its progenitor, a CCSN event is triggered, releasing thermal energy and metals into the surrounding gas particles. Consequently, multiple SN explosions can occur individually within a single stellar cluster, starting with the explosion of the most massive Type~II SN progenitor. Each SN releases an energy of $E_{\rm SN, single}=10^{51}$ erg. The lifetime and stellar yields of each progenitor are calculated on the basis of their mass and metallicity, using the information of the high-mass star previously stored through IMF sampling (\citealp{Portinari1998}). 

We should mention that in the {\sc Burst} and {\sc Indiv} runs, SN feedback is applied instantaneously upon star formation to compensate for the lack of photoionization heating effects. To enable a meaningful comparison with these runs, we implemented the {\sc Simf} runs with no delay time for stars in the most massive bin, which have the shortest lifetimes. This means that the first star to explode as an SN is triggered immediately after star formation, with no delay between star formation and SN feedback. Subsequent stars, however, explode with a time delay corresponding to their respective lifetimes. This adjustment allows us to isolate the effects of discrete temporal SN feedback in comparison with the other two methods. As in the other runs, {\sc Simf} also omits photoionization heating for massive stars that explode after a delay. We discuss the effect of delay time in Section~\ref{sec:4}.

Third, using individual IMF sampling, a Type~II SN event happens whenever an individual massive Pop~II star ($8-40\msun$) forms, releasing an energy of $E_{\rm SN, single}=10^{51}$ erg along with the corresponding metals into the ISM. Unlike the {\sc Burst} and {\sc Simf} methods, which typically result in 3-4 SN events from a single SSP when Pop II star formation takes place, this individual sampling approach allows prior explosions to regulate star formation more consistently, determining whether subsequent explosions will occur. This means that, although the {\sc Simf} method also considers individual SN explosions, multiple events are set to occur regardless of whether the surrounding medium has already been regulated.

\begin{figure}
  \centering
  \includegraphics[width=85mm]{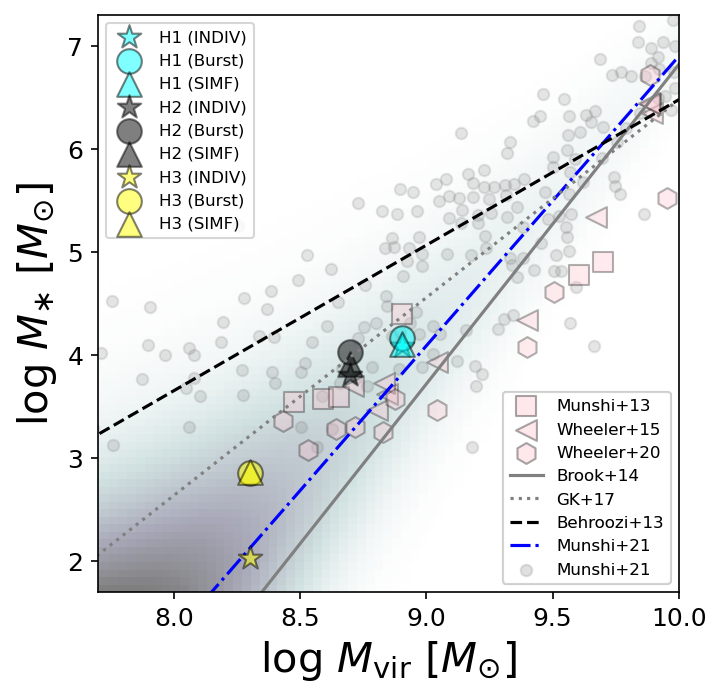}%
   \caption{The relationship between stellar mass and virial mass is depicted for three simulated galaxies, utilizing three distinct IMF sampling methods. These methods are represented by different symbols: circles for the {\sc Burst} runs, triangles for the {\sc Simf} runs, and stars for the {\sc Indiv} runs. Generally, the stellar mass estimates are broadly similar across the three methods, with no clear systematic differences among the {\sc Burst}, {\sc Simf}, and {\sc Indiv} runs.}
   \label{fig5_SMHM}
\end{figure}

Intermediate-mass stars, with masses ranging from $0.8 \msun$ to $8 \msun$, shed mass during their AGB phase. We use the tables from \citet{Marigo2001} to calculate the metal yields from these Pop~II stars at each timestep, releasing metals into the surrounding medium. For Type Ia SNe, which originate from progenitors with masses ranging from $3\msun$ to $8\msun$, we utilize empirical delay time distributions (e.g., \citealp{Barris2006, Forster2006}) and derive metal yields using the spherically symmetric W7 model (\citealp{Thielemann2003}). The energy released by Type Ia SN is imparted as thermal energy to nearby gas particles, although its feedback is relatively weak, since the energy is injected gradually over billions of years. Metals expelled from AGB stars and SNe are evenly distributed among 32 neighboring gas particles, resulting in metallicity $Z_i$ for each gas particle, where $Z_i = m_{\rm metal,i}/(m_{\rm gas}+m_{\rm metal,i})$. Metal transport is carried out through a diffusion scheme, with the mixing efficiency on unresolved scales determined by the physical properties at the SPH smoothing kernel scale (e.g., \citealp{Klessen2003, Greif2009}). We should note that the transfer of ejected material to neighboring gas particles increases their mass. Nevertheless, we assume a mass of $63\msun$ for gas particles when they are converted into star particles. While stars are not formed solely from gas particles with increased mass due to distributed SN remnants, such occurrences could violate mass conservation. This issue will be explored in greater detail in future studies with a more refined implementation.

\begin{figure*}
    \centering
    \includegraphics[width=175mm]{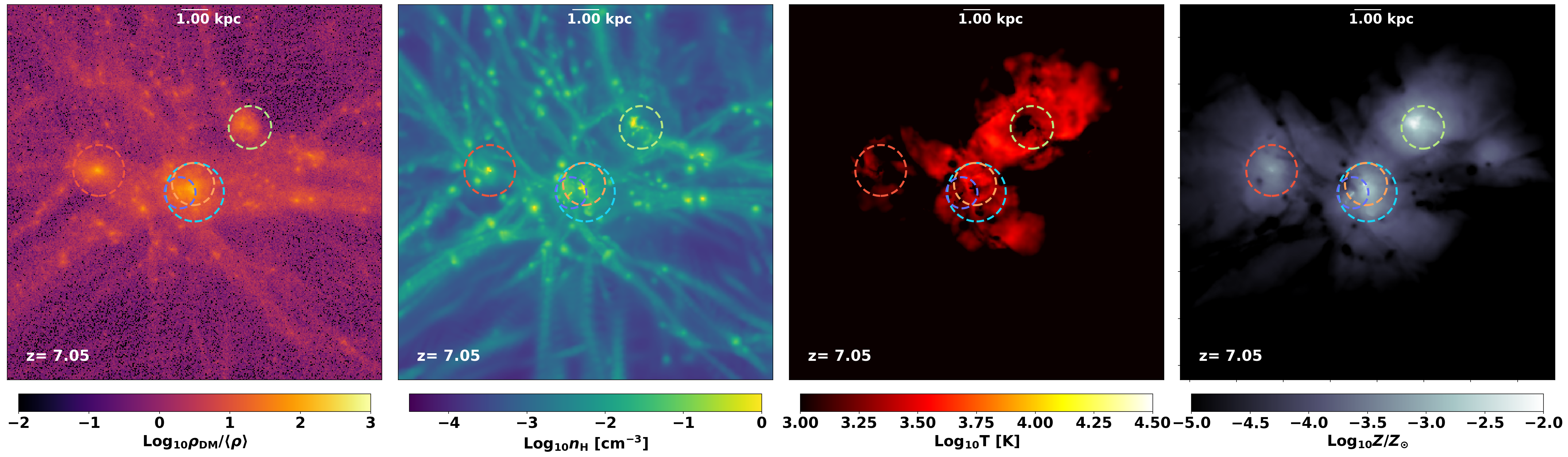}
    \caption{The morphology of the simulated UFD analog for {\sc Halo2} at $z\approx7$ is illustrated. From left to right, the panels show dark matter overdensity, hydrogen number density, gas temperature, and gas metallicity, respectively, averaged along the line of sight within a cubical region of 80 kpc comoving size for the {\sc Burst}. Colored circles indicate the virial radii of the progenitor haloes. The visualization clearly demonstrates that star formation occurs in multiple progenitor haloes rather than predominantly in the most massive main progenitor halo. This is attributed to the high resolution of our simulations, which enables the tracing of star formation from Pop~III stars in minihaloes with masses $M_{\rm vir}\approx10^{5-6}\msun$ at $z>14$.}%
    \label{fig6_morph}
\end{figure*}

\section{Results}
\label{sec:3}
This section presents the key findings from our series of simulations. In particular, Section~\ref{sec:3.1} investigates how various IMF sampling and SN injecting methods influence both the burstiness of star formation and the resulting SFHs of the simulated UFD analogs. Section~\ref{sec:3.2} analyzes the effects of different methods on the resulting metallicity, focusing on the MDF and the MZR.

\subsection{Mass assembly history}
\label{sec:3.1}
As shown in Fig.~\ref{fig4}, each target halo starts as a minihalo with a virial mass of $M_{\rm vir}=[1\sim8]\times10^5\msun$ at $z \gtrsim 14$. These haloes eventually evolve into galaxies with a total mass of $M_{\rm vir}=[2\sim8]\times10^8\msun$ and a stellar mass of $M_{\ast}\approx10^3-10^4\msun$ at $z=0$. Here, we only present the mass assembly histories of galaxies in {\sc Burst} runs, as those from the {\sc Simf} and {\sc Indiv} runs are identical. It should be mentioned that we conduct our runs until $z\approx6$, by which time star formation in these small galaxies ($M_{\rm vir}\lesssim10^9\msun$ at $z=0$) tends to cease due to global heating from reionization and internal feedback (e.g., \citealp{Jeon2017, Jeon2021a}). Despite the absence of hydrodynamic simulations down to redshift $z=0$, we can utilize data from DM-only simulations that reach $z=0$ with the same initial conditions. This enables us to identify the progenitor haloes at $z=6$ for the final assembled halo at $z=0$ in hydrodynamic simulations. Assuming that star formation has already ceased in all progenitor haloes prior to $z = 6$, we identify which stars, formed before $z = 6$, have become members of the final halo at $z=0$.

To achieve this, we first employ DM-only simulations. Specifically, we identify the progenitor haloes at $z=6$ that share more than 80\% of their DM particle IDs with those of the final halo at $z=0$. Next, we match the progenitor haloes at $z=6$ from the DM-only simulations with those from the hydrodynamic runs. This involves using the central positions and virial radii of the progenitor haloes identified at $z=6$ from the DM-only simulations to locate the corresponding haloes in the hydrodynamic runs. During this process, we confirm that an almost 100\% match of the haloes can be achieved. Finally, we analyze the star particles within the progenitor haloes identified in the hydrodynamic runs, since these stars represent the member stars of the final halo at $z=0$.

The derived stellar masses of the UFD analogs at $z=0$ are presented in Fig.~\ref{fig5_SMHM} as a function of their virial masses at $z=0$, illustrating the stellar mass-halo mass relation (SMHM), with the precise values detailed in Table~\ref{table:simul}. An increasing trend in stellar masses is found with increasing halo masses. The three runs, which adopt the same initial conditions but employ different IMF sampling methods, are distinguished by different symbols. The resultant stellar masses range from $M_{\ast}\approx10^2\msun$ to $M_{\ast}\approx1.5\times10^4\msun$ at $z=0$. Since we halt our simulations at $z=6$, the stellar masses at $z=6$ may differ from the final masses at $z=0$ due to mass loss from stellar particles during the AGB phase and Type Ia SNe. To estimate the final stellar masses at $z=0$, we calculate the mass loss of all stellar particles based on the assumed IMF. Mass loss from Type II SN events occurs promptly—immediately in the {\sc Indiv} and {\sc Burst} runs, and within approximately 50 Myr in the {\sc Simf} run, corresponding to the lifetime of an 8 $M_{\odot}$ Pop II star. Our estimates also account for mass loss from AGB winds and Type Ia SNe. These mass loss estimates are validated by comparing them with results from selected hydrodynamic simulations that perform down to $z=0$.

As illustrated in Fig.~\ref{fig5_SMHM}, the stellar mass estimates are generally consistent across the three methods, with no significant systematic differences among the {\sc Burst}, {\sc Simf}, and {\sc Indiv} runs. For Halo3, which exhibits the smallest stellar mass ($M_{\star} \lesssim 700 M_{\odot}$), the {\sc Indiv} method yields a slightly lower mass compared to the {\sc Burst} and {\sc Simf} methods. However, this appears to be a stochastic effect caused by small-number statistics. Additional simulations with different random seeds confirm that there is minimal variation in stellar mass attributable to the sampling method and the associated feedback. This trend will be discussed in detail in Section~\ref{sec:3.1.1}. In Fig.~\ref{fig5_SMHM}, we compare our SMHM relationship estimates with the best-fit results from previous abundance-matching studies (e.g., \citealp{Behroozi2013, Brook2014, Garrison2016, Munshi2021}) and simulated values from other theoretical works (e.g., \citealp{Munshi2013, Wheeler2015, Wheeler2019, Munshi2021}). Notably, \citet{Munshi2021} predicted a steeper slope in the UFD regime, represented by the blue dash-dotted line, where the fit is based on other simulated results shown as gray circles in Fig.~\ref{fig5_SMHM}. They assumed a larger scatter in the SMHM relation toward lower-mass regimes, compared to abundance-matching-based fits, which typically exhibit smaller scatter. Our results fall between these two trends, each reflecting different assumptions about the degree of intrinsic scatter.

\subsubsection{The evolution of the progenitor haloes}
\label{sec:3.1.1}
As we trace the star formation activity within progenitors at high-$z$, our findings reveal that star formation commences across multiple progenitor haloes, rather than being predominantly concentrated in the most massive main progenitor. Fig.~\ref{fig6_morph} illustrates the distribution of progenitor haloes of {\sc Halo2} at $z\approx7$, with each progenitor halo represented by circles of different colors. These haloes eventually merge into a single UFD analog at $z=0$. Each panel, from left to right, presents the DM overdensity, hydrogen number density, gas temperature, and gas metallicity, averaged along the line of sight, within a cubical region of comoving size $\sim$80 kpc. The presence of multiple haloes implies that the member stars of the UFD analog at $z=0$ originate from diverse environments within each progenitor, rather than undergoing a monolithic evolution in a single main progenitor. Consequently, the characteristics of these stars are determined by the specific star formations and metal enrichment histories they experienced within their respective progenitor haloes.

\begin{figure}
  \centering
  \includegraphics[width=80mm]{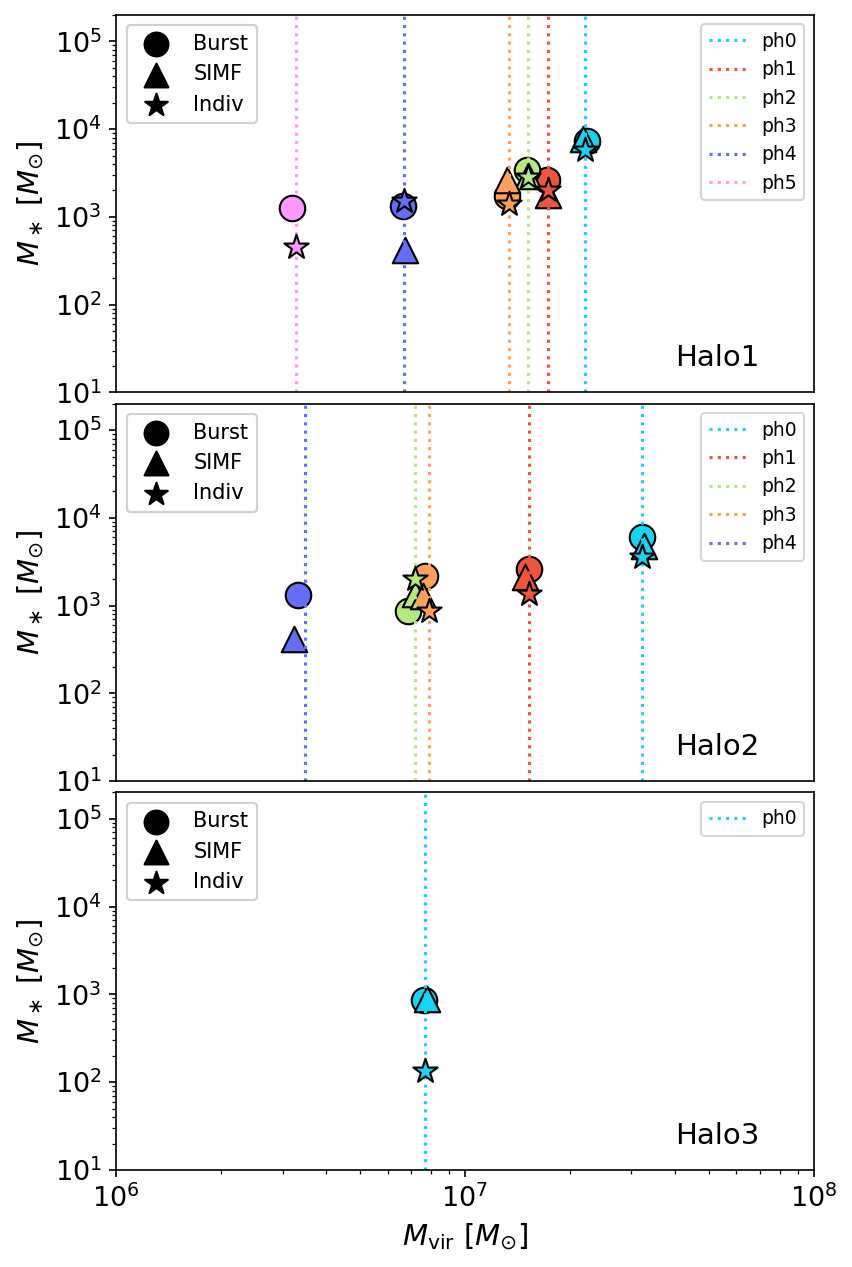}%
   \caption{The relationship between stellar and virial masses within progenitor haloes at $z=6$, which eventually assemble into a single UFD analog by $z=0$. The stellar masses in progenitor haloes are generally consistent across the {\sc Burst}, {\sc Simf}, and {\sc Indiv} runs, with no significant systematic differences. Any observed variations are likely due to stochastic effects, particularly in the lowest-mass haloes, where even a few SNe can have a significant impact on the outcome. In these low-mass haloes, star formation occurs very late, and the effects of reionization result in only 1-2 starbursts. As a result, the occurrence of just 1-2 individual SNe can suppress further star formation, leading to a lower total stellar mass.}
   \label{fig7_mh_smhm}
\end{figure}

\begin{figure*}
  \centering
  \includegraphics[width=165mm]{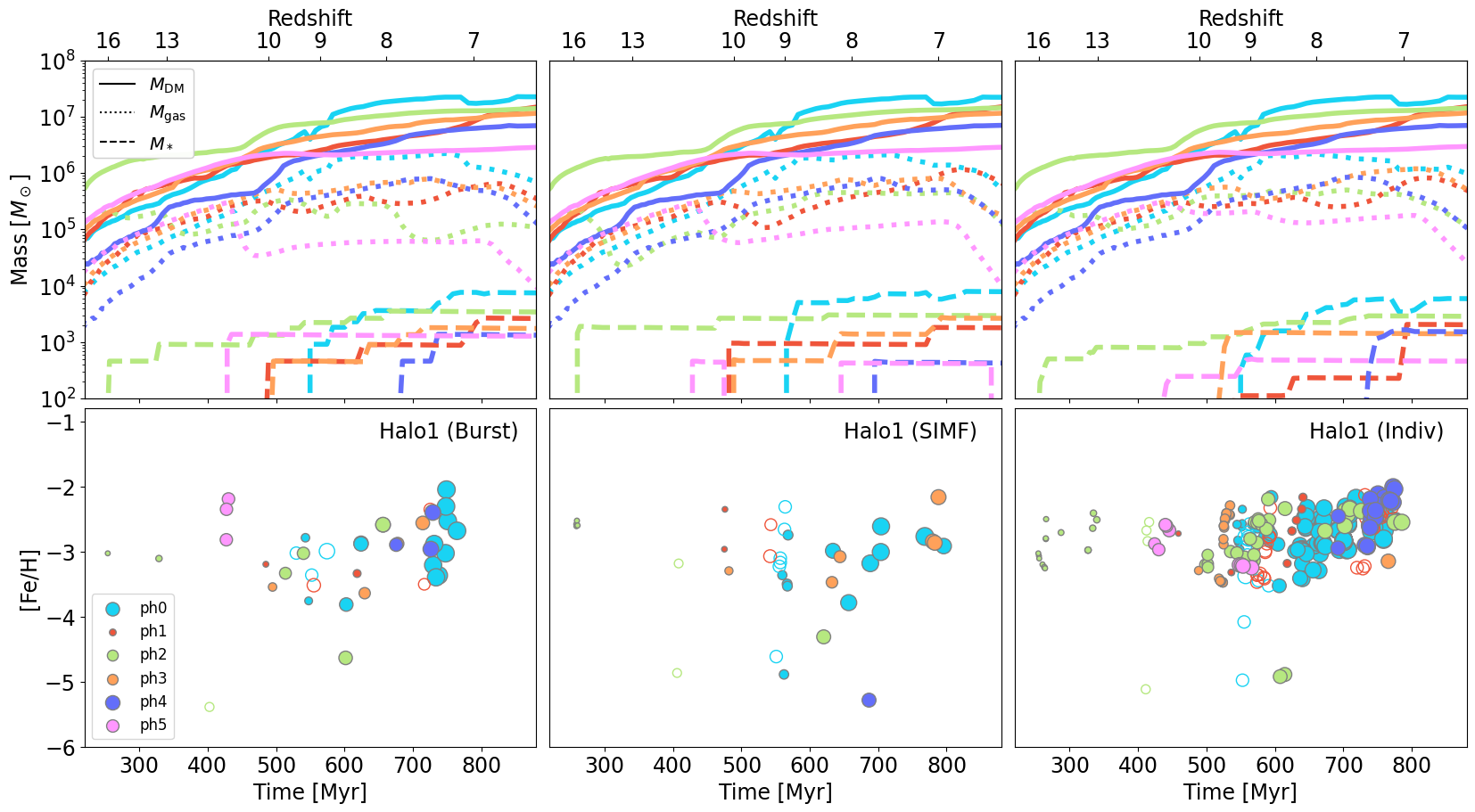}%
   \caption{The top panels illustrate the mass evolution histories, while the bottom panels show the corresponding metallicity of star particles within progenitor haloes identified at $z\approx6$ for {\sc Halo1}, utilizing different IMF sampling methods: {\sc Burst} (left), {\sc Simf} (middle), and {\sc Indiv} (right). In the top panels, solid, dotted, and dashed lines represent the evolution of dark matter, gas, and stars, respectively. In the bottom panels, we differentiate between stars formed through in situ star formation within the virial radius (filled circles) and those that merged after being generated in other progenitor haloes (open circles). While the {\sc Burst} and {\sc Simf} runs show similar trends, the {\sc Indiv} run exhibits more continuous star formation and more frequent metallicity enrichment through self-enrichment, likely because individually resolved SN feedback exerts weaker suppression compared to the other runs.} 
   \label{fig9_feh_mul}
\end{figure*}

Fig.~\ref{fig7_mh_smhm} shows the SMHM relations for progenitor haloes at $z=6$ in {\sc Halo1}, {\sc Halo2}, and {\sc Halo3}, generally demonstrating an upward trend in stellar mass with increasing halo mass. Note that the stellar masses are calculated by summing the values of the star particles within the virial radius of each progenitor halo at $z=6$. In addition, the order of the progenitor haloes is arranged from the most massive ({\sc ph0}) to the least massive haloes ({\sc ph5}), specifically, at $z=6$, thus this does not imply that {\sc ph0} is always the most massive during $z\gtrsim6$. We find that in each progenitor, the stellar masses are generally consistent across the different IMF sampling and SN injection methods, showing no distinct systematic trends. This indicates that within the level of variation in our simulations, the choice of method does not strongly affect the final stellar mass. However, the absence of a trend may simply reflect the small stellar masses involved ($M_{\star} \lesssim 10^{4} M_{\odot}$), where method-dependent differences are expected to be modest and easily overshadowed by stochastic effects. Particularly, in the lowest-mass haloes ($M_{\text{vir}, \text{ph}} \le 10^{7} M_{\odot}$), the much smaller stellar masses make the results more susceptible to stochasticity rather than the method chosen, as even a few SNe can cause significant variations in stellar masses.

The detailed evolution of each progenitor halo of {\sc Halo1} is displayed in Fig.~\ref{fig9_feh_mul}, with results shown from left to right for the {\sc Burst}, {\sc Simf}, and {\sc Indiv} runs, each adopting different IMF sampling and SN feedback injection methods. The top panels depict the evolution of the DM (solid line), gas (dotted line), and stellar (dashed line) masses, respectively, while the bottom panels present the corresponding stellar metallicity, expressed as $\rm [Fe/H]$, for each progenitor. The open circles in the bottom panels, sharing the same color, indicate stars that originated in different haloes other than the selected progenitors but eventually merged into one of these progenitor haloes. In addition, the size of each circle reflects the relative mass of the progenitor halo at the time the star entered it. Since we compare runs with identical initial conditions but different IMF sampling and SN feedback injection methods, the DM evolution of each progenitor halo remains consistent across all runs, whereas the gas and stellar evolution vary with the chosen method and the stochastic nature of star formation. In general, most progenitor haloes experience a gas loss of 1-2 orders of magnitude between $z=7$ and $z=6$, during which the effect of reionization is introduced. 

\subsubsection{Comparison between {\sc Busrt}, {\sc Simf}, and {\sc Indiv} runs}
\label{sec:3.1.2}

In this subsection, we present a comparative analysis of the simulation results for three scenarios: {\sc Burst}, where multiple SNe are triggered simultaneously; {\sc Simf}, where SN events are staggered based on the lifetimes of massive stars, resulting in temporally discrete SN feedback; and {\sc Indiv}, where star particles are separated into two types, with massive stars sampled individually to achieve both spatially and temporally discrete SN feedback. It is important to note that in the {\sc Burst} scenario, the number of SNe in an SSP particle of mass $m_{\rm PopII}=500\msun$ is fixed at 3.4, whereas in the {\sc Simf} case, this number varies due to stochastic effects.

Overall, the trend in stellar mass evolution is evident: while progenitor haloes with different methods initiate star formation at similar times, their SFHs differ. In general, star formation in the {\sc Indiv} scenario occurs more continuously compared to the more episodic patterns in the {\sc Burst} and {\sc Simf} methods. Despite these differences, the final stellar masses remain comparable across the methods. For example, all {\sc ph0} haloes, indicated by cyan lines and circles in Fig.~\ref{fig9_feh_mul}, initiate star formation around $z \approx 9.2$, with final stellar masses of 7380, 7800, and 6370$\msun$ for the {\sc Burst}, {\sc Simf}, and {\sc Indiv} runs, respectively. This pattern is maintained for {\sc ph1} through {\sc ph3}, but deviates in the two least massive haloes, {\sc ph4} and {\sc ph5}, where their small stellar masses make them highly sensitive to even minor SN feedback, leading to stochastic variations between runs.

In {\sc ph0}, the different SFHs arising from the choice of IMF sampling and SN feedback scheme lead to variations in the epoch of most significant stellar-mass growth: around $z\sim7.2$ for {\sc Burst} ($\sim4000\msun$), $z\sim9$ for {\sc Simf} ($\sim5000\msun$), and $z\lesssim9$ for {\sc Indiv} ($\sim5000\msun$). In {\sc Simf} and {\sc Burst} runs, another progenitor hosting a massive gas clump forms a dense clump outside {\sc ph0}, which later merges with it. This external clump participates in the subsequent starburst, but whether the burst occurs before or after the merger differs among runs. In the {\sc Simf} run, this clump begins forming stars while still outside the {\sc ph0}, resulting in the majority of stars being formed ex-situ. In contrast, in the {\sc Burst} run, the onset of star formation in the dense clump is delayed until $z\sim6.5$, occurring after the merger. Thus, both methods produce bursty star formation, but at different epochs, likely as a result of stochastic variations in gas dynamics and feedback. Meanwhile, the {\sc Indiv} run yields a more continuous and extended SFH, contrasting with the more bursty, concentrated episodes seen in the other two methods, where the star particle is characterized by $500\msun$ SSP. This suggests that individually sampling SN progenitors facilitates smoother star-formation histories.

In {\sc ph1}, the timing of SN feedback in the {\sc Burst} and {\sc Simf} runs plays a central role in shaping the SFH. In both scenarios, in situ star formation begins at $z\sim10$ (indicated by a red circle) after gas enrichment from Pop~III SNe. In the {\sc Burst} run, the strong energy releasefrom a single $500 \msun$ SSP temporarily halts star formation until $z\sim8.4$ (about 60 Myr later), after which the dense gas recovers and star formation resumes. In contrast, in the {\sc Simf} run, the first Pop II SSP forms at $z\sim10.2$, followed by a second one just 1 Myr later. Because {\sc Simf} staggers the SNe---the most massive sampled star explodes immediately, while the remaining SNe occur after their respective lifetimes---the gas can partially recover between SN events. This allows the second SSP to form before the delayed SNe from the first cluster explode. However, over the next $\sim$50 Myr, the cumulative SN energy from the two SSPs in {\sc Simf} exceeds that of the single SSP in {\sc Burst}, ultimately quenching in situ star formation in the {\sc Simf} scenario.

As a result, the differences in SN injection schemes naturally lead to variations in stellar-metallicity evolution. In both the {\sc Burst} and {\sc Simf} methods, stars form as $500~\msun$ SSPs, causing SN energy to be injected in a relatively clustered manner---simultaneously in {\sc Burst} and temporally staggered yet spatially co-located in {\sc Simf}. As discussed above, if the gas is not suppressed by the first SN in the {\sc Simf} run, another SSPs with $m_{\rm PopII}=500\msun$ may form before additional SNe from the preceding SSP, producing a rapid increase in stellar mass followed by strong SN feedback from multiple SSPs. Therefore, the clustered feedback in {\sc Burst} and {\sc Simf} is more suppressive, limiting the persistence of dense gas and resulting in lower stellar metallicities. In contrast, the {\sc Indiv} method forms stars in smaller increments (typically $\lesssim 100~\msun$) with spatially and temporally dispersed SNe. This dispersion allows the gas to recover between SN events, enabling star formation from gas that is promptly enriched by preceding explosions. As a result, self-enrichment occurs more frequently, leading to higher $\mathrm{[Fe/H]}$ values. We explore these qualitative differences in detail in Section~\ref{sec:3.2.1}.

Finally, we find that low-metallicity stars with $\rm [Fe/H]\lesssim-5$ result from external metal enrichment. For instance, in {\sc ph2}, shown in green, although {\sc ph2} ranks as the third most massive halo at $z\approx6$, its virial mass surpasses others at $z\gtrsim9$, allowing it to be the first to initiate star formation at $z\approx16$. Interestingly, in {\sc ph2}, we find that the initial ex situ star particles at $z\approx11.3$, marked by an open circle, have significantly low metallicities of $\rm [Fe/H]=-5.1$ in the {\sc Burst} run and $\rm [Fe/H]=-4.9$ in the {\sc Simf} run. This is due to external metal enrichment as follows: at $z=15.7$, metals are ejected by Pop~III and Pop~II stars, represented by filled circles, formed at a density peak about 1 kpc away, beyond the virial radius of {\sc ph2} ($r_{\rm vir}\approx200$ pc). Over the next 150 Myr, these externally sourced metals enrich the gas within the virial radius of {\sc ph2}, leading to the formation of metal-poor Pop~II stars with $\rm [Fe/H] \lesssim-4.9$ in {\sc ph2} for both runs. In the relatively low-mass progenitors ($M_{\rm vir}\lesssim6\times10^6\msun$ at $z=6$), randomness increases due to the small number of star particles formed. This is because star formation not only begins later but is also likely to be quickly quenched by the reionization effect. We also provide the mass evolution histories and the corresponding stellar metallicities formed in progenitors of  {\sc Halo2} in Appendix~\ref{sec:A}, where a similar trend to {\sc Halo1} is observed.

\begin{figure}
    \centering
    \includegraphics[width=80mm]{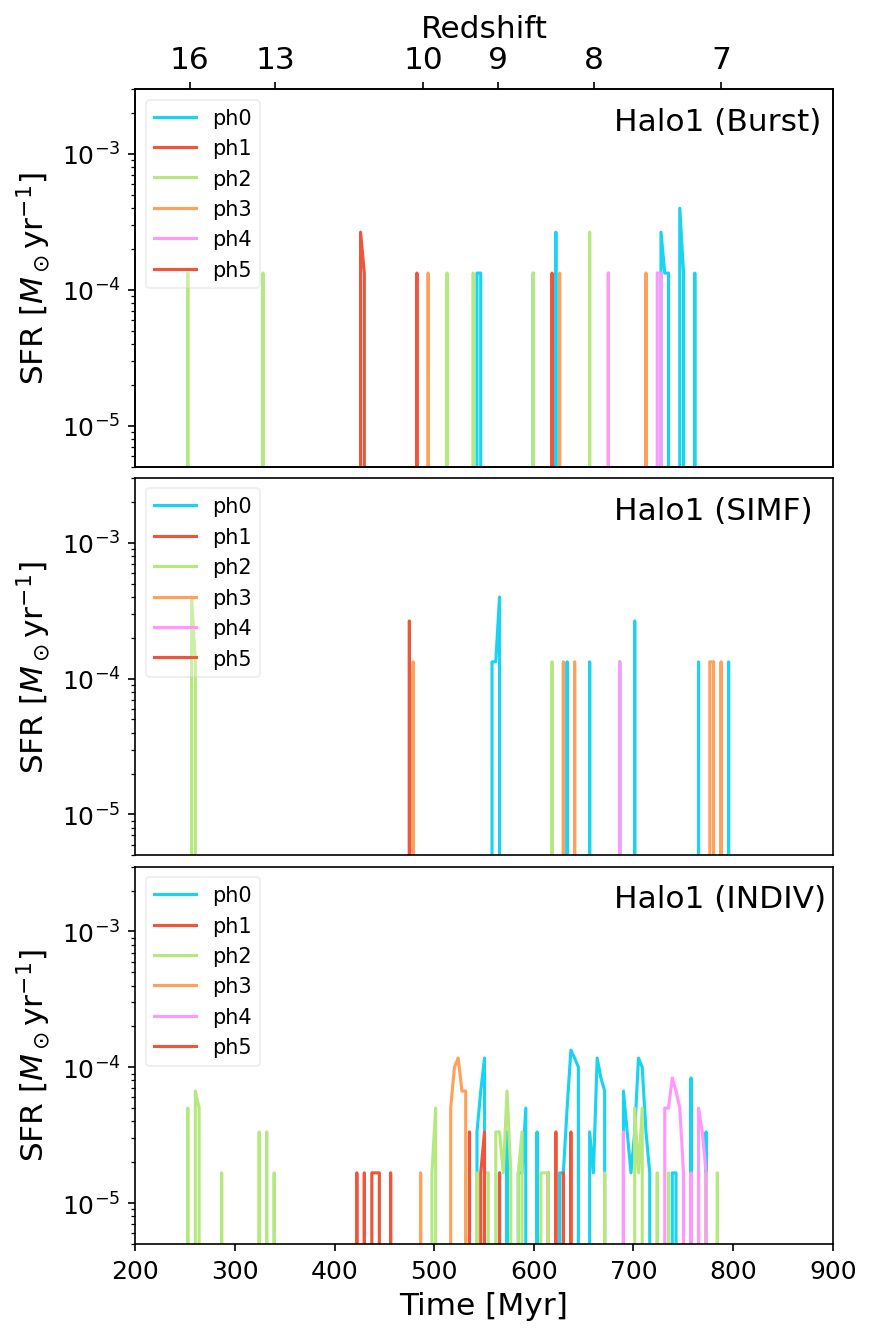}
    \caption{Star formation rates as a function of cosmic time are shown for the simulated UFD analogs, with different colors representing each progenitor halo. Note that the estimates are evaluated from all stars found in each halo, using a time bin of 3 Myr. This indicates that stars form in various environments rather than a single main progenitor. It also reveals that the IMF sampling method influences whether star formation will be discrete or continuous.}%
    \label{fig8_sfh_mul}%
\end{figure}

\subsubsection{Star formation histories and burstiness}
\label{sec:3.1.3}
Fig.~\ref{fig8_sfh_mul} depicts total star formation rates (SFRs), with each progenitor halo denoted by a different color. From top to bottom, the panels correspond to the {\sc Burst}, {\sc Simf}, and {\sc Indiv} runs for {\sc Halo1}. The SFRs, computed in 3.5~Myr time bins, range from $10^{-6} \msun/\rm yr$ to a few $10^{-4} \msun/\rm yr$. The {\sc Burst} and {\sc Simf} show bursty star formation behavior, whereas the {\sc Indiv} run exhibits more continuous SF, particularly in relatively massive progenitor haloes ($M_{\rm vir}\gtrsim$ a few $10^7\msun$ at $z\approx6$). As SF becomes more bursty, the intervals between bursts increase, producing extended quiescent periods with little or no star formation. Therefore, the {\sc Burst} and {\sc Simf} runs display highly bursty SFR patterns, while the {\sc Indiv} run maintains a smoother, more continuous trend.

\begin{figure}
    \centering
    \includegraphics[width=75mm]{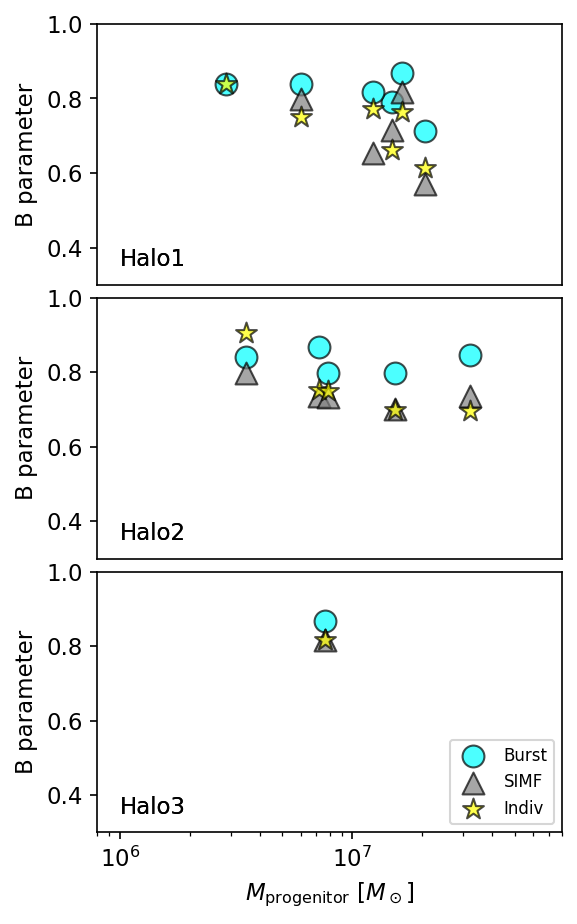}
    \caption{The B-parameter is used to measure the stochasticity of progenitor haloes in terms of SN rate, rather than SFRs. Specifically, in {\sc Simf}, although stars form simultaneously as a SSP, the SNe explode sequentially over time, resulting in a continuous SN rate. A B-parameter closer to 0 indicates a more continuous distribution. Unlike SFRs, the {\sc Simf} runs exhibit a more continuous trend, similar to or even more so than the {\sc Indiv} runs.}%
    \label{fig10_Bparams}%
\end{figure}

\begin{figure}
    \centering
    \includegraphics[width=86mm]{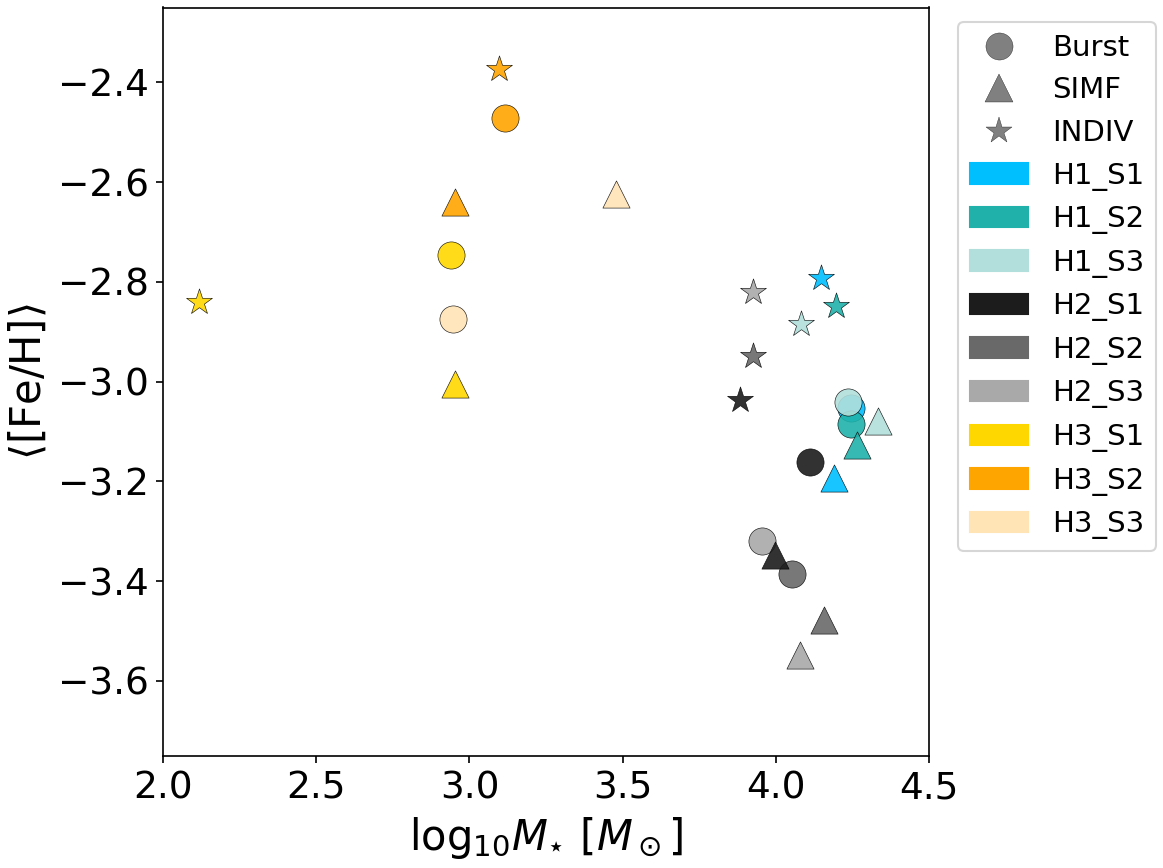}
    \caption{Stellar mass–metallicity relation for 27 simulations across three methods; each halo–method combination is repeated three times with different random seeds. For {\sc Halo1} and {\sc Halo2}, the {\sc Indiv} runs consistently yield higher stellar metallicities than the {\sc Burst} and {\sc SIMF} runs. This trend is not apparent for the least-massive system, {\sc Halo3}, where the very small stellar mass and strong sensitivity to feedback obscure any clear pattern.}
    \label{fig_MZR_all}
\end{figure}

However, regarding the frequency of SN events, the {\sc Simf} runs are expected to be comparably continuous to the {\sc Indiv} runs, since individual SNe occur with delay times in the {\sc Simf} scheme. To quantify this burstiness of SN events, we adopt the B-parameter introduced by \citet{Applebaum2020}, defined as $\rm B=(\sigma/\mu - 1)/(\sigma/\mu + 1)$, where $\sigma$ is the standard deviation of the SN rate and $\mu$ is the mean SN rate. The SN rate is computed in 3~Myr time bins, yielding values between -1 and 1, with larger values (closer to 1) indicating stronger burstiness. In the {\sc Burst} run, multiple SNe are triggered simultaneously each time a star particle with a mass of $500\msun$ forms. In contrast, in the {\sc Simf} run, although stars are sampled from a single SSP particle, individual SNe occur at different times depending on stellar lifetimes. When computing the B-parameter, these time delays are explicitly accounted for to capture the temporal distribution of SN events.

As demonstrated in Fig.~\ref{fig10_Bparams}, the most {\sc Burst} runs tend to show high B-parameter values above 0.8, indicating high burstiness, whereas the values in the {\sc Simf} and {\sc Indiv} runs are lower than those of the {\sc Burst} run, implying a more continuous pattern. To sum up, in terms of SFRs, both the {\sc Burst} and {\sc Simf} runs exhibit bursty patterns. However, when considering the SN rate, the {\sc Simf} runs are as continuous as the {\sc Indiv} runs. Furthermore, there is a trend that the values of the B parameter decrease as the progenitor masses increase, which is pronounced in the {\sc Halo1} case. This means that as the mass of the progenitor halo increases, burstiness decreases, and star formation occurs more continuously. This result concurs with the expectation that smaller haloes are more susceptible to feedback, allowing them to quench star formation easily.

\begin{figure*}
    \centering
    \includegraphics[width=180mm]{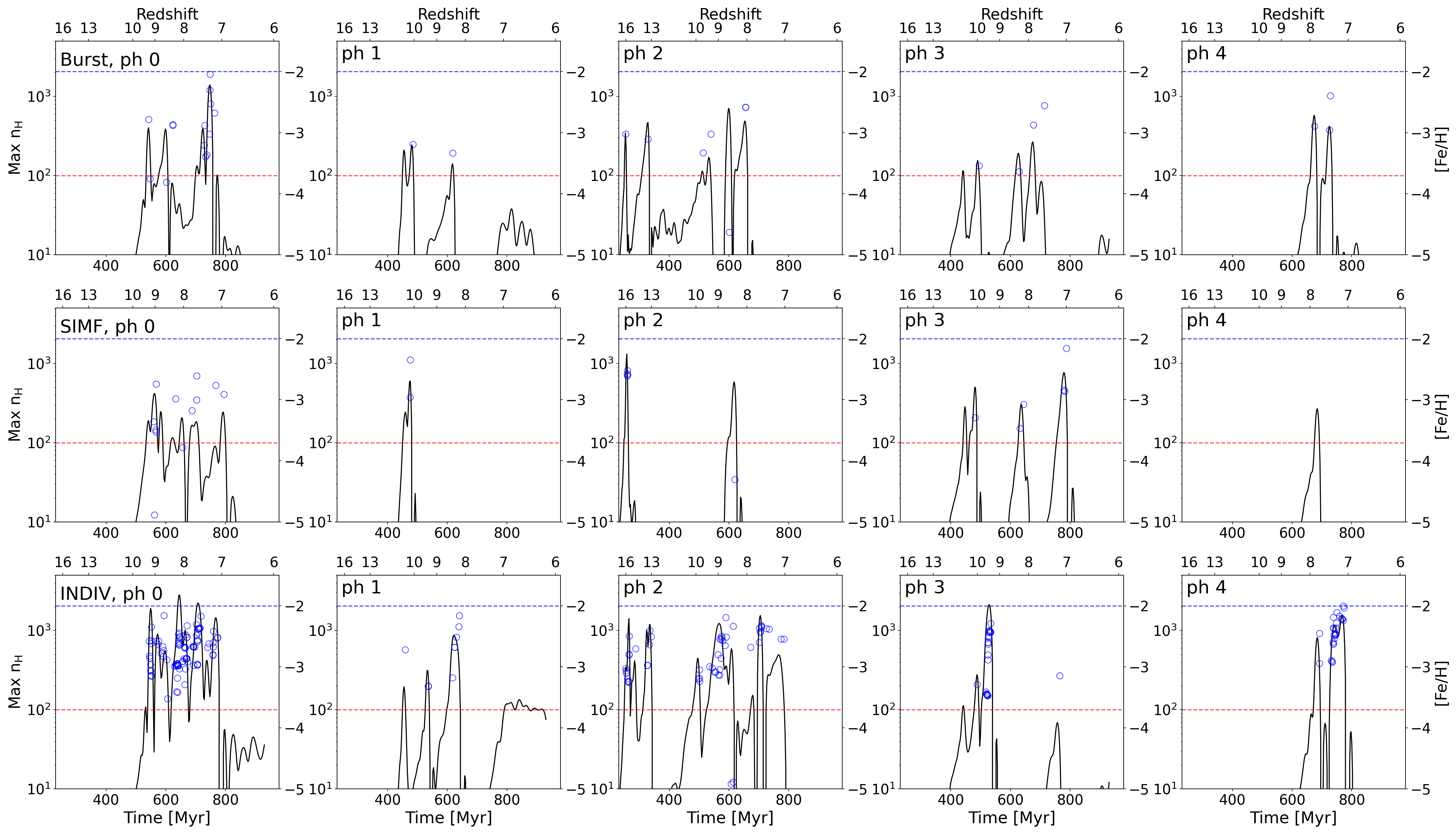}
    \caption{Time evolution of the maximum hydrogen number density within each progenitor halo of {\sc Halo1}. Each column represents progenitors from {\sc ph0} to {\sc ph4}, while each row shows results based on three different approaches: {\sc Burst} (top), {\sc Simf} (middle), and {\sc Indiv} (bottom). The black curves show the maximum hydrogen number density ($n_{\rm H}$) as a function of cosmic time. Blue open circles mark the formation times and stellar metallicities of individual \textit{in situ} stars, with the metallicity values indicated on the right y-axis ($[\mathrm{Fe/H}]$). The red dashed horizontal line marks the star-formation threshold at $n_{\rm H}=100~\mathrm{cm^{-3}}$, and the blue dashed horizontal line indicates a relatively low metallicity of $[\mathrm{Fe/H}]=-2$ for reference, with no stars exceeding this value.}%
    \label{fig_density_evolution}
\end{figure*}

\begin{figure*}
    \centering
    \includegraphics[width=180mm]{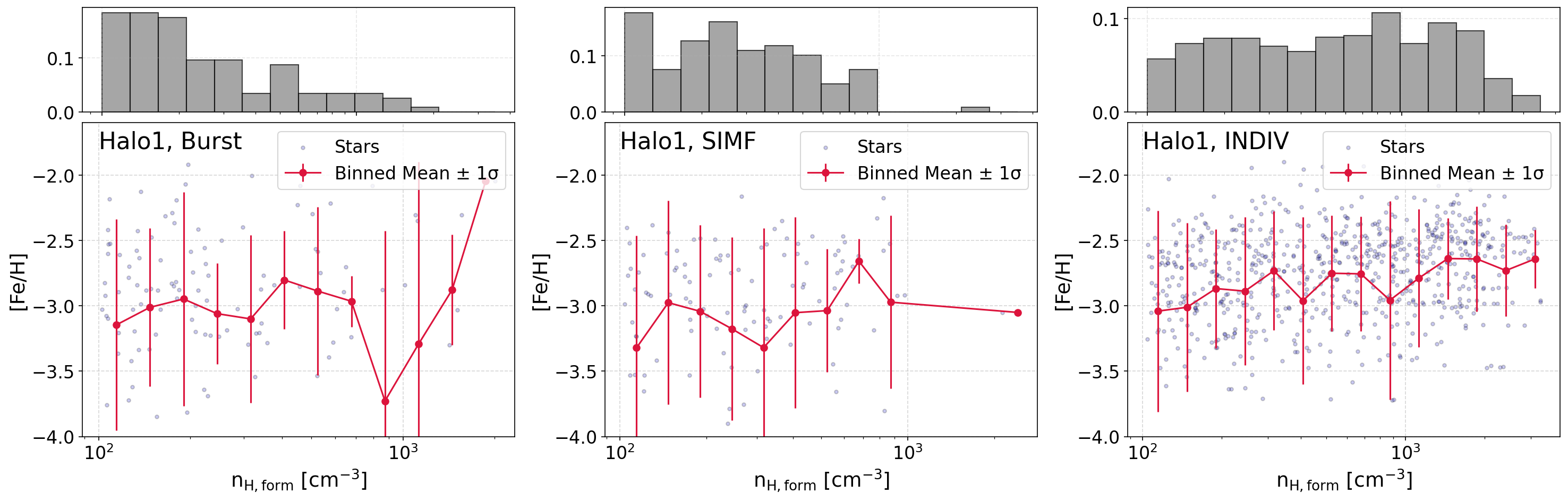}
    \caption{Correlation between the density of star-forming gas at birth ($n_{\rm H, \text{form}}$) and stellar metallicity for {\sc Halo1} across three different schemes (from left to right: {\sc Burst}, {\sc Simf}, {\sc Indiv}). The gas densities are measured just before star formation. Blue points with low opacity represent individual Pop~II star particles from simulations with various random-seed realizations. Red symbols depict the binned mean $\pm 1\sigma$ within logarithmic bins of gas density at birth. The panels above each plot show a normalized histogram of star counts versus $n_{\rm H, \text{form}}$. In the {\sc Burst} and {\sc Simf} runs, 80\% of stars form from relatively low-density gas ($n_{\rm H, \text{form}} \lesssim 547~\mathrm{cm^{-3}}$ and $n_{\rm H, \text{form}} \lesssim 557~\mathrm{cm^{-3}}$, respectively). By contrast, in the {\sc Indiv} run, even the median birth density already exceeds $585~\mathrm{cm^{-3}}$, indicating that a substantial fraction of stars form in denser environments. Consequently, stars formed in the {\sc Indiv} run also exhibit systematically higher metallicities, with the mean [Fe/H] among the upper 50\% of stars reaching $\mathrm{[Fe/H]=-2.75}$. In contrast, the mean $\mathrm{[Fe/H]}$ of the lower 80\% of stars in the {\sc Burst} and {\sc Simf} runs remains at $\mathrm{[Fe/H]=-3.01}$ and $\mathrm{[Fe/H]=-3.17}$, respectively.}
    \label{fig_nh_feh_stars}
\end{figure*}

\begin{figure*}
    \centering
    \includegraphics[width=130mm]{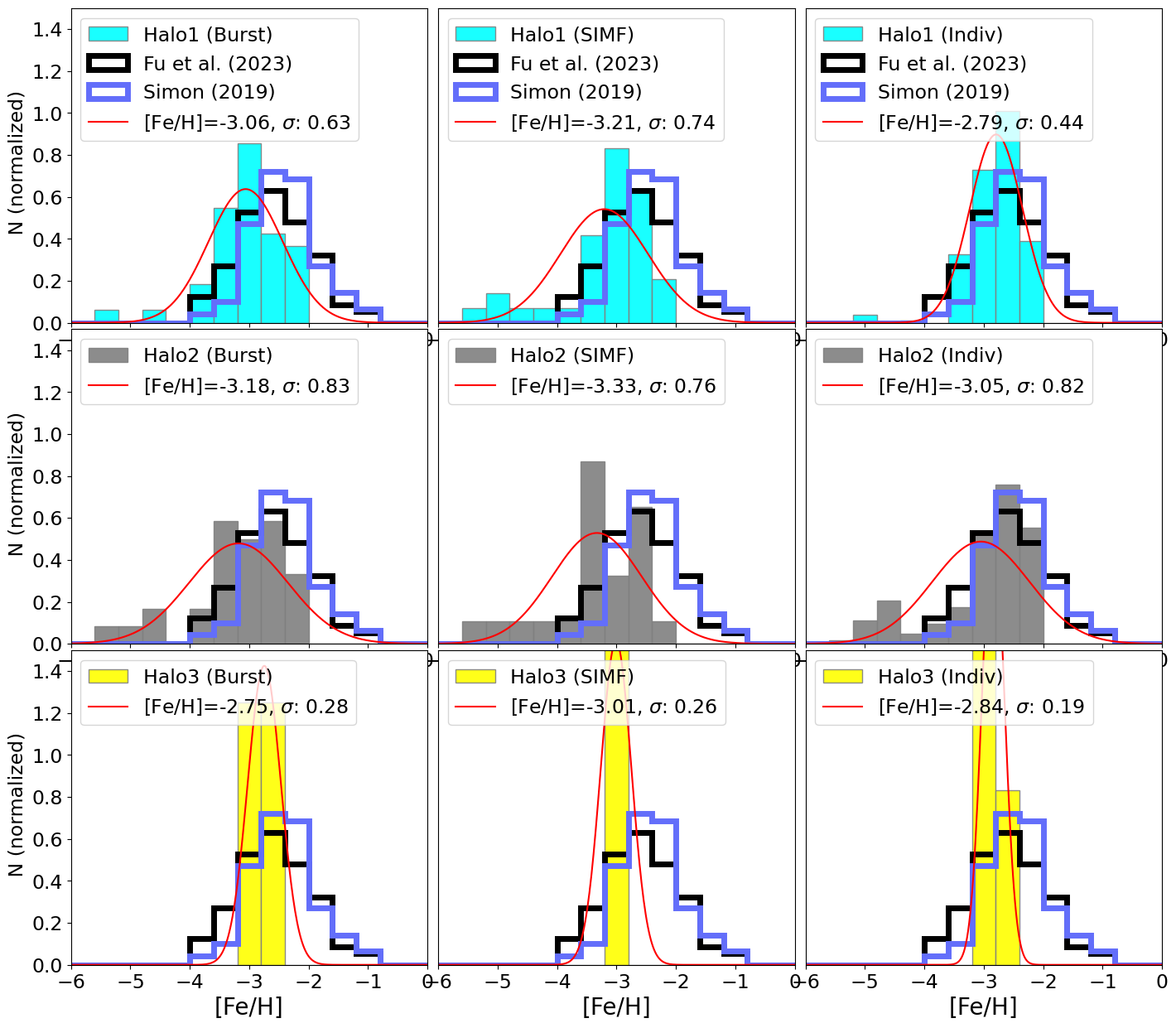}
    \caption{The metallicity distribution function for the simulated UFD analogs for {\sc Halo1} (top panels), {\sc Halo2} (middle panels), and {\sc Halo3} (bottom panels), compared with the composite MDFs of the observed MW UFDs, depicted by blue (\citealp{Simon2019}) and black (\citealp{Fu2023}) histogram contours. Also, we provide a Gaussian fit (red solid line) characterized by its mean and dispersion values. Note that the metallicities in {\sc Halo3} predominantly fall within the range of $\rm [Fe/H]=-3.0$ to $\rm [Fe/H]=-2.0$. This characteristic is attributed to one or two starbursts occurring at a later time, closer to the epoch of reionization.}%
    \label{fig:MDF_mul_fit}%
\end{figure*}

\subsection{Stellar abundance}
\label{sec:3.2}
Thus far, we have shown that SFHs are influenced by both stochastic effects and methodological differences, particularly due to differences in the implementation of SN feedback. In this section, we explore how these differences affect stellar metallicity variations. Specifically, in Section~\ref{sec:3.2.1}, we present the stellar mass-metallicity relationship and discuss potential reasons for the observed differences among the three approaches. Then, in Section~\ref{sec:3.2.2}, we examine the metallicity distribution function (MDF) of stars, comparing it with observational data. Finally, in Section~\ref{sec:3.2.3}, we compare our mass-metallicity relation with those from other studies.

\subsubsection{Stellar mass-metallicity relation}
\label{sec:3.2.1}

The stellar mass-metallicity relation (MZR) across all simulation sets is shown in Fig.~\ref{fig_MZR_all}, which shows that the {\sc Indiv} runs consistently produce higher stellar metallicities than the {\sc Burst} and {\sc Simf} runs, particularly in {\sc Halo1} and {\sc Halo2}. To understand the physical origin of this difference, we investigate the evolution of the maximum hydrogen gas density ($n_{\rm H,max}$) in each progenitor halo, as illustrated in Fig.~\ref{fig_density_evolution}. We find that both stellar metallicity and $n_{\rm H,max}$ are systematically higher in the {\sc Indiv} runs, and these two quantities are strongly correlated. In {\sc Indiv}, $n_{\rm H,max}$ typically exceeds $10^{3}\ \mathrm{cm^{-3}}$, while in the non-{\sc Indiv} runs it remains around $400\ \mathrm{cm^{-3}}$. This correlation suggests that the enhanced metallicities in {\sc Indiv} are linked to the denser star-forming environments.

Although the causal direction is not entirely straightforward, the different nature of SN feedback provides a plausible explanation. In {\sc Indiv}, SN explosions are temporally and spatially dispersed, which mitigates their cumulative impact on the surrounding medium. This smoother feedback allows the gas to re-accrete and cool between SN events, enabling the density to rebound to high values ($n_{\rm H,max}\sim2\times10^{3}\ \mathrm{cm^{-3}}$). Stars formed in such dense environments explode in relatively high-density regions, keeping the ambient gas dense enough for subsequent star formation to occur. This continuous cycle promotes sustained star formation and frequent self-enrichment, leading to higher stellar metallicities. Similar behavior is observed across other progenitors in the {\sc Indiv} set, such as {\sc ph1} and {\sc ph3}, as shown in Fig.~\ref{fig_density_evolution}.

In contrast, in the {\sc Burst} and {\sc Simf} runs, SN feedback is stronger and more concentrated. The cumulative energy from co-located SNe originating in massive SSPs efficiently evacuates the surrounding gas, lowering the ambient density to much smaller values. In {\sc Simf}, for instance, sequential SNe explode in the same location, injecting feedback energy into an already rarefied medium, which further suppresses gas recovery. As a result, the gas fails to reach the density required for sustained star formation, and the growth of stellar metallicity is hindered.

These differences are quantitatively illustrated in Fig.~\ref{fig_nh_feh_stars}, which presents stellar metallicity as a function of the gas density at star formation ($n_{\rm H, form}$). In the {\sc Burst} and {\sc Simf} runs, about 80\% of stars form from relatively low-density gas ($n_{\rm H,form}\lesssim550\ \mathrm{cm^{-3}}$), yielding mean metallicities of $\mathrm{[Fe/H]}\approx-3.1$ and $-3.2$, respectively. By contrast, in {\sc Indiv}, stars form at systematically higher densities ($n_{\rm H,form}\gtrsim600\ \mathrm{cm^{-3}}$) and show higher metallicities ($\mathrm{[Fe/H]}\approx-2.7$). This clearly demonstrates that spatially and temporally discrete SN feedback in the {\sc Indiv} method enables star formation in denser gas, fostering efficient metal retention and frequent re-enrichment.

Finally, the higher metallicity in {\sc Indiv} is further aided by its improved star-particle resolution. In the {\sc Burst} and {\sc Simf} runs, all star particles have a fixed mass of $500\ \mathrm{M_{\odot}}$, which limits localized enrichment. In {\sc Indiv}, by contrast, star particles can be as small as $60\ \mathrm{M_{\odot}}$, allowing individual SNe to enrich the surrounding gas multiple times before the next star forms. Therefore, both the discrete feedback implementation and the higher mass resolution of star particles work together to promote efficient, localized metal recycling in the {\sc Indiv} runs.

\subsubsection{Metallicity distribution function}
\label{sec:3.2.2}
Fig.~\ref{fig:MDF_mul_fit} shows the MDFs of stars formed in {\sc Halo1} (top panels), {\sc Halo2} (middle panels), and {\sc Halo3} (bottom panels), in each case. These results are compared with the composite MDFs of observed MW UFDs, represented by black lines (\citealp{Fu2023}), and those of UFDs in the Local Group (LG), shown as blue lines (\citealp{Simon2019}). To ensure consistency, we adopt the method of \citet{Fu2023} to fit the MDF. This involves modeling the MDF of the simulated UFD analogs with a Gaussian distribution. We determine the best parameters for the mean and the dispersion by maximizing the two-parameter Gaussian likelihood function (\citealp{Walker2006}), utilizing the {\sc emcee} to sample the posterior distribution. The resulting fits are illustrated as red lines in Fig.~\ref{fig:MDF_mul_fit}.

The MDFs of the simulated UFD analogs exhibit the following characteristics: Firstly, for the same initial conditions, the MDF peak in the {\sc Indiv} runs tends to be at higher metallicities. For instance, in {\sc Halo1} and {\sc Halo2}, the {\sc Indiv} runs share the same peak as the observational data. Moreover, their MDFs closely match the observed data, whereas the MDFs for the {\sc Burst} and {\sc Simf} runs are skewed towards lower metallicities due to the absence of relatively high metallicities ($\rm [Fe/H]\gtrsim-2$). We find that in the {\sc Burst} and {\sc Simf} runs, there are no high-metallicity stars above $\rm [Fe/H]=-2.1$.

Secondly, unlike the observational data, which do not show stars with low-metallicity below $\rm [Fe/H]=-4$, our results do feature such stars, which contribute to lowering the average metallicity. As discussed in Section~\ref{sec:3.1.2}, these low-metallicity stars tend to originate from external metal-enrichment, which means the source of enrichment lies outside the star-forming region, with $\sim$~kpc scale featuring distance among minihaloes. Another scenario that can produce low-metallicity stars involves star formation occurring at a density peak located on the outskirts of the progenitor halo, although still within the virial radius, thus classifying them as in situ stars by definition. However, since this density peak is not self-enriching but instead polluted by metals from the inner regions, it leads to the formation of low-metallicity stars. The fraction of low-metallicity stars ($\rm [Fe/H]\lesssim-4$) in {\sc Indiv} is 13\%, which is significantly lower compared to 34\% in {\sc Burst} and 31\% in {\sc Simf} for {\sc Halo1}. This occurs because, in {\sc Indiv} runs, although low-metallicity SSPs with $m_{\rm PopII}\approx63\msun$ may initially form because of external metal enrichment, subsequent in situ stars continuously emerge from polluted gas. Meanwhile, in {\sc Burst} and {\sc Simf}, a relatively large mass of stars ($m_{\rm PopII}\approx500\msun$) are born with low-metallicity, which increases the fraction of such low-metallicity stars.

On the other hand, when in situ star formation takes place, we find that the initial metallicity for Pop II stars, set by Pop III stars, is approximately \textcolor{blue}{$\rm [Fe/H]\approx-3.3$}. This happens because, to ensure consistency in our comparisons using different IMF sampling and SN feedback injection methods, we set the mass of Pop III stars to $m_{\rm PopIII}=20\msun$. This setting typically establishes a metallicity floor of $\rm [Fe/H]\approx-3.3$ in the gas where a Pop III SN explodes. However, this metallicity floor is quickly altered by metals ejected from subsequent Pop II SNe. Finally, relatively high-metallicity stars ($\rm [Fe/H] \gtrsim -2$) are absent in our UFD analogs, creating a gap with observations.

\begin{figure}
    \centering
    \includegraphics[width=75mm]{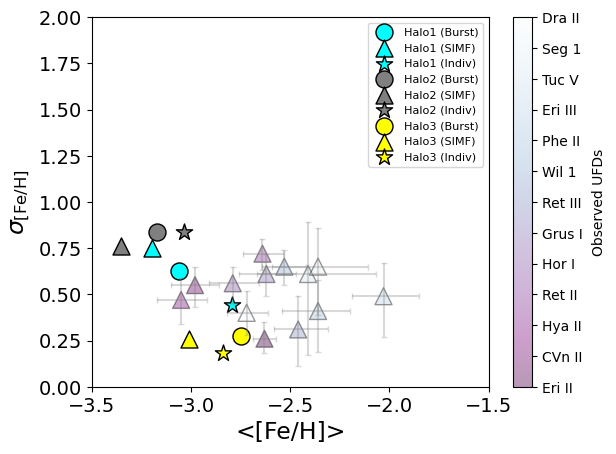}
    \includegraphics[width=75mm]{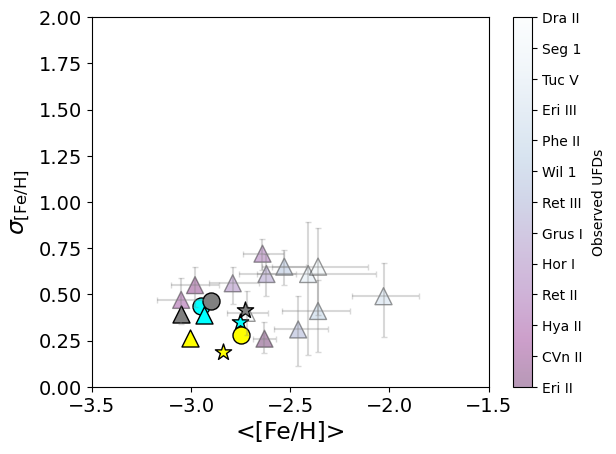}
    \caption{The best parameters for the mean and dispersion of MDFs of the simulated UFD analogs, based on all stars (top panel) and stars with metallicities above $\rm [Fe/H]=-4$ (bottom panel). These parameters are then compared to the MDFs of the observed 13 UFDs reported by \citet{Fu2023}. Excluding low-metallicity stars below $\rm [Fe/H]=-4$ notably decreases the dispersions of the MDFs, bringing them into closer alignment with the observational findings.}
    \label{fig:Fe_H_sigma}%
\end{figure}

Fig.~\ref{fig:Fe_H_sigma} compares the estimates for the mean and dispersion of the MDFs of the simulated UFD analogs with those of the 13 MW UFDs, as reported by \citet{Fu2023}. The observed UFDs are color coded according to their magnitudes, ranging from Draco~II, with a low magnitude of $\rm M_v = -0.8$, to Eridanus II, with a high magnitude of around $\rm M_v = -7.1$. In the upper panel, all stars are included in fitting the MDFs, while the lower panel presents values derived after excluding stars with $\rm [Fe/H]\lesssim-4$. This exclusion is due to the metallicity measured in \citet{Fu2023}, using the MESA Stellar Isochrones and Tracks (MIST) models (\citealp{Choi2016, Dotter2016}), which are based on a metallicity grid ranging from $\rm [Fe/H]=-4$ to $\rm [Fe/H]=-1$. The results indicate that when no low-metallicity cut is applied, {\sc Halo1} shows a dispersion value of $\sigma_{\rm [Fe/H]}=0.44-0.75$, which agrees with the upper limit of the observed dispersion values, and the average metallicity remains within the observed range. However, for {\sc Burst} and {\sc Simf} {\sc Halo2} runs, the average metallicity is about $0.5-1$ dex lower than the observed ranges ($\rm [Fe/H]\approx[-3.0,-2.0]$), and the dispersion is nearly twice as large.

By excluding low-metallicity stars, as shown in the lower panel of Fig.~\ref{fig:Fe_H_sigma}, a clear trend emerges: the estimated mean metallicities of simulated UFD analogs, $\langle \rm [Fe/H]\rangle$, decrease, closely matching observational values. This exclusion also leads to a reduced dispersion in the MDF, which is associated with a shorter overall duration of star formation. This trend is particularly noticeable in {\sc Halo2}, where the average dispersion in three different IMF sampling and SN feedback injection methods decreases from $\langle\sigma_{\rm [Fe/H]}\rangle=$0.81 to 0.42. Similarly, the average dispersions in the absence of low-metallicity stars are reduced to $\langle\sigma_{\rm [Fe/H]}\rangle=0.39$ in {\sc Halo1} and $\langle\sigma_{\rm [Fe/H]}\rangle=0.24$ in {\sc Halo3}, showing good agreement with the observational findings.

\subsubsection{MZR comparison to other work}
\label{sec:3.2.3}
\begin{figure}
    \centering
    \includegraphics[width=85mm]{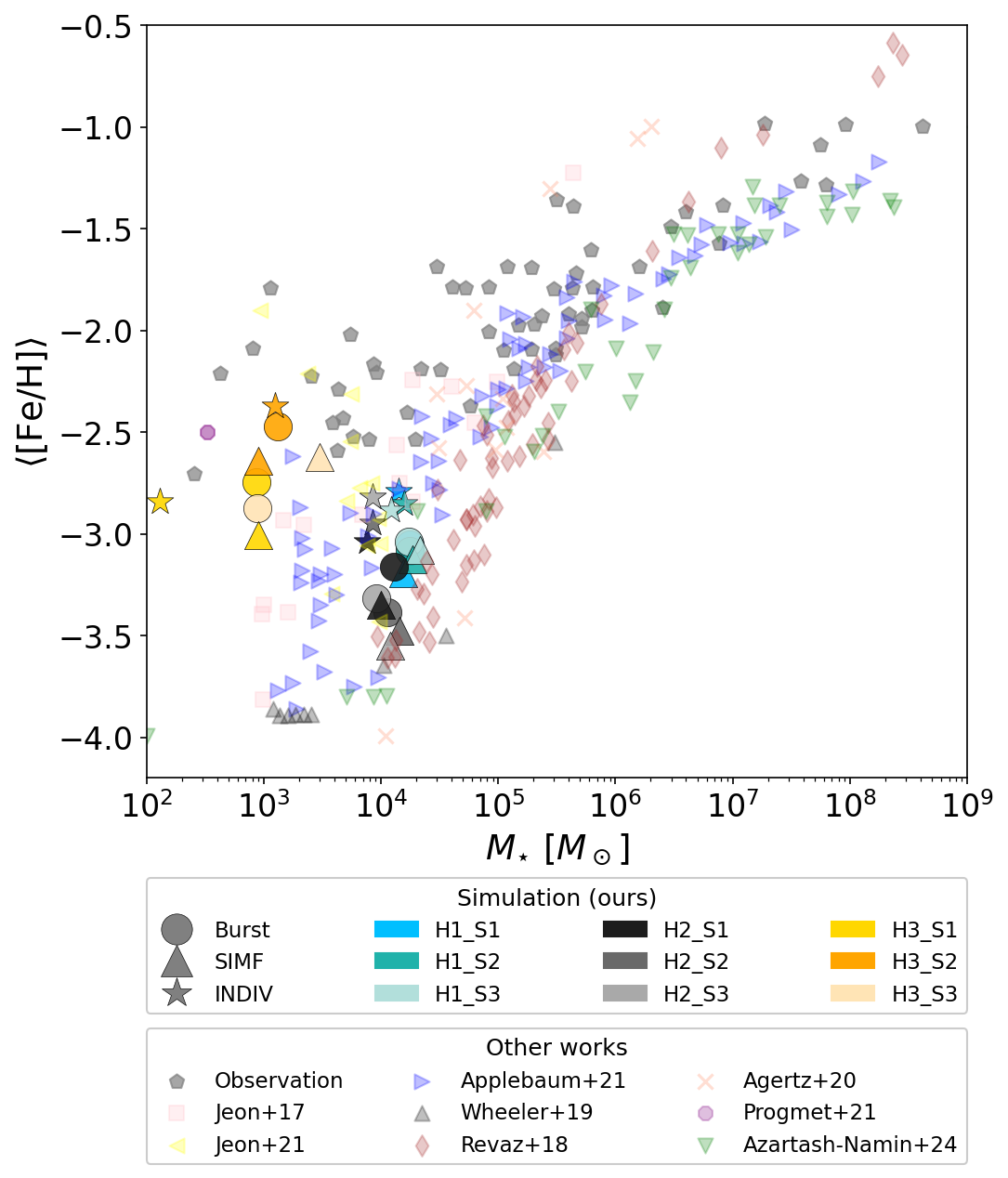}
    \caption{The stellar mass$-\rm [Fe/H]$ relation of the UFD analogs obtained from the current simulations is compared to estimates derived from observations (\citealp{Kirby2013, Fu2023}) and other theoretical studies (\citealp{Jeon2017, Wheeler2019, Agertz2020, Applebaum2021, Revaz2016, Azartash-Namin2024}). It is evident that the estimated metallicities from the {\sc Indiv} runs generally close to the observed range, while the values from the {\sc Burst} and {\sc Simf} runs are lower by $\sim$1 dex compared to those of observed MW dwarf galaxies with stellar masses of $M_\ast\lesssim 10^5\msun$. In particular, the SN feedback in the {\sc Burst} and {
    \sc Simf} runs might be too strong, causing metals to be expelled from the galaxies and reducing the metallicity of the gas clouds from which stars form.}
    \label{fig:MZR_all_27}
\end{figure}

We present the stellar mass-$\rm [Fe/H]$ relationship in Fig.~\ref{fig:MZR_all_27} based on our current simulations, compared with data from other high-resolution simulations, where the gas mass $m_{\rm gas}$ is between a few $10\msun$ and $100\msun$ (\citealp{Wheeler2019, Agertz2020, Azartash-Namin2024}), as well as with observational data (\citealp{Kirby2013, Fu2023}). The metallicity of observed MW dwarfs with stellar masses $M_{\rm\ast}\lesssim 10^4\msun$ falls within the range of $\rm -2.5<[Fe/H]<-1.7$ (\citealp{Kirby2013, Fu2023}). As noted by \citet{Fu2023}, the observed UFD regime shows a plateau rather than a decreasing trend with diminishing luminosity. This indicates that the correlation between stellar masses and metallicity is relatively weak compared to that found in more massive dwarf galaxies. Our results also support this observation, showing, for instance, that the lowest mass halo, {\sc Halo3}, exhibits higher average metallicity values compared to the more massive halo, by 0.26 dex ({\sc Halo1}) and 0.48 dex ({\sc Halo2}). Specifically, in both scenarios—{\sc Burst} and {\sc Simf}—where the burstiness in star formation is high, the correlation weakens as the stochasticity increases.

As demonstrated in Fig.~\ref{fig:MZR_all_27}, our results show that the metallicities of the {\sc Indiv} runs show good agreement with the observed values, while the {\sc Burst} and {\sc Simf} runs exhibit metallicities that are $0.5-1.0$ dex lower than the observed lower limit. This discrepancy occurs because, in the {\sc Indiv} runs, the metallicity of gas enriched by earlier SNe is more rapidly incorporated into the newly formed stars. However, during {\sc Burst} runs, where multiple SNe occur simultaneously within an SSP, the intense feedback significantly reduces the metallicity of stars formed afterward. In the {\sc Simf} approach, unlike {\sc Indiv}, if the initial SN of an SSPs does not halt star formation, subsequent SSPs form rapidly, leading to a starburst over a short period. This results in strong cumulative SN effects followed by a quenching phase. These powerful SN effects enhance metal diffusion, resulting in stars with lower metallicity compared to those formed in the {\sc Indiv} run.

Previous theoretical studies have often struggled to reproduce the high-metallicity observed in the UFD regime. \citet{Wheeler2019} suggested that lower metallicity compared to observations might be due to the absence of Pop~III stars. However, as shown in our study, the contribution of Pop~III stars is minimal because of a rapid transition from Pop~III to Pop~II star formation at early times. Likewise, \citet{Sanati2023} highlighted the challenge of increasing metallicity, even when considering metal contributions from Pop~III SNe. As noted by \citet{Jaacks2018a}, pre-enrichment by Pop~III stars appears insufficient to raise the metallicity floor beyond $\rm [Fe/H]\gtrsim -4$, especially in low-density regions.

One possible factor leading to lower metallicity in theoretical models is the presence of excessively strong outflows that hinder the re-incorporation of ejected metals. \citet{Agertz2020} supports this idea, highlighting that strong feedback from SNe can expel enriched gas from a galaxy, thus decreasing the metallicity in star-forming clouds. The research by \citet{Agertz2020} also emphasizes the role of photoionization heating by stars, suggesting that heated gas is more likely to retain metals within the ISM. As a result, this enriched gas can give rise to stars with metallicities similar to observed values. Conversely, photoionization heating might lower metallicity by making SN feedback more effective in expelling metal-rich gas from a galaxy (e.g., \citealp{Jeon2014}; \citealp{Smith2019}). The effects of stellar radiation
on the metallicity of UFD analogs will be explored in future work.

Also, \citet{Prgomet2022} proposed a metallicity-dependent IMF as a solution to the discrepancy, where the IMF becomes more top-heavy with decreasing metallicity. This results in a higher fraction of massive stars in low-metallicity environments, enhancing SN feedback and photoionization heating, which suppress star formation more effectively. Additionally, increased massive star production boosts metal output, reducing the stellar mass in UFD analogs while achieving metallicity levels comparable to those with a standard IMF. From an observational perspective, if we exclude low-metallicity stars ($\rm [Fe/H]<-4)$, as is done in observational studies, the average metallicity would increase by approximately 0.22 dex in {\sc Halo1}, 0.39 dex in {\sc Halo2},  potentially narrowing the gap between our results and the observations.

\section{Comparison to other work}
\label{sec:4}

A number of studies have explored the impact of IMF sampling and related SN injection schemes on galaxy properties (e.g., \citealp{Applebaum2020, Smith2021a, Keller2022, Hu2023}), often comparing stochastic with IMF-averaged (continuous) implementations. In this discussion, we primarily focus on IMF sampling and its associated SN effects on small dwarf galaxies. For example, \citet{Applebaum2020} performed high-resolution simulations of isolated dwarf galaxies (and MW-mass galaxies) using a stochastic IMF. The simulations involved masses of about ${\sim}1.0\times10^4\msun$ ($1.8\times10^4\msun$) for DM, ${\sim}1.4\times10^3\msun$ ($3.3\times10^3\msun$) for gas, and ${\sim}420\msun$ ($990\msun$) for stars. They compared a scheme where SNe were discretized based on the sampled stellar masses (stochastic IMF sampling) with one where SN feedback followed the IMF-averaged rate (continuous feedback). They discovered that halos with $M_{\rm halo}\lesssim10^{8.5}\msun$ formed fewer stars in the stochastic scenario than with the IMF-averaged approach. They attributed the lower stellar masses in the stochastic-IMF simulations to more clustered, burst-like SN events compared to the continuous IMF run, which couples more efficiently to the ISM, resulting in stronger heating, and thus suppressing further star formation compared to the IMF-averaged feedback that distributes energy over time.

It is important to note that our {\sc Simf} method is conceptually in line with stochastic IMF sampling but diverges in its treatment of delay times. In the study by \citet{Applebaum2020}, the delay times for SNe are determined based on the stellar lifetimes. On the other hand, our approach assumes that the most massive SN in each star particle explodes immediately. This ensures a consistent onset of feedback across three methods in our work while maintaining temporal discreteness for subsequent SNe in the {\sc SIMF} run. This choice enables us to isolate the impact of SNe's temporal discreteness without introducing any delay time effect between star formation and the first SN explosion. As a result, direct comparisons with \citet{Applebaum2020} should be made with caution, particularly given the differences in numerical resolution and delay-time prescriptions. Furthermore, \citet{Keller2022} emphasized the importance of the delay time, demonstrating that even with the same total SN energy budget, changing the delay between star formation and the first SN can greatly influence the ability of SNe to regulate star formation and drive outflows. For example, extended delays (over 20 Myr) before the onset of SN feedback can amplify the SN feedback effect due to the increased clustering of star formation.

As previously mentioned, the reason we implement instantaneous SN feedback for {\sc Burst} and {\sc Indiv}, and for the first (most massive) SN in {\sc Simf}, is partly to compensate for the absence of photoionization heating from Pop III and Pop II stars, often implemented with radiative transfer (RT). If photoionization heating were included, it could suppress subsequent star formation even before the first SN occurs. This is particularly relevant for Pop II stars, which have a lower emission rate of ionizing photons per second but a much longer lifespan compared to Pop III stars. Applying long delay times without considering photoionization heating effects could artificially boost star formation. For instance, \citet{Agertz2020} suggested that photoionization heating by stars could reduce the stellar mass by a factor of $5-10$ for dwarf galaxies with a mass of $M_{\rm vir} \approx 10^9\msun$. To evaluate the impact of ignoring delay times and photoionization heating effects, we conducted two additional {\sc Indiv} simulations for {\sc Halo1}: (i) without RT but with SN delays, and (ii) with both RT and SN delays, comparing these scenarios to our default no-RT, no-delay run. The resulting stellar masses were about ${\sim}3.7\times10^4,\msun$ (noRT+delay), ${\sim}2.7\times10^4,\msun$ (RT+delay), and ${\sim}1.4\times10^4,\msun$ (noRT+noDelay). These experiments show that including delays without RT could increase stellar mass compared to our default no-RT, no-delay setup, while adding RT partially mitigates this increase.

In exploring the effects of photoionization heating in conjunction with the IMF sampling method, \citet{Smith2021a} examined IMF implementations in an isolated dwarf galaxy with a virial mass of $M_{\rm vir}=10^{10}\msun$, using a baryonic mass resolution of $20\msun$. They compared continuous and stochastic IMF sampling approaches, where the continuous IMF also utilized individual SN feedback via Poisson sampling from the IMF-averaged rate. Their study showed that when SN feedback alone is used as stellar feedback, and both approaches adopt the same SN delay time (i.e., waiting for massive stars to evolve before exploding) and the same star particle mass, the results are identical between the continuous IMF and the stochastic sampling simulations.

However, \citet{Smith2021a} pointed out that there is a notable difference between stochastic and continuous IMF sampling when photoionization heating is applied, leading to more effective regulation of star formation with IMF-averaged rates compared to stochastic IMF sampling. This is because, in the IMF-averaged approach, each star generates H II regions with ionizing photons, whereas the stochastic IMF sampling results in fewer H II regions linked to the randomly selected stars.
They also cautioned that the impact of IMF sampling depends on star particle resolution and feedback channels. Given that the galaxies they studied are about 100 times more massive than our UFD analogs, their systems are naturally less sensitive to burstiness caused by IMF sampling. Extending such comparisons to the UFD regime with explicit RT calculations remains a crucial area for future research.

\section{Caveat: Adjustment of star formation efficiency}
\label{sec:5}

The star formation efficiency (SFE) represents the fraction of gas converted into stars per free-fall time and is often treated as a free parameter in simulations. While SFEs have been widely used in galaxy formation studies, there is no consensus on how they should scale with numerical resolution, especially when simulating UFD galaxies. For instance, \citet{Agertz2020} used $\epsilon_{\rm ff}=0.1$ for $300\ \mathrm{M_\odot}$ gas particles, \citet{Andersson2024} adopted the same value at $200\ \mathrm{M_\odot}$ resolution, and \citet{Brauer2025} employed $\epsilon_{\rm ff}=0.02$ for similar particle masses.

In our simulations, all models applied a uniform SFE regardless of their star-particle mass. However, the {\sc Indiv} runs form star particles, including both individual and SSP particles, with an average mass of $63\ \msun$, whereas the {\sc Burst} and {\sc Simf} runs form SSPs of $500\ \msun$. This nearly eightfold difference in mass resolution suggests that different effective efficiencies may be required. To assess the impact of such scaling, we introduce a correction to the star formation probability, proportional to the ratio of gas-particle mass to the newly formed star-particle mass:

\begin{equation}
P = \frac{m_{\rm gas}}{\Delta m_\ast}\frac{\Delta t}{\tau_\ast}
= \frac{63}{500}\frac{\Delta t}{\tau_\ast},
\end{equation}
where $\tau_\ast = \tau_{\rm ff}/\epsilon_{\rm ff}$ is the star formation timescale determined by the local free-fall time $\tau_{\rm ff}$ and the efficiency per free-fall $\epsilon_{\rm ff}$. This correction effectively reduces the efficiency by a factor of eight for the {\sc Burst} and {\sc SIMF} models.

To evaluate its influence, we additionally perform the {\sc Burst} model of {\sc Halo1} three times with the reduced efficiency and different random seeds. The results of these modified runs are presented in Appendix \ref{sec:B}, showing the maximum hydrogen density evolution (Fig. \ref{figB1}) and the MZR (Fig. \ref{figB2}). Notably, the stellar metallicity, along with their formation times, exhibits more continuous behavior compared to the original {\sc Burst} and {\sc Simf} runs with {\sc Halo1} (see Fig. \ref{fig_density_evolution}). Furthermore, it shows similarly higher maximum gas density evolution, qualitatively resembling the {\sc Indiv} model. This is enabled by lowering the efficiency, with delayed star formation enhancing the star-forming densities. Moreover, the modified runs result in significantly higher stellar masses—about 2.2 times greater than the fiducial {\sc Burst} case and 1.7 times higher than the {\sc Indiv} runs—as well as higher stellar metallicities. These results are illustrated in the $n_{\rm H,max}$ evolution and the MZR in Appendix \ref{sec:B}.

These results demonstrate that variations in $\epsilon_{\rm ff}$ can substantially affect both the star formation and enrichment histories. The reduced-efficiency runs reproduce some qualitative features of the {\sc Indiv} model—such as higher gas densities and comparable stellar metallicities—but produce about 2.2 times larger stellar masses. This outcome indicates that simply lowering the efficiency does not lead to convergence toward the {\sc Indiv} results. The lack of convergence suggests that the outcomes are highly sensitive to how SN feedback couples to the surrounding gas, rather than being solely governed by the choice of $\epsilon_{\rm ff}$. Therefore, systematic convergence tests on $\epsilon_{\rm ff}$, together with careful calibration of feedback implementation, are essential for developing physically consistent and resolution-independent star formation models.

It is important to note that as UFD simulations achieve higher resolutions where stellar particle mass often exceeds gas particle mass, a fixed star formation efficiency—commonly set at $\epsilon_{\rm ff}=0.01-0.1$ —is often employed without accounting for the mass ratio factor in star formation from gas. Despite utilizing varying stellar-to-gas mass ratios and efficiencies, different simulations frequently compare outcomes, such as stellar mass and metallicity, directly across simulations with differing efficiencies. Our findings highlight the necessity for careful consideration of these factors.

\section{Summary and Conclusions}
\label{sec:6}
In our study, we examined the impact of various Initial Mass Function (IMF) sampling and supernova (SN) injection methods on the Star Formation Histories (SFHs) and metal enrichment histories of Ultra-Faint Dwarf (UFD) analogs using cosmological hydrodynamic zoom-in simulations. Achieving a high mass resolution ($m_{\rm gas}\approx 63\msun$) is essential for accurately resolving UFD analogs, the faintest galaxy systems in the Universe ($L < 10^5\lsun$) (e.g., \citealp{Simon2019}). As mass resolution increases, it becomes necessary to revise existing sub-grid models, originally developed to represent baryonic physics in massive galaxies. Importantly, the IMF sampling method and SN injection play a vital role in simulating UFD analogs, as it changes star particle resolution and behavior of SN feedback, thus shaping the SFHs of the simulated galaxies.

Specifically, in this study, we conducted a series of hydrodynamic simulations on UFD analogs characterized by a virial mass of $M_{\rm vir} \approx 2-8 \times 10^8\msun$ and stellar masses of $M_{\ast} \lesssim 10^2-10^{4}\msun$, comparing three different methods. The first method, called {\sc Burst}, models stars to form continuously according to a given IMF but releases the entire SN energy all at once, significantly impacting the surrounding medium. The second method, named {\sc Simf} (Stochastic IMF), involves stochastically selecting the masses of stars, with SNe exploding one after another with a time lag during a single starburst, releasing the SN energy discretely. This method better captures the discrete nature of SN events. The final method, {\sc Indiv}, treats massive stars that will undergo SN explosions as individual star particles, while smaller stars ($m_{*}<8\msun$) are represented as a single stellar population (SSP). This method more realistically represents the stochastic nature of SNe, as each SN explodes individually, unlike the {\sc Simf}, which can result in multiple SNe originating from a SSP particle.

The main results are summarized as follows.

\begin{itemize}

    \item Unlike massive galaxies, which primarily form stars in situ within their main progenitor halo, we found that stars in UFD analogs originate from multiple progenitor haloes—typically 4 to 8 at $z=6$ for $M_{\rm vir} \approx 5-8 \times 10^8 \msun$ at $z=0$. These progenitor haloes eventually merge to form a single UFD analog at a later stage. Star formation in these progenitors begins when the virial mass is around $M_{\rm vir} \approx 10^5-10^6 \msun$ at $z \gtrsim 14$ and ceases before reionization. Consequently, the identity of the primary halo can change, as no single halo consistently remains more massive than the others at high redshifts ($z\gtrsim6)$. 
    
    \item Our findings indicate that despite variations in sampling and SN feedback injection schemes, the final stellar masses are generally similar across different methods. However, star formation is more continuous in the {\sc Indiv} scenario. This continuity arises because SN feedback in the {\sc Indiv} runs is more spatially and temporally dispersed, resulting in less suppression. Consequently, the gas can re-accrete and cool between SN events, leading to a rapid rebound in density that facilitates star formation. This process achieves significantly higher star-forming densities—by a factor of 10 to 100—compared to the {\sc Burst} and {\sc Simf} runs.

    \item Our analysis indicates that the systematically higher stellar metallicities observed in the {\sc Indiv} runs are also due to less suppressive SN feedback, which allows the gas to sustain higher densities. This maintained high-density environment promotes in situ self-enrichment. In contrast, clustered SNe in the non-{\sc Indiv} runs tend to strongly evacuate gas, significantly lower densities, and suppress subsequent star formation. Furthermore, the lower star particle masses in the {\sc Indiv} runs allow for more precise resolution of localized metal recycling, further enhancing the metallicity of newly formed stars. These factors collectively lead to a metallicity enhancement in the simulated galaxies within the {\sc Indiv} scenario, emphasizing the critical role of IMF sampling and feedback implementation in shaping the chemical evolution of UFD analogs.

    \item When comparing the metallicity distribution function (MDF) of observed UFDs with simulated UFD analogs, our results exhibit in the {\sc Indiv} runs, the MDF peaks at higher metallicities, closely matching observational data for {\sc Halo1} and {\sc Halo2}, unlike the {\sc Burst} and {\sc Simf} runs, which skew towards lower metallicities. This is attributed to the absence of stars with relatively high metallicities ($\rm [Fe/H]\gtrsim-2$) in these runs. Also, our results exhibit a low-metallicity tail ($\rm [Fe/H]\lesssim-4$) that is rarely found in observations, reducing average metallicity. We find that these stars originate from external metal enrichment or form at density peaks polluted by metals. Excluding these low-metallicity stars ($\rm [Fe/H]\lesssim-4$) brings the mean metallicities of simulated UFDs into good agreement with observational values and reduces the dispersion in the MDFs.

    \item In the context of the stellar mass-metallicity relation, the average metallicity of the simulated UFD analogs, particularly in the {\sc Indiv} runs, shows a good agreement with observed values. However, estimates in the {\sc Burst} and {\sc Simf} runs tend to show lower $\rm \langle[Fe/H]\rangle$ values by about 1 dex for stellar masses of $M_\ast\lesssim 10^5\msun$ compared to observations. This difference can be attributed to the continuous nature of star formation in {\sc Indiv}, where the gas metallicity shaped by previous SN events is rapidly reflected in the stellar metallicity, since stars are sampled individually. Moreover, the relatively weak energy from a single SN event in {\sc Indiv} allows for continuous star formation, leading to stars with relatively high-metallicity ($\rm [Fe/H]\gtrsim-2$).

    \item In this study, we apply the same star formation efficiency across three methods. However, caution is needed regarding efficiency based on star sampling. As UFD simulations reach higher resolutions, where stellar particle mass can exceed gas particle mass, a fixed star formation efficiency, typically $\epsilon_{\rm ff}=0.01 - 0.1$, is used without considering the mass ratio. Despite varying stellar-to-gas mass ratios and efficiencies, simulations often directly compare galaxy properties like stellar mass and metallicity across different simulations. It is essential to adjust efficiency according to star sampling, although most UFD galaxy simulations overlook these factors.
\\

\end{itemize}

UFDs are essential components in the hierarchical structure of larger galaxies and present invaluable opportunities to investigate the underlying mechanisms of galaxy formation and evolution. Despite being the faintest and most rudimentary systems, their physical properties are shaped by a complex interplay of baryonic processes, including star formation, stellar feedback, metal enrichment, and the dynamics of dark matter halo assembly. The properties of UFD analogs can vary considerably based on the sub-grid models employed to simulate these baryonic physics. Thus, it is imperative to advance theoretical research to develop and implement more sophisticated and realistic numerical models to enhance our understanding of UFDs. Alongside these theoretical developments, forthcoming observational data from the Vera C. Rubin Observatory's Legacy Survey of Space and Time (LSST, \citealp{LSST2019}) and next-generation telescopes such as the Giant Magellan Telescope (GMT) are anticipated to significantly deepen our comprehension of the universe's smallest galaxy systems.

\section*{acknowledgements}
We sincerely thank the referee for their insightful comments and for pointing out the critical issue, which has contributed significantly to the improvement of our study. The authors extend their gratitude to Chanhee Shin for implementing the stochastic method in the code and for his assistance in writing the manuscript. The authors also thank Volker Springel, Joop Schaye, and Claudio Dalla Vecchia for granting permission to use their versions of \textsc{gadget}. M.~J. acknowledges support from the Korean government-funded National Research Foundation (NRF) grants (MSIT) under the grant numbers 2021R1A2C109491713 and 2022M3K3A1093827. M.G. is supported by the BK21 FOUR program through NRF under Ministry of Education (Kyung Hee University, School of Space Science).

\section*{DATA AVAILABILITY}
The simulation data and results of this paper may be available upon
request.

\bibliography{myrefs2}{}

@ARTICLE{Durier2012,
   author = {{Durier}, F. and {Dalla Vecchia}, C.},
    title = "{Implementation of feedback in smoothed particle hydrodynamics: towards concordance of methods}",
  journal = {MNRAS},
archivePrefix = "arXiv",
   eprint = {1105.3729},
 primaryClass = "astro-ph.CO",
     year = 2012,
    month = jan,
   volume = 419,
    pages = {465-478},
      doi = {10.1111/j.1365-2966.2011.19712.x},
   adsurl = {http://adsabs.harvard.edu/abs/2012MNRAS.419..465D}
}

@ARTICLE{Saitoh2009,
   author = {{Saitoh}, T.~R. and {Makino}, J.},
    title = "{A Necessary Condition for Individual Time Steps in SPH Simulations}",
  journal = {ApJ},
archivePrefix = "arXiv",
   eprint = {0808.0773},
     year = 2009,
    month = jun,
   volume = 697,
    pages = {L99-L102},
      doi = {10.1088/0004-637X/697/2/L99},
   adsurl = {http://adsabs.harvard.edu/abs/2009ApJ...697L..99S}
}

@ARTICLE{planck2016,
       author = {{Planck Collaboration}},
        title = "{Planck 2015 results. XIII. Cosmological parameters}",
      journal = {A\&A},
         year = 2016,
        month = sep,
       volume = {594},
          eid = {A13},
        pages = {A13},
          doi = {10.1051/0004-6361/201525830},
archivePrefix = {arXiv},
       eprint = {1502.01589},
 primaryClass = {astro-ph.CO},
       adsurl = {https://ui.adsabs.harvard.edu/abs/2016A&A...594A..13P}
}

@ARTICLE{Abel2000,
   author = {{Abel}, T. and {Bryan}, G.~L. and {Norman}, M.~L.},
    title = "{The Formation and Fragmentation of Primordial Molecular Clouds}",
  journal = {ApJ},
   eprint = {arXiv:astro-ph/0002135},
     year = 2000,
    month = sep,
   volume = 540,
    pages = {39-44},
      doi = {10.1086/309295},
   adsurl = {http://adsabs.harvard.edu/abs/2000ApJ...540...39A}
}

@ARTICLE{Bromm2001a,
   author = {{Bromm}, V. and {Kudritzki}, R.~P. and {Loeb}, A.},
    title = "{{noopsort{a}}{G}eneric Spectrum and Ionization Efficiency of a Heavy Initial Mass Function for the First Stars}",
  journal = {ApJ},
   eprint = {arXiv:astro-ph/0007248},
     year = 2001,
    month = may,
   volume = 552,
    pages = {464-472},
      doi = {10.1086/320549},
   adsurl = {http://adsabs.harvard.edu/abs/2001ApJ...552..464B}
}

@ARTICLE{Bromm2013,
   author = {{Bromm}, V.},
    title = "{Formation of the first stars}",
  journal = {Rep. Prog. Phys.},
archivePrefix = "arXiv",
   eprint = {1305.5178},
 primaryClass = "astro-ph.CO",
     year = 2013,
    month = nov,
   volume = 76,
   number = 11,
      eid = {112901},
    pages = {112901},
      doi = {10.1088/0034-4885/76/11/112901},
   adsurl = {http://adsabs.harvard.edu/abs/2013RPPh...76k2901B}
}

@ARTICLE{Fan2006,
   author = {{Fan et al.}},
    title = "{A Survey of z{\gt}5.7 Quasars in the Sloan Digital Sky Survey. IV. Discovery of Seven Additional Quasars}",
  journal = {AJ},
   eprint = {arXiv:astro-ph/0512080},
     year = 2006,
    month = mar,
   volume = 131,
    pages = {1203-1209},
      doi = {10.1086/500296},
   adsurl = {http://adsabs.harvard.edu/abs/2006AJ....131.1203F}
}

@ARTICLE{Greif2009,
   author = {{Greif}, T.~H. and {Johnson}, J.~L. and {Klessen}, R.~S. and 
	{Bromm}, V.},
    title = "{The observational signature of the first HII regions}",
  journal = {MNRAS},
archivePrefix = "arXiv",
   eprint = {0905.1717},
 primaryClass = "astro-ph.CO",
     year = 2009,
    month = oct,
   volume = 399,
    pages = {639-649},
      doi = {10.1111/j.1365-2966.2009.15336.x},
   adsurl = {http://adsabs.harvard.edu/abs/2009MNRAS.399..639G}
}

@ARTICLE{Heger2002,
   author = {{Heger}, A. and {Woosley}, S.~E.},
    title = "{The Nucleosynthetic Signature of Population III}",
  journal = {ApJ},
   eprint = {arXiv:astro-ph/0107037},
     year = 2002,
    month = mar,
   volume = 567,
    pages = {532-543},
      doi = {10.1086/338487},
   adsurl = {http://adsabs.harvard.edu/abs/2002ApJ...567..532H}
}

@ARTICLE{Heger2003,
   author = {{Heger}, A. and {Fryer}, C.~L. and {Woosley}, S.~E. and {Langer}, N. and 
	{Hartmann}, D.~H.},
    title = "{How Massive Single Stars End Their Life}",
  journal = {ApJ},
   eprint = {arXiv:astro-ph/0212469},
     year = 2003,
    month = jul,
   volume = 591,
    pages = {288-300},
      doi = {10.1086/375341},
   adsurl = {http://adsabs.harvard.edu/abs/2003ApJ...591..288H}
}

@ARTICLE{Heger2010,
   author = {{Heger}, A. and {Woosley}, S.~E.},
    title = "{Nucleosynthesis and Evolution of Massive Metal-free Stars}",
  journal = {ApJ},
archivePrefix = "arXiv",
   eprint = {0803.3161},
     year = 2010,
    month = nov,
   volume = 724,
    pages = {341-373},
      doi = {10.1088/0004-637X/724/1/341},
   adsurl = {http://adsabs.harvard.edu/abs/2010ApJ...724..341H}
}

@ARTICLE{Komatsu2011,
   author = {{Komatsu}, E. and {Smith}, K.~M. and {Dunkley}, J. and {Bennett}, C.~L. and 
	{Gold}, B. and {Hinshaw}, G. and {Jarosik}, N. and {Larson}, D. and 
	{Nolta}, M.~R. and {Page}, L. and {Spergel}, et al.},
    title = "{Seven-year Wilkinson Microwave Anisotropy Probe (WMAP) Observations: Cosmological Interpretation}",
  journal = {ApJS},
archivePrefix = "arXiv",
   eprint = {1001.4538},
 primaryClass = "astro-ph.CO",
     year = 2011,
    month = feb,
   volume = 192,
      eid = {18},
    pages = {18},
      doi = {10.1088/0067-0049/192/2/18},
   adsurl = {http://adsabs.harvard.edu/abs/2011ApJS..192...18K}
}

@ARTICLE{Omukai2000,
   author = {{Omukai}, K.},
    title = "{Protostellar Collapse with Various Metallicities}",
  journal = {ApJ},
   eprint = {arXiv:astro-ph/0003212},
     year = 2000,
    month = may,
   volume = 534,
    pages = {809-824},
      doi = {10.1086/308776},
   adsurl = {http://adsabs.harvard.edu/abs/2000ApJ...534..809O}
}

@ARTICLE{Springel2001,
   author = {{Springel}, V. and {White}, S.~D.~M. and {Tormen}, G. and {Kauffmann}, G.
	},
    title = "{Populating a cluster of galaxies - I. Results at [formmu2]z=0}",
  journal = {MNRAS},
   eprint = {arXiv:astro-ph/0012055},
     year = 2001,
    month = dec,
   volume = 328,
    pages = {726-750},
      doi = {10.1046/j.1365-8711.2001.04912.x},
   adsurl = {http://adsabs.harvard.edu/abs/2001MNRAS.328..726S}
}

@ARTICLE{Springel2005,
   author = {{Springel}, V.},
    title = "{The cosmological simulation code GADGET-2}",
  journal = {MNRAS},
   eprint = {arXiv:astro-ph/0505010},
     year = 2005,
    month = dec,
   volume = 364,
    pages = {1105-1134},
      doi = {10.1111/j.1365-2966.2005.09655.x},
   adsurl = {http://adsabs.harvard.edu/abs/2005MNRAS.364.1105S}
}

@ARTICLE{Yoon2012,
   author = {{Yoon}, S.-C. and {Dierks}, A. and {Langer}, N.},
    title = "{Evolution of massive Population III stars with rotation and magnetic fields}",
  journal = {A\&A},
archivePrefix = "arXiv",
   eprint = {1201.2364},
 primaryClass = "astro-ph.SR",
     year = 2012,
    month = jun,
   volume = 542,
      eid = {A113},
    pages = {A113},
      doi = {10.1051/0004-6361/201117769},
   adsurl = {http://adsabs.harvard.edu/abs/2012A%26A...542A.113Y}
}

@ARTICLE{Klessen2003,
   author = {{Klessen}, R.~S. and {Lin}, D.~N.},
    title = "{Diffusion in supersonic turbulent compressible flows}",
  journal = {Physical Review E},
   eprint = {astro-ph/0302527},
     year = 2003,
    month = apr,
   volume = 67,
   number = 4,
      eid = {046311},
    pages = {046311},
      doi = {10.1103/PhysRevE.67.046311},
   adsurl = {http://adsabs.harvard.edu/abs/2003PhRvE..67d6311K}
}

@ARTICLE{Hirano2014,
   author = {{Hirano}, S. and {Hosokawa}, T. and {Yoshida}, N. and {Umeda}, H. and 
	{Omukai}, K. and {Chiaki}, G. and {Yorke}, H.~W.},
    title = "{One Hundred First Stars: Protostellar Evolution and the Final Masses}",
  journal = {ApJ},
archivePrefix = "arXiv",
   eprint = {1308.4456},
 primaryClass = "astro-ph.CO",
     year = 2014,
    month = feb,
   volume = 781,
      eid = {60},
    pages = {60},
      doi = {10.1088/0004-637X/781/2/60},
   adsurl = {http://adsabs.harvard.edu/abs/2014ApJ...781...60H}
}

@ARTICLE{Schneider2010,
   author = {{Schneider}, R. and {Omukai}, K.},
    title = "{Metals, dust and the cosmic microwave background: fragmentation of high-redshift star-forming clouds}",
  journal = {MNRAS},
archivePrefix = "arXiv",
   eprint = {0910.3665},
 primaryClass = "astro-ph.CO",
     year = 2010,
    month = feb,
   volume = 402,
    pages = {429-435},
      doi = {10.1111/j.1365-2966.2009.15891.x},
   adsurl = {http://adsabs.harvard.edu/abs/2010MNRAS.402..429S}
}

@ARTICLE{Haardt2012,
   author = {{Haardt}, F. and {Madau}, P.},
    title = "{Radiative Transfer in a Clumpy Universe. IV. New Synthesis Models of the Cosmic UV/X-Ray Background}",
  journal = {ApJ},
archivePrefix = "arXiv",
   eprint = {1105.2039},
 primaryClass = "astro-ph.CO",
     year = 2012,
    month = feb,
   volume = 746,
      eid = {125},
    pages = {125},
      doi = {10.1088/0004-637X/746/2/125},
   adsurl = {http://adsabs.harvard.edu/abs/2012ApJ...746..125H}
}

@ARTICLE{Glover2007,
  author = {{Glover}, S.~C.~O. and {Jappsen}, A.-K.},
    title = "{Star Formation at Very Low Metallicity. I. Chemistry and Cooling at Low Densities}",
  journal = {ApJ},
archivePrefix = "arXiv",
   eprint = {0705.0182},
     year = 2007,
    month = sep,
   volume = 666,
    pages = {1-19},
      doi = {10.1086/519445},
   adsurl = {http://adsabs.harvard.edu/abs/2007ApJ...666....1G}
}

@ARTICLE{Schaye2010,
   author = {{Schaye}, J. and {Dalla Vecchia}, C. and {Booth}, C.~M. and 
	{Wiersma}, R.~P.~C. and {Theuns}, T. and {Haas}, M.~R. and {Bertone}, S. and 
	{Duffy}, A.~R. and {McCarthy}, I.~G. and {van de Voort}, F.},
    title = "{The physics driving the cosmic star formation history}",
  journal = {MNRAS},
archivePrefix = "arXiv",
   eprint = {0909.5196},
 primaryClass = "astro-ph.CO",
     year = 2010,
    month = mar,
   volume = 402,
    pages = {1536-1560},
      doi = {10.1111/j.1365-2966.2009.16029.x},
   adsurl = {http://adsabs.harvard.edu/abs/2010MNRAS.402.1536S}
}

@ARTICLE{Kirby2013,
   author = {{Kirby}, E.~N. and {Cohen}, J.~G. and {Guhathakurta}, P. and 
	{Cheng}, L. and {Bullock}, J.~S. and {Gallazzi}, A.},
    title = "{The Universal Stellar Mass-Stellar Metallicity Relation for Dwarf Galaxies}",
  journal = {ApJ},
archivePrefix = "arXiv",
   eprint = {1310.0814},
 primaryClass = "astro-ph.GA",
     year = 2013,
    month = dec,
   volume = 779,
      eid = {102},
    pages = {102},
      doi = {10.1088/0004-637X/779/2/102},
   adsurl = {http://adsabs.harvard.edu/abs/2013ApJ...779..102K}
}

@ARTICLE{Tolstoy2009,
   author = {{Tolstoy}, E. and {Hill}, V. and {Tosi}, M.},
    title = "{Star-Formation Histories, Abundances, and Kinematics of Dwarf Galaxies in the Local Group}",
  journal = {ARA\&A},
archivePrefix = "arXiv",
   eprint = {0904.4505},
 primaryClass = "astro-ph.CO",
     year = 2009,
    month = sep,
   volume = 47,
    pages = {371-425},
      doi = {10.1146/annurev-astro-082708-101650},
   adsurl = {http://adsabs.harvard.edu/abs/2009ARA%26A..47..371T}
}

@ARTICLE{Bovill2009,
   author = {{Bovill}, M.~S. and {Ricotti}, M.},
    title = "{Pre-Reionization Fossils, Ultra-Faint Dwarfs, and the Missing Galactic Satellite Problem}",
  journal = {ApJ},
archivePrefix = "arXiv",
   eprint = {0806.2340},
     year = 2009,
    month = mar,
   volume = 693,
    pages = {1859-1870},
      doi = {10.1088/0004-637X/693/2/1859},
   adsurl = {http://adsabs.harvard.edu/abs/2009ApJ...693.1859B}
}

@ARTICLE{Simpson2013,
   author = {{Simpson}, C.~M. and {Bryan}, G.~L. and {Johnston}, K.~V. and 
	{Smith}, B.~D. and {Mac Low}, M.-M. and {Sharma}, S. and {Tumlinson}, J.
	},
    title = "{The effect of feedback and reionization on star formation in low-mass dwarf galaxy haloes}",
  journal = {MNRAS},
archivePrefix = "arXiv",
   eprint = {1211.1071},
 primaryClass = "astro-ph.GA",
     year = 2013,
    month = jul,
   volume = 432,
    pages = {1989-2011},
      doi = {10.1093/MNRAS/stt474},
   adsurl = {http://adsabs.harvard.edu/abs/2013MNRAS.432.1989S}
}

@ARTICLE{Behroozi2013,
   author = {{Behroozi}, P.~S. and {Wechsler}, R.~H. and {Conroy}, C.},
    title = "{The Average Star Formation Histories of Galaxies in Dark Matter Halos from z = 0-8}",
  journal = {ApJ},
archivePrefix = "arXiv",
   eprint = {1207.6105},
 primaryClass = "astro-ph.CO",
     year = 2013,
    month = jun,
   volume = 770,
      eid = {57},
    pages = {57},
      doi = {10.1088/0004-637X/770/1/57},
   adsurl = {http://adsabs.harvard.edu/abs/2013ApJ...770...57B}
}

@ARTICLE{Vecchia2012,
   author = {{Dalla Vecchia}, C. and {Schaye}, J.},
    title = "{Simulating galactic outflows with thermal supernova feedback}",
  journal = {MNRAS},
archivePrefix = "arXiv",
   eprint = {1203.5667},
     year = 2012,
    month = oct,
   volume = 426,
    pages = {140-158},
      doi = {10.1111/j.1365-2966.2012.21704.x},
   adsurl = {http://adsabs.harvard.edu/abs/2012MNRAS.426..140D}
}

@ARTICLE{Marigo2001,
   author = {{Marigo}, P.},
    title = "{Chemical yields from low- and intermediate-mass stars: Model predictions and basic observational constraints}",
  journal = {A\&A},
   eprint = {astro-ph/0012181},
     year = 2001,
    month = apr,
   volume = 370,
    pages = {194-217},
      doi = {10.1051/0004-6361:20000247},
   adsurl = {http://adsabs.harvard.edu/abs/2001A%26A...370..194M}
}

@ARTICLE{Portinari1998,
   author = {{Portinari}, L. and {Chiosi}, C. and {Bressan}, A.},
    title = "{Galactic chemical enrichment with new metallicity dependent stellar yields}",
  journal = {A\&A},
   eprint = {astro-ph/9711337},
     year = 1998,
    month = jun,
   volume = 334,
    pages = {505-539},
   adsurl = {http://adsabs.harvard.edu/abs/1998A%26A...334..505P}
}

@ARTICLE{Schmidt1959,
   author = {{Schmidt}, M.},
    title = "{The Rate of Star Formation.}",
  journal = {ApJ},
     year = 1959,
    month = mar,
   volume = 129,
    pages = {243},
      doi = {10.1086/146614},
   adsurl = {http://adsabs.harvard.edu/abs/1959ApJ...129..243S}
}

@ARTICLE{Weisz2014,
   author = {{Weisz}, D.~R. and {Dolphin}, A.~E. and {Skillman}, E.~D. and 
	{Holtzman}, J. and {Gilbert}, K.~M. and {Dalcanton}, J.~J. and 
	{Williams}, B.~F.},
    title = "{The Star Formation Histories of Local Group Dwarf Galaxies. II. Searching For Signatures of Reionization}",
  journal = {ApJ},
archivePrefix = "arXiv",
   eprint = {1405.3281},
     year = 2014,
    month = jul,
   volume = 789,
      eid = {148},
    pages = {148},
      doi = {10.1088/0004-637X/789/2/148},
   adsurl = {http://adsabs.harvard.edu/abs/2014ApJ...789..148W}
}

@ARTICLE{Hirano2015,
   author = {{Hirano}, S. and {Hosokawa}, T. and {Yoshida}, N. and {Omukai}, K. and 
	{Yorke}, H.~W.},
    title = "{Primordial star formation under the influence of far ultraviolet radiation: 1540 cosmological haloes and the stellar mass distribution}",
  journal = {MNRAS},
archivePrefix = "arXiv",
   eprint = {1501.01630},
     year = 2015,
    month = mar,
   volume = 448,
    pages = {568-587},
      doi = {10.1093/MNRAS/stv044},
   adsurl = {http://adsabs.harvard.edu/abs/2015MNRAS.448..568H}
}

@ARTICLE{Hahn2011,
   author = {{Hahn}, O. and {Abel}, T.},
    title = "{Multi-scale initial conditions for cosmological simulations}",
  journal = {MNRAS},
archivePrefix = "arXiv",
   eprint = {1103.6031},
     year = 2011,
    month = aug,
   volume = 415,
    pages = {2101-2121},
      doi = {10.1111/j.1365-2966.2011.18820.x},
   adsurl = {http://adsabs.harvard.edu/abs/2011MNRAS.415.2101H}
}

@ARTICLE{Ferland1998,
   author = {{Ferland}, G.~J. and {Korista}, K.~T. and {Verner}, D.~A. and 
	{Ferguson}, J.~W. and {Kingdon}, J.~B. and {Verner}, E.~M.},
    title = "{CLOUDY 90: Numerical Simulation of Plasmas and Their Spectra}",
  journal = {PASP},
     year = 1998,
    month = jul,
   volume = 110,
    pages = {761-778},
      doi = {10.1086/316190},
   adsurl = {http://adsabs.harvard.edu/abs/1998PASP..110..761F}
}

@ARTICLE{Barris2006,
   author = {{Barris}, B.~J. and {Tonry}, J.~L.},
    title = "{The Rate of Type Ia Supernovae at High Redshift}",
  journal = {ApJ},
   eprint = {astro-ph/0509655},
     year = 2006,
    month = jan,
   volume = 637,
    pages = {427-438},
      doi = {10.1086/498292},
   adsurl = {http://adsabs.harvard.edu/abs/2006ApJ...637..427B}
}

@ARTICLE{Forster2006,
   author = {{F{\"o}rster}, F. and {Wolf}, C. and {Podsiadlowski}, P. and 
	{Han}, Z.},
    title = "{Constraints on Type Ia supernova progenitor time delays from high-z supernovae and the star formation history}",
  journal = {MNRAS},
   eprint = {astro-ph/0601454},
     year = 2006,
    month = jun,
   volume = 368,
    pages = {1893-1904},
      doi = {10.1111/j.1365-2966.2006.10258.x},
   adsurl = {http://adsabs.harvard.edu/abs/2006MNRAS.368.1893F}
}

@ARTICLE{Safranek2016,
   author = {{Safranek-Shrader}, C. and {Montgomery}, M.~H. and {Milosavljevi{\'c}}, M. and 
	{Bromm}, V.},
    title = "{Star formation in the first galaxies - III. Formation, evolution, and characteristics of the first metal-enriched stellar cluster}",
  journal = {\mnras},
archivePrefix = "arXiv",
   eprint = {1501.03212},
     year = 2016,
    month = jan,
   volume = 455,
    pages = {3288-3302},
      doi = {10.1093/mnras/stv2545},
   adsurl = {http://adsabs.harvard.edu/abs/2016MNRAS.455.3288S}
}

@ARTICLE{Munshi2013,
   author = {{Munshi}, F. and {Governato}, F. and {Brooks}, A.~M. and {Christensen}, C. and 
	{Shen}, S. and {Loebman}, S. and {Moster}, B. and {Quinn}, T. and 
	{Wadsley}, J.},
    title = "{Reproducing the Stellar Mass/Halo Mass Relation in Simulated {$\Lambda$}CDM Galaxies: Theory versus Observational Estimates}",
  journal = {ApJ},
archivePrefix = "arXiv",
   eprint = {1209.1389},
     year = 2013,
    month = mar,
   volume = 766,
      eid = {56},
    pages = {56},
      doi = {10.1088/0004-637X/766/1/56},
   adsurl = {http://adsabs.harvard.edu/abs/2013ApJ...766...56M}
}

@ARTICLE{Garrison2016,
       author = {{Garrison-Kimmel}, Shea and {Bullock}, James S. and {Boylan-Kolchin}, Michael and {Bardwell}, Emma},
        title = "{Organized chaos: scatter in the relation between stellar mass and halo mass in small galaxies}",
      journal = {\mnras},
         year = 2017,
        month = jan,
       volume = {464},
       number = {3},
        pages = {3108-3120},
          doi = {10.1093/mnras/stw2564},
archivePrefix = {arXiv},
       eprint = {1603.04855},
 primaryClass = {astro-ph.GA},
       adsurl = {https://ui.adsabs.harvard.edu/abs/2017MNRAS.464.3108G}
}

@ARTICLE{Gunn1965,
   author = {{Gunn}, J.~E. and {Peterson}, B.~A.},
    title = "{On the Density of Neutral Hydrogen in Intergalactic Space.}",
  journal = {ApJ},
     year = 1965,
    month = nov,
   volume = 142,
    pages = {1633-1641},
      doi = {10.1086/148444},
   adsurl = {http://adsabs.harvard.edu/abs/1965ApJ...142.1633G}
}

@INPROCEEDINGS{Thielemann2003,
   author = {{Thielemann}, F.-K. and {Argast}, D. and {Brachwitz}, F. and 
	{Hix}, W.~R. and {H{\"o}flich}, P. and {Liebend{\"o}rfer}, M. and 
	{Martinez-Pinedo}, G. and {Mezzacappa}, A. and {Nomoto}, K. and 
	{Panov}, I.},
    title = "{Supernova Nucleosynthesis and Galactic Evolution}",
booktitle = {From Twilight to Highlight: The Physics of Supernovae},
     year = 2003,
   editor = {{Hillebrandt}, W. and {Leibundgut}, B.},
    pages = {331},
      doi = {10.1007/10828549_46},
   adsurl = {http://adsabs.harvard.edu/abs/2003fthp.conf..331T}
}

@ARTICLE{Onorbe2015,
   author = {{O{\~n}orbe}, J. and {Boylan-Kolchin}, M. and {Bullock}, J.~S. and 
	{Hopkins}, P.~F. and {Kere{\v s}}, D. and {Faucher-Gigu{\`e}re}, C.-A. and 
	{Quataert}, E. and {Murray}, N.},
    title = "{Forged in FIRE: cusps, cores and baryons in low-mass dwarf galaxies}",
  journal = {MNRAS},
archivePrefix = "arXiv",
   eprint = {1502.02036},
     year = 2015,
    month = dec,
   volume = 454,
    pages = {2092-2106},
      doi = {10.1093/MNRAS/stv2072},
   adsurl = {http://adsabs.harvard.edu/abs/2015MNRAS.454.2092O}
}

@ARTICLE{Wheeler2015,
   author = {{Wheeler}, C. and {O{\~n}orbe}, J. and {Bullock}, J.~S. and 
	{Boylan-Kolchin}, M. and {Elbert}, O.~D. and {Garrison-Kimmel}, S. and 
	{Hopkins}, P.~F. and {Kere{\v s}}, D.},
    title = "{Sweating the small stuff: simulating dwarf galaxies, ultra-faint dwarf galaxies, and their own tiny satellites}",
  journal = {MNRAS},
archivePrefix = "arXiv",
   eprint = {1504.02466},
     year = 2015,
    month = oct,
   volume = 453,
    pages = {1305-1316},
      doi = {10.1093/MNRAS/stv1691},
   adsurl = {http://adsabs.harvard.edu/abs/2015MNRAS.453.1305W}
}

@ARTICLE{Sawala2011,
   author = {{Sawala}, T. and {Scannapieco}, C. and {Maio}, U. and {White}, S.
	},
    title = "{Formation of isolated dwarf galaxies with feedback}",
  journal = {MNRAS},
archivePrefix = "arXiv",
   eprint = {0902.1754},
     year = 2010,
    month = mar,
   volume = 402,
    pages = {1599-1613},
      doi = {10.1111/j.1365-2966.2009.16035.x},
   adsurl = {http://adsabs.harvard.edu/abs/2010MNRAS.402.1599S}
}

@ARTICLE{Brook2014,
   author = {{Brook}, C.~B. and {Di Cintio}, A. and {Knebe}, A. and {Gottl{\"o}ber}, S. and 
	{Hoffman}, Y. and {Yepes}, G. and {Garrison-Kimmel}, S.},
    title = "{The Stellar-to-halo Mass Relation for Local Group Galaxies}",
  journal = {ApJL},
archivePrefix = "arXiv",
   eprint = {1311.5492},
     year = 2014,
    month = mar,
   volume = 784,
      eid = {L14},
    pages = {L14},
      doi = {10.1088/2041-8205/784/1/L14},
   adsurl = {http://adsabs.harvard.edu/abs/2014ApJ...784L..14B}
}

@ARTICLE{Revaz2016,
   author = {{Revaz}, Y. and {Arnaudon}, A. and {Nichols}, M. and {Bonvin}, V. and 
	{Jablonka}, P.},
    title = "{Computational issues in chemo-dynamical modelling of the formation and evolution of galaxies}",
  journal = {A\&A},
archivePrefix = "arXiv",
   eprint = {1601.02017},
     year = 2016,
    month = apr,
   volume = 588,
      eid = {A21},
    pages = {A21},
      doi = {10.1051/0004-6361/201526438},
   adsurl = {http://adsabs.harvard.edu/abs/2016A%26A...588A..21R}
}

@ARTICLE{Brown2014,
   author = {{Brown}, T.~M. and {Tumlinson}, J. and {Geha}, M. and {Simon}, J.~D. and 
	{Vargas}, L.~C. and {VandenBerg}, D.~A. and {Kirby}, E.~N. and 
	{Kalirai}, J.~S. and {Avila}, R.~J. and {Gennaro}, M. and {Ferguson}, H.~C. and 
	{Mu{\~n}oz}, R.~R. and {Guhathakurta}, P. and {Renzini}, A.},
    title = "{The Quenching of the Ultra-faint Dwarf Galaxies in the Reionization Era}",
  journal = {ApJ},
archivePrefix = "arXiv",
   eprint = {1410.0681},
     year = 2014,
    month = dec,
   volume = 796,
      eid = {91},
    pages = {91},
      doi = {10.1088/0004-637X/796/2/91},
   adsurl = {http://adsabs.harvard.edu/abs/2014ApJ...796...91B}
}

@ARTICLE{Wetzel2015,
   author = {{Wetzel}, A.~R. and {Deason}, A.~J. and {Garrison-Kimmel}, S.
	},
    title = "{Satellite Dwarf Galaxies in a Hierarchical Universe: Infall Histories, Group Preprocessing, and Reionization}",
  journal = {ApJ},
archivePrefix = "arXiv",
   eprint = {1501.01972},
     year = 2015,
    month = jul,
   volume = 807,
      eid = {49},
    pages = {49},
      doi = {10.1088/0004-637X/807/1/49},
   adsurl = {http://adsabs.harvard.edu/abs/2015ApJ...807...49W}
}

@ARTICLE{McConnachie2012,
   author = {{McConnachie}, A.~W.},
    title = "{The Observed Properties of Dwarf Galaxies in and around the Local Group}",
  journal = {AJ},
archivePrefix = "arXiv",
   eprint = {1204.1562},
     year = 2012,
    month = jul,
   volume = 144,
      eid = {4},
    pages = {4},
      doi = {10.1088/0004-6256/144/1/4},
   adsurl = {http://adsabs.harvard.edu/abs/2012AJ....144....4M}
}

@ARTICLE{Jeon2017,
	author = {{Jeon}, M. and {Besla}, G. and {Bromm}, V.},
	       title = "{Connecting the First Galaxies with Ultrafaint Dwarfs in the Local Group: Chemical Signatures of Population III Stars}",
	         journal = {ApJ},
		 archivePrefix = "arXiv",
		    eprint = {1702.07355},
		          year = 2017,
			      month = oct,
			         volume = 848,
				       eid = {85},
				           pages = {85},
					         doi = {10.3847/1538-4357/aa8c80},
						    adsurl = {http://adsabs.harvard.edu/abs/2017ApJ...848...85J}
}

@ARTICLE{Stinson2007,
   author = {{Stinson}, G.~S. and {Dalcanton}, J.~J. and {Quinn}, T. and
        {Kaufmann}, T. and {Wadsley}, J.},
    title = "{Breathing in Low-Mass Galaxies: A Study of Episodic Star Formation}",
  journal = {ApJ},
archivePrefix = "arXiv",
   eprint = {0705.4494},
     year = 2007,
    month = sep,
   volume = 667,
    pages = {170-175},
      doi = {10.1086/520504},
   adsurl = {http://adsabs.harvard.edu/abs/2007ApJ...667..170S}
}

@ARTICLE{Jaacks2018a,
       author = {{Jaacks}, Jason and {Finkelstein}, Steven L. and {Bromm}, Volker},
        title = "{Legacy of star formation in the pre-reionization universe}",
      journal = {\mnras},
         year = 2019,
        month = sep,
       volume = {488},
       number = {2},
        pages = {2202-2221},
          doi = {10.1093/mnras/stz1529},
archivePrefix = {arXiv},
       eprint = {1804.07372},
 primaryClass = {astro-ph.GA},
       adsurl = {https://ui.adsabs.harvard.edu/abs/2019MNRAS.488.2202J}
}

@ARTICLE{Simon2019,
       author = {{Simon}, Joshua D.},
        title = "{The Faintest Dwarf Galaxies}",
      journal = {ARA\&A},
         year = 2019,
        month = aug,
       volume = {57},
        pages = {375-415},
          doi = {10.1146/annurev-astro-091918-104453},
archivePrefix = {arXiv},
       eprint = {1901.05465},
 primaryClass = {astro-ph.GA},
       adsurl = {https://ui.adsabs.harvard.edu/abs/2019ARA&A..57..375S}
}

@ARTICLE{Bullock2000,
       author = {{Bullock}, James S. and {Kravtsov}, Andrey V. and {Weinberg}, David H.},
        title = "{Reionization and the Abundance of Galactic Satellites}",
      journal = {ApJ},
         year = 2000,
        month = aug,
       volume = {539},
       number = {2},
        pages = {517-521},
          doi = {10.1086/309279},
archivePrefix = {arXiv},
       eprint = {astro-ph/0002214},
 primaryClass = {astro-ph},
       adsurl = {https://ui.adsabs.harvard.edu/abs/2000ApJ...539..517B}
}

@ARTICLE{Rey2019,
       author = {{Rey}, Martin P. and {Pontzen}, Andrew and {Agertz}, Oscar and
         {Orkney}, Matthew D.~A. and {Read}, Justin I. and
         {Saintonge}, Am{\'e}lie and {Pedersen}, Christian},
        title = "{EDGE: The Origin of Scatter in Ultra-faint Dwarf Stellar Masses and Surface Brightnesses}",
      journal = {ApJL},
         year = 2019,
        month = nov,
       volume = {886},
       number = {1},
          eid = {L3},
        pages = {L3},
          doi = {10.3847/2041-8213/ab53dd},
archivePrefix = {arXiv},
       eprint = {1909.04664},
 primaryClass = {astro-ph.GA},
       adsurl = {https://ui.adsabs.harvard.edu/abs/2019ApJ...886L...3R}
}

@ARTICLE{Wheeler2019,
       author = {{Wheeler}, Coral and {Hopkins}, Philip F. and {Pace}, Andrew B. and
         {Garrison-Kimmel}, Shea and {Boylan-Kolchin}, Michael and
         {Wetzel}, Andrew and {Bullock}, James S. and
         {Kere{\v{s}}}, Du{\v{s}}an and {Faucher-Gigu{\`e}re}, Claude-Andr{\'e} and
         {Quataert}, Eliot},
        title = "{Be it therefore resolved: cosmological simulations of dwarf galaxies with 30 solar mass resolution}",
      journal = {MNRAS},
         year = 2019,
        month = dec,
       volume = {490},
       number = {3},
        pages = {4447-4463},
          doi = {10.1093/mnras/stz2887},
archivePrefix = {arXiv},
       eprint = {1812.02749},
 primaryClass = {astro-ph.GA},
       adsurl = {https://ui.adsabs.harvard.edu/abs/2019MNRAS.490.4447W}
}

@ARTICLE{Jeon2014,
   author = {{Jeon}, M. and {Pawlik}, A.~H. and {Bromm}, V. and {Milosavljevi{\'c}}, M.
        },
    title = "{Recovery from Population III supernova explosions and the onset of second-generation star formation}",
  journal = {MNRAS},
archivePrefix = "arXiv",
   eprint = {1407.0034},
     year = 2014,
    month = nov,
   volume = 444,
    pages = {3288-3300},
      doi = {10.1093/MNRAS/stu1980},
   adsurl = {http://adsabs.harvard.edu/abs/2014MNRAS.444.3288J}
}

@ARTICLE{Carlin2019,
       author = {{Carlin}, Jeffrey L. and {Garling}, Christopher T. and
         {Peter}, Annika H.~G. and {Crnojevi{\'c}}, Denija and
         {Forbes}, Duncan A. and {Hargis}, Jonathan R. and
         {Mutlu-Pakdil}, Bur{\c{c}}in and {Pucha}, Ragadeepika and
         {Romanowsky}, Aaron J. and {Sand}, David J. and {Spekkens}, Kristine and
         {Strader}, Jay and {Willman}, Beth},
        title = "{Tidal Destruction in a Low-mass Galaxy Environment: The Discovery of Tidal Tails around DDO 44}",
      journal = {ApJ},
         year = 2019,
        month = dec,
       volume = {886},
       number = {2},
          eid = {109},
        pages = {109},
          doi = {10.3847/1538-4357/ab4c32},
archivePrefix = {arXiv},
       eprint = {1906.08260},
 primaryClass = {astro-ph.GA},
       adsurl = {https://ui.adsabs.harvard.edu/abs/2019ApJ...886..109C}
}

@ARTICLE{Rey2020,
       author = {{Rey}, Martin P. and {Pontzen}, Andrew and {Agertz}, Oscar and {Orkney}, Matthew D.~A. and {Read}, Justin I. and {Rosdahl}, Joakim},
        title = "{EDGE: from quiescent to gas-rich to star-forming low-mass dwarf galaxies}",
      journal = {\mnras},
         year = 2020,
        month = sep,
       volume = {497},
       number = {2},
        pages = {1508-1520},
          doi = {10.1093/mnras/staa1640},
archivePrefix = {arXiv},
       eprint = {2004.09530},
 primaryClass = {astro-ph.GA},
       adsurl = {https://ui.adsabs.harvard.edu/abs/2020MNRAS.497.1508R}
}

@ARTICLE{Agertz2020,
       author = {{Agertz}, Oscar and {Pontzen}, Andrew and {Read}, Justin I. and
         {Rey}, Martin P. and {Orkney}, Matthew and {Rosdahl}, Joakim and
         {Teyssier}, Romain and {Verbeke}, Robbert and {Kretschmer}, Michael and
         {Nickerson}, Sarah},
        title = "{EDGE: the mass-metallicity relation as a critical test of galaxy formation physics}",
      journal = {MNRAS},
         year = 2020,
        month = jan,
       volume = {491},
       number = {2},
        pages = {1656-1672},
          doi = {10.1093/mnras/stz3053},
archivePrefix = {arXiv},
       eprint = {1904.02723},
 primaryClass = {astro-ph.GA},
       adsurl = {https://ui.adsabs.harvard.edu/abs/2020MNRAS.491.1656A}
}

@ARTICLE{Applebaum2020,
       author = {{Applebaum}, Elaad and {Brooks}, Alyson M. and {Quinn}, Thomas R. and {Christensen}, Charlotte R.},
        title = "{A stochastically sampled IMF alters the stellar content of simulated dwarf galaxies}",
      journal = {\mnras},
         year = 2020,
        month = feb,
       volume = {492},
       number = {1},
        pages = {8-21},
          doi = {10.1093/mnras/stz3331},
archivePrefix = {arXiv},
       eprint = {1811.00022},
 primaryClass = {astro-ph.GA},
       adsurl = {https://ui.adsabs.harvard.edu/abs/2020MNRAS.492....8A}
}

@ARTICLE{LSST2019,
       author = {{LSST Collaboration}},
        title = "{LSST: From Science Drivers to Reference Design and Anticipated Data Products}",
      journal = {ApJ},
         year = 2019,
        month = mar,
       volume = {873},
       number = {2},
          eid = {111},
        pages = {111},
          doi = {10.3847/1538-4357/ab042c},
archivePrefix = {arXiv},
       eprint = {0805.2366},
 primaryClass = {astro-ph},
       adsurl = {https://ui.adsabs.harvard.edu/abs/2019ApJ...873..111I}
}

@ARTICLE{Haas2010,
       author = {{Haas}, M.~R. and {Anders}, P.},
        title = "{Variations in integrated galactic initial mass functions due to sampling method and cluster mass function}",
      journal = {\aap},
         year = 2010,
        month = mar,
       volume = {512},
          eid = {A79},
        pages = {A79},
          doi = {10.1051/0004-6361/200912967},
archivePrefix = {arXiv},
       eprint = {1001.2009},
 primaryClass = {astro-ph.GA},
       adsurl = {https://ui.adsabs.harvard.edu/abs/2010A&A...512A..79H}
}

@ARTICLE{Salpeter1955,
       author = {{Salpeter}, Edwin E.},
        title = "{The Luminosity Function and Stellar Evolution.}",
      journal = {\apj},
         year = 1955,
        month = jan,
       volume = {121},
        pages = {161},
          doi = {10.1086/145971},
       adsurl = {https://ui.adsabs.harvard.edu/abs/1955ApJ...121..161S}
}

@ARTICLE{Applebaum2021,
       author = {{Applebaum}, Elaad and {Brooks}, Alyson M. and {Christensen}, Charlotte R. and {Munshi}, Ferah and {Quinn}, Thomas R. and {Shen}, Sijing and {Tremmel}, Michael},
        title = "{Ultrafaint Dwarfs in a Milky Way Context: Introducing the Mint Condition DC Justice League Simulations}",
      journal = {\apj},
         year = 2021,
        month = jan,
       volume = {906},
       number = {2},
          eid = {96},
        pages = {96},
          doi = {10.3847/1538-4357/abcafa},
archivePrefix = {arXiv},
       eprint = {2008.11207},
 primaryClass = {astro-ph.GA},
       adsurl = {https://ui.adsabs.harvard.edu/abs/2021ApJ...906...96A}
}

@ARTICLE{Simpson2018,
       author = {{Simpson}, Christine M. and {Grand}, Robert J.~J. and {G{\'o}mez}, Facundo A. and {Marinacci}, Federico and {Pakmor}, R{\"u}diger and {Springel}, Volker and {Campbell}, David J.~R. and {Frenk}, Carlos S.},
        title = "{Quenching and ram pressure stripping of simulated Milky Way satellite galaxies}",
      journal = {\mnras},
         year = 2018,
        month = jul,
       volume = {478},
       number = {1},
        pages = {548-567},
          doi = {10.1093/mnras/sty774},
archivePrefix = {arXiv},
       eprint = {1705.03018},
 primaryClass = {astro-ph.GA},
       adsurl = {https://ui.adsabs.harvard.edu/abs/2018MNRAS.478..548S}
}

@ARTICLE{Su2018,
       author = {{Su}, Kung-Yi and {Hopkins}, Philip F. and {Hayward}, Christopher C. and {Ma}, Xiangcheng and {Boylan-Kolchin}, Michael and {Kasen}, Daniel and {Kere{\v{s}}}, Du{\v{s}}an and {Faucher-Gigu{\`e}re}, Claude-Andr{\'e} and {Orr}, Matthew E. and {Wheeler}, Coral},
        title = "{Discrete effects in stellar feedback: Individual Supernovae, Hypernovae, and IMF Sampling in Dwarf Galaxies}",
      journal = {\mnras},
         year = 2018,
        month = oct,
       volume = {480},
       number = {2},
        pages = {1666-1675},
          doi = {10.1093/mnras/sty1928},
archivePrefix = {arXiv},
       eprint = {1712.02795},
 primaryClass = {astro-ph.GA},
       adsurl = {https://ui.adsabs.harvard.edu/abs/2018MNRAS.480.1666S}
}

@ARTICLE{Hu2017,
       author = {{Hu}, Chia-Yu and {Naab}, Thorsten and {Glover}, Simon C.~O. and {Walch}, Stefanie and {Clark}, Paul C.},
        title = "{Variable interstellar radiation fields in simulated dwarf galaxies: supernovae versus photoelectric heating}",
      journal = {\mnras},
         year = 2017,
        month = oct,
       volume = {471},
       number = {2},
        pages = {2151-2173},
          doi = {10.1093/mnras/stx1773},
archivePrefix = {arXiv},
       eprint = {1701.08779},
 primaryClass = {astro-ph.GA},
       adsurl = {https://ui.adsabs.harvard.edu/abs/2017MNRAS.471.2151H}
}

@ARTICLE{Jeon2021a,
       author = {{Jeon}, Myoungwon and {Bromm}, Volker and {Besla}, Gurtina and {Yoon}, Jinmi and {Choi}, Yumi},
        title = "{The role of faint population III supernovae in forming CEMP stars in ultra-faint dwarf galaxies}",
      journal = {\mnras},
         year = 2021,
        month = mar,
       volume = {502},
       number = {1},
        pages = {1-14},
          doi = {10.1093/mnras/staa4017},
archivePrefix = {arXiv},
       eprint = {2012.10012},
 primaryClass = {astro-ph.GA},
       adsurl = {https://ui.adsabs.harvard.edu/abs/2021MNRAS.502....1J}
}

@ARTICLE{Jeon2021b,
       author = {{Jeon}, Myoungwon and {Besla}, Gurtina and {Bromm}, Volker},
        title = "{Highly r-process enhanced stars in ultra-faint dwarf galaxies}",
      journal = {\mnras},
         year = 2021,
        month = sep,
       volume = {506},
       number = {2},
        pages = {1850-1861},
          doi = {10.1093/mnras/stab1771},
archivePrefix = {arXiv},
       eprint = {2106.13383},
 primaryClass = {astro-ph.GA},
       adsurl = {https://ui.adsabs.harvard.edu/abs/2021MNRAS.506.1850J}
}

@ARTICLE{Akins2021,
       author = {{Akins}, Hollis B. and {Christensen}, Charlotte R. and {Brooks}, Alyson M. and {Munshi}, Ferah and {Applebaum}, Elaad and {Engelhardt}, Anna and {Chamberland}, Lucas},
        title = "{Quenching Timescales of Dwarf Satellites around Milky Way-mass Hosts}",
      journal = {\apj},
         year = 2021,
        month = mar,
       volume = {909},
       number = {2},
          eid = {139},
        pages = {139},
          doi = {10.3847/1538-4357/abe2ab},
archivePrefix = {arXiv},
       eprint = {2008.02805},
 primaryClass = {astro-ph.GA},
       adsurl = {https://ui.adsabs.harvard.edu/abs/2021ApJ...909..139A}
}

@ARTICLE{Emerick2019,
       author = {{Emerick}, Andrew and {Bryan}, Greg L. and {Mac Low}, Mordecai-Mark},
        title = "{Simulating an isolated dwarf galaxy with multichannel feedback and chemical yields from individual stars}",
      journal = {\mnras},
         year = 2019,
        month = jan,
       volume = {482},
       number = {1},
        pages = {1304-1329},
          doi = {10.1093/mnras/sty2689},
archivePrefix = {arXiv},
       eprint = {1807.07182},
 primaryClass = {astro-ph.GA},
       adsurl = {https://ui.adsabs.harvard.edu/abs/2019MNRAS.482.1304E}
}

@ARTICLE{Prgomet2022,
       author = {{Prgomet}, Mateo and {Rey}, Martin P. and {Andersson}, Eric P. and {Segovia Otero}, Alvaro and {Agertz}, Oscar and {Renaud}, Florent and {Pontzen}, Andrew and {Read}, Justin I.},
        title = "{EDGE: The sensitivity of ultra-faint dwarfs to a metallicity-dependent initial mass function}",
      journal = {\mnras},
         year = 2022,
        month = jun,
       volume = {513},
       number = {2},
        pages = {2326-2334},
          doi = {10.1093/mnras/stac1074},
archivePrefix = {arXiv},
       eprint = {2107.00663},
 primaryClass = {astro-ph.GA},
       adsurl = {https://ui.adsabs.harvard.edu/abs/2022MNRAS.513.2326P}
}

@ARTICLE{Smith2021a,
       author = {{Smith}, Matthew C.},
        title = "{The sensitivity of stellar feedback to IMF averaging versus IMF sampling in galaxy formation simulations}",
      journal = {\mnras},
         year = 2021,
        month = apr,
       volume = {502},
       number = {4},
        pages = {5417-5437},
          doi = {10.1093/mnras/stab291},
archivePrefix = {arXiv},
       eprint = {2010.10533},
 primaryClass = {astro-ph.GA},
       adsurl = {https://ui.adsabs.harvard.edu/abs/2021MNRAS.502.5417S}
}

@ARTICLE{Smith2019,
       author = {{Smith}, Matthew C. and {Sijacki}, Debora and {Shen}, Sijing},
        title = "{Cosmological simulations of dwarfs: the need for ISM physics beyond SN feedback alone}",
      journal = {\mnras},
         year = 2019,
        month = may,
       volume = {485},
       number = {3},
        pages = {3317-3333},
          doi = {10.1093/mnras/stz599},
archivePrefix = {arXiv},
       eprint = {1807.04288},
 primaryClass = {astro-ph.GA},
       adsurl = {https://ui.adsabs.harvard.edu/abs/2019MNRAS.485.3317S}
}

@ARTICLE{Smith2018,
       author = {{Smith}, Matthew C. and {Sijacki}, Debora and {Shen}, Sijing},
        title = "{Supernova feedback in numerical simulations of galaxy formation: separating physics from numerics}",
      journal = {\mnras},
         year = 2018,
        month = jul,
       volume = {478},
       number = {1},
        pages = {302-331},
          doi = {10.1093/mnras/sty994},
archivePrefix = {arXiv},
       eprint = {1709.03515},
 primaryClass = {astro-ph.GA},
       adsurl = {https://ui.adsabs.harvard.edu/abs/2018MNRAS.478..302S}
}

@ARTICLE{Sanati2023,
       author = {{Sanati}, Mahsa and {Jeanquartier}, Fabien and {Revaz}, Yves and {Jablonka}, Pascale},
        title = "{How much metal did the first stars provide to the ultra-faint dwarfs?}",
      journal = {\aap},
         year = 2023,
        month = jan,
       volume = {669},
          eid = {A94},
        pages = {A94},
          doi = {10.1051/0004-6361/202244309},
archivePrefix = {arXiv},
       eprint = {2206.11351},
 primaryClass = {astro-ph.GA},
       adsurl = {https://ui.adsabs.harvard.edu/abs/2023A&A...669A..94S}
}

@ARTICLE{Goater2024,
       author = {{Goater}, Alex and {Read}, Justin I. and {No{\"e}l}, Noelia E.~D. and {Orkney}, Matthew D.~A. and {Kim}, Stacy Y. and {Rey}, Martin P. and {Andersson}, Eric P. and {Agertz}, Oscar and {Pontzen}, Andrew and {Vieliute}, Roberta and {Kataria}, Dhairya and {Jeneway}, Kiah},
        title = "{EDGE: The direct link between mass growth history and the extended stellar haloes of the faintest dwarf galaxies}",
      journal = {\mnras},
         year = 2024,
        month = jan,
       volume = {527},
       number = {2},
        pages = {2403-2412},
          doi = {10.1093/mnras/stad3354},
archivePrefix = {arXiv},
       eprint = {2307.05130},
 primaryClass = {astro-ph.GA},
       adsurl = {https://ui.adsabs.harvard.edu/abs/2024MNRAS.527.2403G}
}

@ARTICLE{Gutcke2021,
       author = {{Gutcke}, Thales A. and {Pakmor}, R{\"u}diger and {Naab}, Thorsten and {Springel}, Volker},
        title = "{LYRA - I. Simulating the multiphase ISM of a dwarf galaxy with variable energy supernovae from individual stars}",
      journal = {\mnras},
         year = 2021,
        month = mar,
       volume = {501},
       number = {4},
        pages = {5597-5615},
          doi = {10.1093/mnras/staa3875},
archivePrefix = {arXiv},
       eprint = {2010.07311},
 primaryClass = {astro-ph.GA},
       adsurl = {https://ui.adsabs.harvard.edu/abs/2021MNRAS.501.5597G}
}

@ARTICLE{Gallart2021,
       author = {{Gallart}, C. and {Monelli}, M. and {Ruiz-Lara}, T. and {Calamida}, A. and {Cassisi}, S. and {Cignoni}, M. and {Anderson}, J. and {Battaglia}, G. and {Bermejo-Climent}, J.~R. and {Bernard}, E.~J. and {Mart{\'\i}nez-V{\'a}zquez}, C.~E. and {Mayer}, L. and {Salvadori}, S. and {Monachesi}, A. and {Navarro}, J.~F. and {Shen}, S. and {Surot}, F. and {Tosi}, M. and {Bajaj}, V. and {Strinfellow}, G.~S.},
        title = "{The Star Formation History of Eridanus II: On the Role of Supernova Feedback in the Quenching of Ultrafaint Dwarf Galaxies}",
      journal = {\apj},
         year = 2021,
        month = mar,
       volume = {909},
       number = {2},
          eid = {192},
        pages = {192},
          doi = {10.3847/1538-4357/abddbe},
archivePrefix = {arXiv},
       eprint = {2101.04464},
 primaryClass = {astro-ph.GA},
       adsurl = {https://ui.adsabs.harvard.edu/abs/2021ApJ...909..192G}
}

@ARTICLE{Savino2023,
       author = {{Savino}, Alessandro and {Weisz}, Daniel R. and {Skillman}, Evan D. and {Dolphin}, Andrew and {Cole}, Andrew A. and {Kallivayalil}, Nitya and {Wetzel}, Andrew and {Anderson}, Jay and {Besla}, Gurtina and {Boylan-Kolchin}, Michael and {Brown}, Thomas M. and {Bullock}, James S. and {Collins}, Michelle L.~M. and {Cooper}, M.~C. and {Deason}, Alis J. and {Dotter}, Aaron L. and {Fardal}, Mark and {Ferguson}, Annette M.~N. and {Fritz}, Tobias K. and {Geha}, Marla C. and {Gilbert}, Karoline M. and {Guhathakurta}, Puragra and {Ibata}, Rodrigo and {Irwin}, Michael J. and {Jeon}, Myoungwon and {Kirby}, Evan N. and {Lewis}, Geraint F. and {Mackey}, Dougal and {Majewski}, Steven R. and {Martin}, Nicolas and {McConnachie}, Alan and {Patel}, Ekta and {Rich}, R. Michael and {Simon}, Joshua D. and {Sohn}, Sangmo Tony and {Tollerud}, Erik J. and {van der Marel}, Roeland P.},
        title = "{The Hubble Space Telescope Survey of M31 Satellite Galaxies. II. The Star Formation Histories of Ultrafaint Dwarf Galaxies}",
      journal = {\apj},
         year = 2023,
        month = oct,
       volume = {956},
       number = {2},
          eid = {86},
        pages = {86},
          doi = {10.3847/1538-4357/acf46f},
archivePrefix = {arXiv},
       eprint = {2305.13360},
 primaryClass = {astro-ph.GA},
       adsurl = {https://ui.adsabs.harvard.edu/abs/2023ApJ...956...86S}
}

@ARTICLE{Sales2022,
       author = {{Sales}, Laura V. and {Wetzel}, Andrew and {Fattahi}, Azadeh},
        title = "{Baryonic solutions and challenges for cosmological models of dwarf galaxies}",
      journal = {Nature Astronomy},
         year = 2022,
        month = jun,
       volume = {6},
        pages = {897-910},
          doi = {10.1038/s41550-022-01689-w},
archivePrefix = {arXiv},
       eprint = {2206.05295},
 primaryClass = {astro-ph.GA},
       adsurl = {https://ui.adsabs.harvard.edu/abs/2022NatAs...6..897S}
}

@ARTICLE{Munshi2021,
       author = {{Munshi}, Ferah and {Brooks}, Alyson M. and {Applebaum}, Elaad and {Christensen}, Charlotte R. and {Quinn}, T. and {Sligh}, Serena},
        title = "{Quantifying Scatter in Galaxy Formation at the Lowest Masses}",
      journal = {\apj},
         year = 2021,
        month = dec,
       volume = {923},
       number = {1},
          eid = {35},
        pages = {35},
          doi = {10.3847/1538-4357/ac0db6},
archivePrefix = {arXiv},
       eprint = {2101.05822},
 primaryClass = {astro-ph.GA},
       adsurl = {https://ui.adsabs.harvard.edu/abs/2021ApJ...923...35M}
}

@ARTICLE{Fu2023,
       author = {{Fu}, Sal Wanying and {Weisz}, Daniel R. and {Starkenburg}, Else and {Martin}, Nicolas and {Savino}, Alessandro and {Boylan-Kolchin}, Michael and {C{\^o}t{\'e}}, Patrick and {Dolphin}, Andrew E. and {Ji}, Alexander P. and {Longeard}, Nicolas and {Mateo}, Mario L. and {Patel}, Ekta and {Sandford}, Nathan R.},
        title = "{Metallicity Distribution Functions of 13 Ultra-faint Dwarf Galaxy Candidates from Hubble Space Telescope Narrowband Imaging}",
      journal = {\apj},
         year = 2023,
        month = dec,
       volume = {958},
       number = {2},
          eid = {167},
        pages = {167},
          doi = {10.3847/1538-4357/ad0030},
archivePrefix = {arXiv},
       eprint = {2306.06260},
 primaryClass = {astro-ph.GA},
       adsurl = {https://ui.adsabs.harvard.edu/abs/2023ApJ...958..167F}
}

@ARTICLE{Walker2006,
       author = {{Walker}, Matthew G. and {Mateo}, Mario and {Olszewski}, Edward W. and {Bernstein}, Rebecca and {Wang}, Xiao and {Woodroofe}, Michael},
        title = "{Internal Kinematics of the Fornax Dwarf Spheroidal Galaxy}",
      journal = {\aj},
         year = 2006,
        month = apr,
       volume = {131},
       number = {4},
        pages = {2114-2139},
          doi = {10.1086/500193},
archivePrefix = {arXiv},
       eprint = {astro-ph/0511465},
 primaryClass = {astro-ph},
       adsurl = {https://ui.adsabs.harvard.edu/abs/2006AJ....131.2114W}
}

@ARTICLE{Azartash-Namin2024,
       author = {{Azartash-Namin}, Bianca and {Engelhardt}, Anna and {Munshi}, Ferah and {Keller}, B.~W. and {Brooks}, Alyson M. and {Van Nest}, Jordan and {Christensen}, Charlotte R. and {Quinn}, Tom and {Wadsley}, James},
        title = "{Bursting with Feedback: The Relationship between Feedback Model and Bursty Star Formation Histories in Dwarf Galaxies}",
      journal = {\apj},
         year = 2024,
        month = jul,
       volume = {970},
       number = {1},
          eid = {40},
        pages = {40},
          doi = {10.3847/1538-4357/ad49a5},
archivePrefix = {arXiv},
       eprint = {2401.06041},
 primaryClass = {astro-ph.GA},
       adsurl = {https://ui.adsabs.harvard.edu/abs/2024ApJ...970...40A}
}

@ARTICLE{Deng2024,
       author = {{Deng}, Yunwei and {Li}, Hui and {Liu}, Boyuan and {Kannan}, Rahul and {Smith}, Aaron and {Bryan}, Greg L.},
        title = "{RIGEL: Simulating dwarf galaxies at solar mass resolution with radiative transfer and feedback from individual massive stars}",
      journal = {arXiv e-prints},
         year = 2024,
        month = may,
          eid = {arXiv:2405.08869},
        pages = {arXiv:2405.08869},
          doi = {10.48550/arXiv.2405.08869},
archivePrefix = {arXiv},
       eprint = {2405.08869},
 primaryClass = {astro-ph.GA},
       adsurl = {https://ui.adsabs.harvard.edu/abs/2024arXiv240508869D}
}

@ARTICLE{Leroy2008,
       author = {{Leroy}, Adam K. and {Walter}, Fabian and {Brinks}, Elias and {Bigiel}, Frank and {de Blok}, W.~J.~G. and {Madore}, Barry and {Thornley}, M.~D.},
        title = "{The Star Formation Efficiency in Nearby Galaxies: Measuring Where Gas Forms Stars Effectively}",
      journal = {\aj},
         year = 2008,
        month = dec,
       volume = {136},
       number = {6},
        pages = {2782-2845},
          doi = {10.1088/0004-6256/136/6/2782},
archivePrefix = {arXiv},
       eprint = {0810.2556},
 primaryClass = {astro-ph},
       adsurl = {https://ui.adsabs.harvard.edu/abs/2008AJ....136.2782L}
}

@ARTICLE{Klessen2023,
       author = {{Klessen}, Ralf S. and {Glover}, Simon C.~O.},
        title = "{The First Stars: Formation, Properties, and Impact}",
      journal = {\araa},
         year = 2023,
        month = aug,
       volume = {61},
        pages = {65-130},
          doi = {10.1146/annurev-astro-071221-053453},
archivePrefix = {arXiv},
       eprint = {2303.12500},
 primaryClass = {astro-ph.CO},
       adsurl = {https://ui.adsabs.harvard.edu/abs/2023ARA&A..61...65K}
}

@ARTICLE{Choi2016,
       author = {{Choi}, Jieun and {Dotter}, Aaron and {Conroy}, Charlie and {Cantiello}, Matteo and {Paxton}, Bill and {Johnson}, Benjamin D.},
        title = "{Mesa Isochrones and Stellar Tracks (MIST). I. Solar-scaled Models}",
      journal = {\apj},
         year = 2016,
        month = jun,
       volume = {823},
       number = {2},
          eid = {102},
        pages = {102},
          doi = {10.3847/0004-637X/823/2/102},
archivePrefix = {arXiv},
       eprint = {1604.08592},
 primaryClass = {astro-ph.SR},
       adsurl = {https://ui.adsabs.harvard.edu/abs/2016ApJ...823..102C}
}

@ARTICLE{Dotter2016,
       author = {{Dotter}, Aaron},
        title = "{MESA Isochrones and Stellar Tracks (MIST) 0: Methods for the Construction of Stellar Isochrones}",
      journal = {\apjs},
         year = 2016,
        month = jan,
       volume = {222},
       number = {1},
          eid = {8},
        pages = {8},
          doi = {10.3847/0067-0049/222/1/8},
archivePrefix = {arXiv},
       eprint = {1601.05144},
 primaryClass = {astro-ph.SR},
       adsurl = {https://ui.adsabs.harvard.edu/abs/2016ApJS..222....8D}
}

@ARTICLE{Andersson2023,
       author = {{Andersson}, Eric P. and {Agertz}, Oscar and {Renaud}, Florent and {Teyssier}, Romain},
        title = "{INFERNO: Galactic winds in dwarf galaxies with star-by-star simulations including runaway stars}",
      journal = {\mnras},
         year = 2023,
        month = may,
       volume = {521},
       number = {2},
        pages = {2196-2214},
          doi = {10.1093/mnras/stad692},
archivePrefix = {arXiv},
       eprint = {2209.06218},
 primaryClass = {astro-ph.GA},
       adsurl = {https://ui.adsabs.harvard.edu/abs/2023MNRAS.521.2196A}
}

@ARTICLE{Brauer2024,
       author = {{Brauer}, Kaley and {Emerick}, Andrew and {Mead}, Jennifer and {Ji}, Alexander P. and {Wise}, John H. and {Bryan}, Greg L. and {Mac Low}, Mordecai-Mark and {Cote}, Benoit and {Andersson}, Eric P. and {Frebel}, Anna},
        title = "{AEOS: Star-by-Star Cosmological Simulations of Early Chemical Enrichment and Galaxy Formation}",
      journal = {arXiv e-prints},
         year = 2024,
        month = oct,
          eid = {arXiv:2410.16366},
        pages = {arXiv:2410.16366},
          doi = {10.48550/arXiv.2410.16366},
archivePrefix = {arXiv},
       eprint = {2410.16366},
 primaryClass = {astro-ph.GA},
       adsurl = {https://ui.adsabs.harvard.edu/abs/2024arXiv241016366B}
}

@ARTICLE{Hopkins2014,
       author = {{Hopkins}, Philip F. and {Kere{\v{s}}}, Du{\v{s}}an and {O{\~n}orbe}, Jos{\'e} and {Faucher-Gigu{\`e}re}, Claude-Andr{\'e} and {Quataert}, Eliot and {Murray}, Norman and {Bullock}, James S.},
        title = "{Galaxies on FIRE (Feedback In Realistic Environments): stellar feedback explains cosmologically inefficient star formation}",
      journal = {\mnras},
         year = 2014,
        month = nov,
       volume = {445},
       number = {1},
        pages = {581-603},
          doi = {10.1093/mnras/stu1738},
archivePrefix = {arXiv},
       eprint = {1311.2073},
 primaryClass = {astro-ph.CO},
       adsurl = {https://ui.adsabs.harvard.edu/abs/2014MNRAS.445..581H}
}

@ARTICLE{Katz1992,
       author = {{Katz}, Neal},
        title = "{Dissipational Galaxy Formation. II. Effects of Star Formation}",
      journal = {\apj},
         year = 1992,
        month = jun,
       volume = {391},
        pages = {502},
          doi = {10.1086/171366},
       adsurl = {https://ui.adsabs.harvard.edu/abs/1992ApJ...391..502K}
}

@ARTICLE{Navarro1993,
       author = {{Navarro}, J.~F. and {White}, S.~D.~M.},
        title = "{Simulations of Dissipative Galaxy Formation in Hierarchically Clustering Universes - Part One - Tests of the Code}",
      journal = {\mnras},
         year = 1993,
        month = nov,
       volume = {265},
        pages = {271},
          doi = {10.1093/mnras/265.2.271},
       adsurl = {https://ui.adsabs.harvard.edu/abs/1993MNRAS.265..271N}
}

@ARTICLE{Stinson2010,
       author = {{Stinson}, G.~S. and {Bailin}, J. and {Couchman}, H. and {Wadsley}, J. and {Shen}, S. and {Nickerson}, S. and {Brook}, C. and {Quinn}, T.},
        title = "{Cosmological galaxy formation simulations using smoothed particle hydrodynamics}",
      journal = {\mnras},
         year = 2010,
        month = oct,
       volume = {408},
       number = {2},
        pages = {812-826},
          doi = {10.1111/j.1365-2966.2010.17187.x},
archivePrefix = {arXiv},
       eprint = {1004.0675},
 primaryClass = {astro-ph.CO},
       adsurl = {https://ui.adsabs.harvard.edu/abs/2010MNRAS.408..812S}
}

@ARTICLE{Zhang2024,
       author = {{Zhang}, Eric and {Sales}, Laura V. and {Marinacci}, Federico and {Torrey}, Paul and {Vogelsberger}, Mark and {Springel}, Volker and {Li}, Hui and {Pakmor}, R{\"u}diger and {Gutcke}, Thales A.},
        title = "{Bursty Star Formation in Dwarfs is Sensitive to Numerical Choices in Supernova Feedback Models}",
      journal = {\apj},
         year = 2024,
        month = nov,
       volume = {975},
       number = {2},
          eid = {229},
        pages = {229},
          doi = {10.3847/1538-4357/ad7f57},
archivePrefix = {arXiv},
       eprint = {2406.10338},
 primaryClass = {astro-ph.GA},
       adsurl = {https://ui.adsabs.harvard.edu/abs/2024ApJ...975..229Z}
}

@ARTICLE{Collins2022,
       author = {{Collins}, Michelle L.~M. and {Read}, Justin I.},
        title = "{Observational constraints on stellar feedback in dwarf galaxies}",
      journal = {Nature Astronomy},
         year = 2022,
        month = may,
       volume = {6},
        pages = {647-658},
          doi = {10.1038/s41550-022-01657-4},
archivePrefix = {arXiv},
       eprint = {2205.06825},
 primaryClass = {astro-ph.GA},
       adsurl = {https://ui.adsabs.harvard.edu/abs/2022NatAs...6..647C}
}

@ARTICLE{Keller2022,
       author = {{Keller}, Benjamin W. and {Kruijssen}, J.~M. Diederik},
        title = "{Uncertainties in supernova input rates drive qualitative differences in simulations of galaxy evolution}",
      journal = {\mnras},
         year = 2022,
        month = may,
       volume = {512},
       number = {1},
        pages = {199-215},
          doi = {10.1093/mnras/stac511},
archivePrefix = {arXiv},
       eprint = {2004.03608},
 primaryClass = {astro-ph.GA},
       adsurl = {https://ui.adsabs.harvard.edu/abs/2022MNRAS.512..199K}
}

@ARTICLE{Hu2023,
       author = {{Hu}, Chia-Yu and {Smith}, Matthew C. and {Teyssier}, Romain and {Bryan}, Greg L. and {Verbeke}, Robbert and {Emerick}, Andrew and {Somerville}, Rachel S. and {Burkhart}, Blakesley and {Li}, Yuan and {Forbes}, John C. and {Starkenburg}, Tjitske},
        title = "{Code Comparison in Galaxy-scale Simulations with Resolved Supernova Feedback: Lagrangian versus Eulerian Methods}",
      journal = {\apj},
         year = 2023,
        month = jun,
       volume = {950},
       number = {2},
          eid = {132},
        pages = {132},
          doi = {10.3847/1538-4357/accf9e},
archivePrefix = {arXiv},
       eprint = {2208.10528},
 primaryClass = {astro-ph.GA},
       adsurl = {https://ui.adsabs.harvard.edu/abs/2023ApJ...950..132H}
}

@ARTICLE{Rey2025,
       author = {{Rey}, Martin P. and {Taylor}, Ethan and {Gray}, Emily I. and {Kim}, Stacy Y. and {Andersson}, Eric P. and {Pontzen}, Andrew and {Agertz}, Oscar and {Read}, Justin I. and {Cadiou}, Corentin and {Yates}, Robert M. and {Orkney}, Matthew D.~A. and {Scholte}, Dirk and {Saintonge}, Am{\'e}lie and {Breneman}, Joseph and {McQuinn}, Kristen B.~W. and {Muni}, Claudia and {Das}, Payel},
        title = "{EDGE: the emergence of dwarf galaxy scaling relations from cosmological radiation-hydrodynamics simulations}",
      journal = {\mnras},
         year = 2025,
        month = aug,
       volume = {541},
       number = {2},
        pages = {1195-1217},
          doi = {10.1093/mnras/staf1058},
archivePrefix = {arXiv},
       eprint = {2503.03813},
 primaryClass = {astro-ph.GA},
       adsurl = {https://ui.adsabs.harvard.edu/abs/2025MNRAS.541.1195R}
}

@ARTICLE{Brauer2025,
       author = {{Brauer}, Kaley and {Emerick}, Andrew and {Mead}, Jennifer and {Ji}, Alexander P. and {Wise}, John H. and {Bryan}, Greg L. and {Mac Low}, Mordecai-Mark and {C{\^o}t{\'e}}, Benoit and {Andersson}, Eric P. and {Frebel}, Anna},
        title = "{AEOS: Star-by-star Cosmological Simulations of Early Chemical Enrichment and Galaxy Formation}",
      journal = {\apj},
         year = 2025,
        month = feb,
       volume = {980},
       number = {1},
          eid = {41},
        pages = {41},
          doi = {10.3847/1538-4357/ada4a1},
archivePrefix = {arXiv},
       eprint = {2410.16366},
 primaryClass = {astro-ph.GA},
       adsurl = {https://ui.adsabs.harvard.edu/abs/2025ApJ...980...41B}
}

@ARTICLE{Andersson2024,
       author = {{Andersson}, Eric P. and {Mac Low}, Mordecai-Mark and {Agertz}, Oscar and {Renaud}, Florent and {Li}, Hui},
        title = "{Pre-supernova feedback sets the star cluster mass function to a power law and reduces the cluster formation efficiency}",
      journal = {\aap},
         year = 2024,
        month = jan,
       volume = {681},
          eid = {A28},
        pages = {A28},
          doi = {10.1051/0004-6361/202347792},
archivePrefix = {arXiv},
       eprint = {2308.12363},
 primaryClass = {astro-ph.GA},
       adsurl = {https://ui.adsabs.harvard.edu/abs/2024A&A...681A..28A}
}
\bibliographystyle{mnras} 

\appendix
\section{Star formation evolution for {\sc Halo2}}
\label{sec:A}

Fig.~\ref{figA1_feh_mul} presents the mass evolution histories (top panels) and the corresponding metallicity of star particles within progenitor halos (bottom panels) for {\sc Halo2}, using different IMF sampling methods: {\sc Burst} (left), {\sc Simf} (middle), and {\sc Indiv} (right).
\begin{figure*}
  \centering
  \includegraphics[width=165mm]{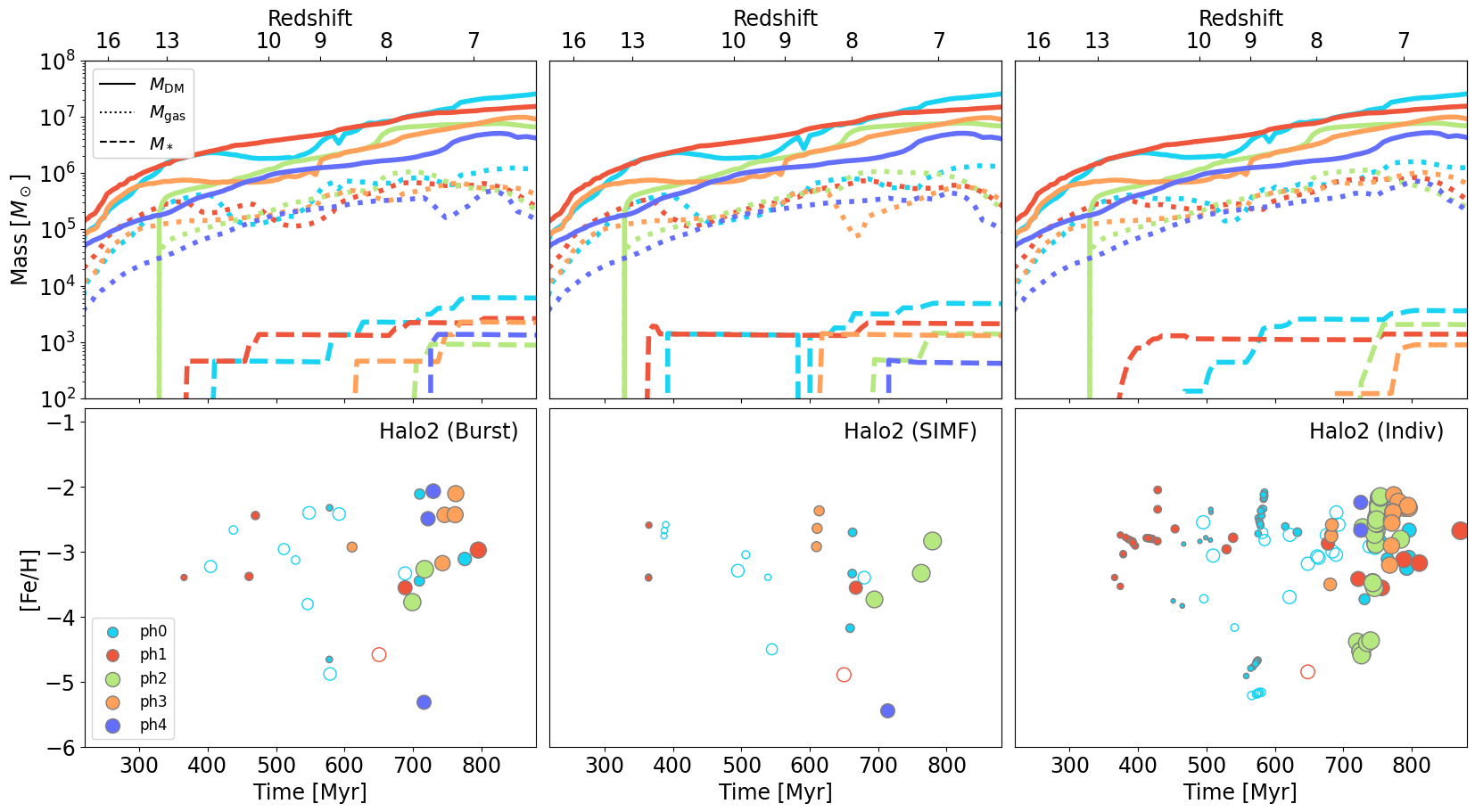}%
   \caption{The same as Fig.~\ref{fig9_feh_mul}, but for {\sc Halo2}. Since {\sc Halo2} ($\rm M_{vir}\approx5\times10^8\msun$ at $z=0$) is less massive than {\sc Halo1} ($\rm M_{vir}\approx8\times10^8\msun$ at $z=0$), star formation in each progenitor halo generally begins later. A similar trend to {\sc Halo1} is observed, with star formation being more continuous in {\sc Indiv} compared to {\sc Burst} and {\sc Simf} runs.}
   \label{figA1_feh_mul}
\end{figure*}

\section{Results of Burst run with low efficiency} 
\label{sec:B}
Here, we exhibit the results of {\sc Burst} run, where we adjust the star formation efficiency by lowering it about eight times. Fig.~\ref{figB2} is the same as Fig.~\ref{fig_density_evolution}, indicating the maximum hydrogen density evolution as a function of cosmic time, with stellar metallicity overlaid with their formation time.

\begin{figure*}
  \centering
  \includegraphics[width=165mm]
  {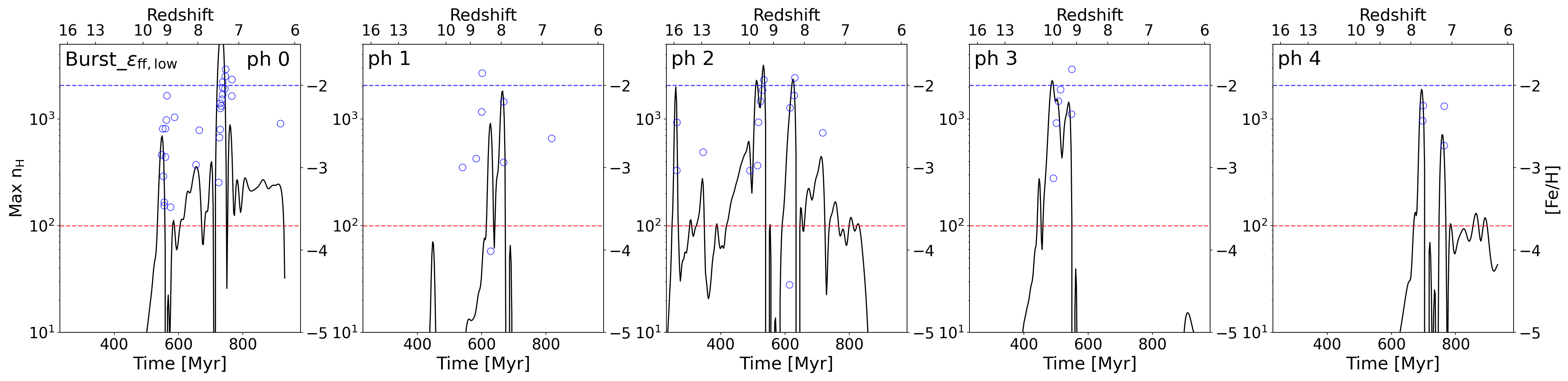}
  \caption{The maximum hydrogen density evolution of {\sc Halo1} with reduced efficiency, as similarly depicted in Fig.~\ref{fig_density_evolution}, demonstrates that the maximum density reaches values exceeding $>10^{3}\rm cm^{-3}$, similar to the results observed in the {\sc Indiv} run.}
   \label{figB1}
\end{figure*}

\begin{figure}
  \centering
  \includegraphics[width=80mm]{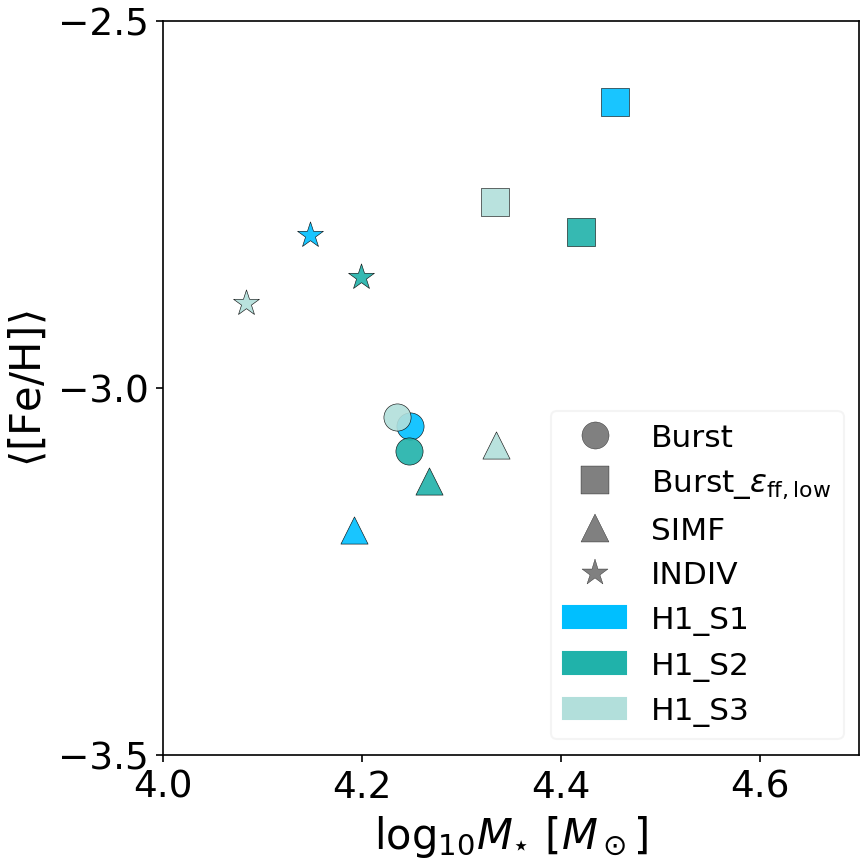}
  \caption{Comparison of the stellar mass–metallicity relation for {\sc Halo1} under different star formation models, which is the same as Fig.~\ref{fig:MZR_all_27} but showing only results for {\sc Halo1}. The {\sc Burst} runs with low efficiency (squares) yield higher stellar masses and metallicities than the fiducial {\sc Burst} cases due to their reduced star formation efficiency.}
  \label{figB2}
\end{figure}








\bsp	
\label{lastpage}
\end{document}